\documentclass{aa}  
\usepackage{natbib}
\usepackage{txfonts}
\usepackage{caption, subcaption}
\usepackage{xcolor}
\usepackage{placeins}
\usepackage{hyperref}
\hypersetup{
    colorlinks=true,           
    linkcolor=blue,             
    citecolor=blue,           
    urlcolor=blue           
}

\usepackage{orcidlink}
\usepackage{booktabs}

\usepackage{graphicx,times}             

\usepackage{amssymb,amsmath}
\usepackage{newtxtext,newtxmath}
\bibpunct{(}{)}{;}{a}{}{,}

\DeclareMathAlphabet\mathbfcal{OMS}{cmsy}{b}{n}
\usepackage{pdflscape}
\usepackage{verbatim}
\usepackage{ulem}
\usepackage{textcomp}

\usepackage[figuresright]{rotating}

\newcommand{\FIG}[1] {Fig. \ref{#1} }

\newcommand{\VEC}[1] {{\rm \bf #1} }
\usepackage{xcolor}

\definecolor{dkblue}{RGB}{54, 86, 169}

\newcommand{\FIGSWITCH}[1]{\ifx \NOTINCLUDEFIG\undefined {#1} \else {\rule{3 in}{2 in}}\fi}

\begin{document}

  \title{The Chinese pulsar timing array data release I
}
    \subtitle{Polarimetry for 56 millisecond pulsars }

\author{
Jiangwei Xu\inst{\ref{doapku},\ref{naoc},\ref{kiaa}}\thanks{E-mail: astroxjw@pku.edu.cn},
Jinchen Jiang\inst{\ref{naoc}}\thanks{E-mail: jiangjinchen@bao.ac.cn},
Heng Xu\inst{\ref{naoc}}\thanks{E-mail: hengxu@bao.ac.cn},
Bojun Wang\inst{\ref{naoc}},
Zihan Xue\inst{\ref{doapku},\ref{kiaa},\ref{naoc}},
Siyuan Chen\inst{\ref{shao}},
Yanjun Guo\inst{\ref{naoc},\ref{klrcas}},
R. Nicolas Caballero,\inst{\ref{hou},\ref{kiaa}},
Kejia Lee\inst{\ref{doapku},\ref{naoc},\ref{ynao},\ref{blaic}},
Jianping Yuan\inst{\ref{xao}},
Yonghua Xu\inst{\ref{ynao}},
Jingbo Wang\inst{\ref{iotlsu}},
Longfei Hao\inst{\ref{ynao}},
Zhixuan Li\inst{\ref{ynao}},
Yuxiang Huang\inst{\ref{ynao}},
Zezhong Xu\inst{\ref{xao}},
Jintao Luo\inst{\ref{ntsc}},
Jinlin Han\inst{\ref{naoc}},
Peng Jiang\inst{\ref{naoc}},
Zhiqiang Shen\inst{\ref{shao}},
Min Wang\inst{\ref{ynao}},
Na Wang\inst{\ref{xao}},
Renxin Xu\inst{\ref{kiaa},\ref{doapku},\ref{klnptpku}},
Xiangping Wu\inst{\ref{naoc}},
Lei Qian\inst{\ref{naoc}},
Youling Yue\inst{\ref{naoc}},
Xin Guan\inst{\ref{naoc}},
Menglin Huang\inst{\ref{naoc}},
Chun Sun\inst{\ref{naoc}}
and Yan Zhu\inst{\ref{naoc}}
}

\institute{
{Department of Astronomy, School of Physics, Peking University, Beijing 100871, P.~R.~China\label{doapku}}\and
{National Astronomical Observatories, Chinese Academy of Sciences, Beijing 100101, P.~R.~China\label{naoc}}\and
{Kavli Institute for Astronomy and Astrophysics, Peking University, Beijing 100871, P.~R.~China\label{kiaa}}\and
{Key Laboratory of Radio Astronomy and Technology, Chinese Academy of Sciences, Beijing 100101, P.~R.~China\label{klrcas}}\and
{Beijing Laser Acceleration Innovation Center, Huairou, Beijing 101400, P.~R.~China\label{blaic}}\and
{Shanghai Astronomical Observatory, Chinese Academy of Sciences, Shanghai 200030, P.~R.~China\label{shao}}\and
{Hellenic Open University, School of Science and Technology, 26335 Patras, Greece\label{hou}}\and
{Xinjiang Astronomical Observatory, Chinese Academy of Sciences, Urumqi 830011, Xinjiang, P.~R.~China\label{xao}}\and
{Yunnan Observatories, Chinese Academy of Sciences, Kunming 650216, Yunnan, P.~R.~China\label{ynao}}\and
{Institute of Optoelectronic Technology, Lishui University, Lishui, Zhejiang, 323000, P.~R.~China\label{iotlsu}}\and
{National Time Service Center, Chinese Academy Of Sciences, Xi'an 710600, P.~R.~China\label{ntsc}}\and
{State Key Laboratory of Nuclear Physics and Technology, School of Physics, Peking University, Beijing 100871, P.~R.~China\label{klnptpku}}\\
}


   \date{Accepted XXX. Received YYY; in original form ZZZ}

\titlerunning{CPTA DR1: Polarization}
\authorrunning{J-W. Xu et al.}

\abstract{We present polarization pulse profiles for 56 millisecond pulsars 
(MSPs) monitored by the Chinese Pulsar Timing Array (CPTA) collaboration using the Five-hundred-meter Aperture Spherical radio Telescope (FAST). The observations centered at 1.25 
GHz with a raw bandwidth of 500 MHz. Due to the high sensitivity ($\sim$16 
K/Jy) of the FAST telescope and our long integration time, the high signal-to-noise 
ratio polarization profiles show features hardly detected before. Among 56 
pulsars, the polarization profiles of PSRs J0406$+$3039, J1327$+$3423, and 
J2022$+$2534 were not previously reported. 80\% of MSPs in the sample show weak components below 3\% of peak flux, 25\% of pulsars show interpulse-like structures, and most pulsars show linear polarization position angle jumps. Six pulsars seem to be emitting for full rotation phase, with another thirteen pulsars being good candidates for such a 360$^\circ$ radiator. We find that the distribution of the polarization percentage in our sample is compatible with the normal pulsar distribution. Our detailed evaluation of the MSP polarization properties suggests that the wave propagation effects in the pulsar magnetosphere are important in shaping the MSP polarization pulse profiles.
}

   \maketitle



\section{Introduction}
\label{sec:intro}
Unlike normal pulsars, millisecond pulsars (MSPs) have significantly
shorter spin periods and lower values of period derivatives. Some MSPs can
work as precise cosmic clocks comparable to atomic clocks in stability
\citep{1991_Taylor,2012_hobbs,2018_verbiest,2020MNRAS_Hobbs}. With the
high timing precision, some MSPs have been utilized to form the pulsar
timing array \citep[PTA;][]{FosterBacker1990} in order to search for
nanohertz gravitational waves \citep[GWs;][]{Sazhin1978,Detweiler1979}.
Recently, several PTAs have presented evidence of nanohertz GW signals
\citep{2023ApJ_Agazie,2023A&A_Antoniadis,2023ApJ_Reardon,2023RAA_Heng}.
To facilitate the detection of GWs, MSPs are monitored regularly for
long time spans, producing extensive datasets.  The obtained polarization
profiles have high signal-to-noise ratios (S/Ns). Such a high-S/N
dataset is beneficial to study the emission properties of MSPs, since
the delicate structures and very weak components can be detected. On the other hand, long-term observation activities can provide valuable insights into
the stability of MSP profiles \citep{Xu2021Atel}, allowing for the
study of the dynamic evolution of MSP magnetospheres.  In addition,
a careful polarization calibration procedure is necessary to minimize
the instrumental effects on pulsar timing. Through a comparison
with the polarization profiles of previous studies, we can perform cross-validation on
the calibration and data processing procedures.

Both MSPs and normal pulsars show similar profile
complexity~\citep{1998ApJ_Kramer}, spectral indexes, and polarization
properties \citep{1999_kramer, 2011_yan, 2015_dai,Gentile_2018,
spiewak_2022,Wahl_2022,2023_Gitika,2024MNRAS_Karastergiou}. Furthermore,
MSPs exhibit similar polarization profile characteristics compared to normal pulsars, which includes linear and circular
polarization components, linear polarization angle swing, interpulse,
orthogonal polarization mode jumps (OPMs), and sense reversal of circular
polarization \citep{1991ApJ_Thorsett,Xilouris_1998,2004_manchester}.
These observational facts indicate that the radiation mechanisms are same
for MSPs and normal pulsars.

On the other hand, the light cylinder ($R_{\rm LC}=P_0c/2\pi$), where
the co-rotation velocity is equal to the light speed, is of smaller
radii for MSPs by a factor of 10 to 1000 compared to normal pulsars.
Consequently, the size of MSP magnetosphere is much smaller. It will inevitably affect the radiation geometry and propagation of radio waves in the magnetosphere \citep{2020MNRAS.498.5003J}. 
In fact, observations had shown that some polarization properties of MSPs differ significantly
from those of normal pulsars. As was
expected geometrically \citep{1970komesaroff}, MSPs have much larger duty cycles and pulse widths. Unlike for normal pulsars, the profile widths and separations between
pulse components in MSPs remain nearly constant across frequencies,
illustrating more compact emission regions \citep{1999_kramer,2015_dai}. 
In addition, a large portion of MSPs contain interpulse in their profiles, which is in contrast to the normal pulsar population \citep{1991ApJ_Thorsett,2004_manchester,2011_yan}. The PA slopes of MSPs are much shallower. Moreover, the linear polarization position angle (PA) curves of most MSPs deviate from the rotating vector model
\citep[RVM;][]{Radhakrishnan1969,1970komesaroff,Xilouris_1998,2004_manchester,Wahl_2022}, which is
generally applicable for normal pulsars\citep{2023_johnston}. The irregular PA curves indicate a much more complex (non-dipolar) magnetic filed structure in MSPs \citep{2011MNRAS.415.1703C}. These differences indicate that the magnetospheres of MSPs are not simply a scaled-down version of that in normal pulsars.

To understand the radiation mechanism and magnetosphere of MSPs better, qualitative polarization measurements of MSPs are essential. 
Polarization properties of normal pulsars have been
extensively studied (e.g., see \cite{2018_Johnston, han2023}). In this paper,
we focus on the polarization studies for MSPs. High-quality polarization profiles for 56 MSPs are presented here.  The data is from the Chinese Pulsar
Timing Array \citep[CPTA;][]{2016ASPC_Lee} observation carried out at FAST \citep{Jiang_2019}.
The large sample size and high-S/N data
allow for a detailed investigation of the polarization properties
of MSPs. In Sect.~\ref{sec:obs}, we describe the observation activities. The
detailed calibration pipeline and data processing are explained in
Sect.~\ref{sec:ana}. In Sects.~\ref{sec:result} and \ref{sec:disc}, we summarize and discuss the properties of MSP polarization. The conclusions are made in Sect.~\ref{sec:conclu}.

\section{Observations}
\label{sec:obs}

We analyzed data for 56 pulsars from the first data release of
CPTA~\citep{2023RAA_Heng}. PSR~J0218$+$4232 was excluded from the polarization studies 
of this paper, because it shows radiation at all rotation phases and the
baseline of pulse profile cannot be uniquely determined. Moreover, the profile presents obvious evolution across different orbital phases. Studies of PSR~J0218$+$4232  will be published elsewhere. Although PSR~J1327$+$3423 is a partially recycled pulsar \citep{2023ApJ_Fiore} rather than a canonical MSP, we included it in the CPTA pulsar list due to its high timing precision and timing stability.

Our data were collected using the central beam of the 19-beam
receiver~\citep{dunning2017design}. The sky coverage of FAST is from
$-14^\circ$ to 66$^\circ$ on declination. The sensitivity and system noise
temperature stay constant at $\sim$ 16 K/Jy and $\sim$ 19 K, respectively,
for the central beam within a zenith angle of 26.4$^{\circ}$~\citep{Jiang_2020}. When the zenith angle of the target is
larger than 26.4$^\circ$, a back illumination strategy is conducted to
avoid ground emission~\citep{Jin2013}. In this case, the telescope gain and system noise temperature 
deteriorate to 11.5~K/Jy and 27~K, when the zenith angle reaches the maximum
of 40$^{\circ}$~\citep{Jiang_2020}. We compare
the difference in polarimetry for small and large zenith angles in
Appendix~\ref{sec:compara}. It induces a difference of less than 0.2\%
in polarization profiles. The effect can
thus be safely neglected for most of the application. Nonetheless, the
polarization pulse profiles in this paper are mainly from data collected
under small zenith angles (<26.4$^\circ$). For about one quarter of the
56 pulsars, we only observed with back illumination mode; that is, for
PSR~J0034$-$0534, J0613$-$0200, J1012$+$5307, J1024$-$0719, J1643$-$1224,
J1744$-$1134, J1832$-$0836, J1843$-$1113, J1911$-$1114, J1918$-$0642,
J2010$-$1323, J2145$-$0750, and J2150$-$0326.

The observation cadence for
all pulsars is approximately once per two weeks, determined by the amount of FAST observing time
allocated for the CPTA project. One exception is PSR J1713$+$0747, which was observed weekly due to its high timing precision. 
The digital backend at FAST saves data in the filterbank format; that is, it
records the signal power as a function of frequency and time. The
observation covers the frequency range of 1-1.5 GHz with a spectral
resolution of 122.07 kHz and takes a time resolution of 49.152 $\mu$s. We dedispersed and folded the filterbank data using \textsc{DSPSR} \citep{2011PASA_Straten} to form 20-minute sub-integration pulsar data. For pulsars in compact binaries, the length of sub-integration time was reduced.
The total observing time and number of observation epochs for all pulsars are listed in Table~\ref{tab:small_summary}. More details of our data collection scheme will be explained by Xu et al. 2024 (in prep.).  

For each observation, we also recorded periodic on-off noise diode signal for the polarization calibration purposes. Before March 2021, we recorded one-to-two-minute noise signals with periods of
1 or 2 second(s) before or after each observation. After that, we changed
the noise injection scheme and the noise was injected to cover the entire observation time span. We note that it can help correct the drift of the electronic system on timescales shorter than 20 minutes. For most pulsars, the pulsar signal will not be affected by this scheme and we can separate the noise calibrator signal from the pulsar signal, thanks to three reasons: 1) the noise level is a factor of 20 lower than the system noise; 2) the noise period does not align with the pulsar period; 3) the noise signal has no dispersion. 
For certain low dispersion measure (DM) pulsars, such as PSR~J1327$+$3423, the noise diode signal cannot be well separated from the pulsar signal, but it has a negligible effect due to the first reason mentioned above.

\section{Data analysis}
\label{sec:ana} 

\subsection{Polarization calibration}
\label{sec:polcal}
The polarization is described by the Stokes parameters,
\begin{equation}
\mathcal{S}=\left[\begin{array}{c}I \\ Q \\ U \\ V\end{array}\right]\,,
\end{equation}
where $I$ is the intensity flux, $Q$ and $U$ denote the linear polarization, and $V$ is the circular polarization. Here, we take the PSR/IEEE convention where the left circular polarization is positive~\citep{straten2010}. Based on the Stokes parameter, we can define the linear polarization position angle ($\Psi$) and the ellipticity angle ($\chi$) as
\begin{equation}
\begin{aligned}
&\tan2\Psi = \frac{U}{Q}\,,\\
&\sin2\chi = \frac{V}{\sqrt{Q^2+U^2+V^2}}\,.
\end{aligned}
\end{equation}

The imperfectness of receiver will contaminate the polarization measurements. In order to obtain the intrinsic Stokes parameters, corrections for the instrumental effects are necessary. All the linear effects affecting Stokes parameters can be empirically expressed as ~\citep{lorimer2012}\footnote{Although the expressions are “empirical” expressions, they are mathematically identical to the polar decomposition \citep{1996A&AS_Hamaker,2000A&AS_Hamaker,2004ApJS_Straten}},
\begin{equation}
\mathcal{S}_{\mathrm{obs}}=\mathcal{M}_{\mathrm{Amp}} \times \mathcal{M}_{\mathrm{CC}} \times \mathcal{M}_{\mathrm{PA}}\times\mathcal{S}_{\mathrm{int}}\,,
\end{equation}
which means that the intrinsic polarization ($\mathcal{S}_{\mathrm{int}}$) is firstly affected by 
effects described by the M\"uller matrix, $\mathcal{M}_{\mathrm{PA}}$, then $\mathcal{M}_{\mathrm{CC}}$ and $\mathcal{M}_{\mathrm{Amp}}$.

The first effect, represented by the M\"uller matrix, $\mathcal{M}_{\mathrm{PA}}$, is caused by feed rotation, which induces a parallactic angle between the feed orientation and the plane of polarization. For FAST, we do not need to perform such a correction, since it is corrected by mechanically rotating the 19-beam receiver~\citep{Jiang_2020} such that one feed always aligns with the northern direction.

The second effect, $\mathcal{M}_{\mathrm{CC}}$,  comes from polarization leakage caused by the cross-coupling between the x and y feeds, such as the non-orthogonality of the instrument polarization basis. It has been shown that such leakage for the central beam of the 19-beam feed is only at a level of $\sim$ 0.034 $\%$ \citep{ching2022}. Since such an effect is negligible for the current work, we leave it for future work. 

The third effect, $\mathcal{M}_{\mathrm{Amp}}$, is from the differential gain and phase imbalance.
In general, the electric signals of the two polarization feeds pass through two separate signal chains with different amplifiers. There will inevitably be differences in time delays and amplifier gains between the two signal chains. Such an effect is described by the following M\"uller matrix~\citep{2001heiles}:
\begin{equation}
\mathcal{M}_{\text {Amp }}=\left[\begin{array}{cccc}1 & \Delta G / 2 & 0 & 0 \\ \Delta G / 2 & 1 & 0 & 0 \\ 0 & 0 & \cos \Delta \phi & -\sin \Delta \phi \\ 0 & 0 & \sin \Delta \phi & \cos \Delta \phi\end{array}\right]\,,
\end{equation}
where  $\Delta G$ and $\Delta\phi$ are the differential gain and phase, respectively. 

\begin{figure*}[ht]
    \centering
    \begin{minipage}[t]{0.5\columnwidth}
    \centering
    \includegraphics[width=0.9\columnwidth]{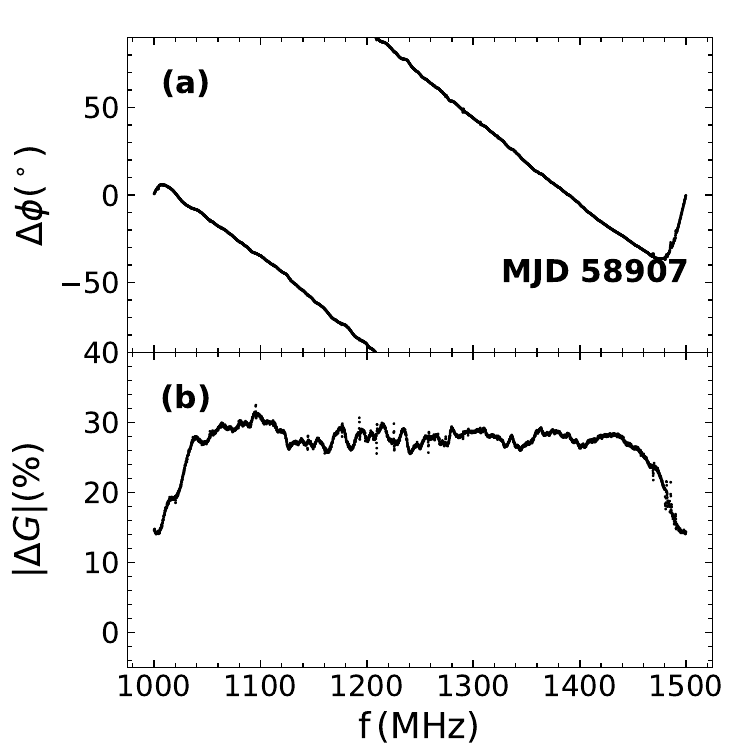}   
    \end{minipage}
    \begin{minipage}[t]{0.5\columnwidth}
    \centering  
    \includegraphics[width=0.9\columnwidth]{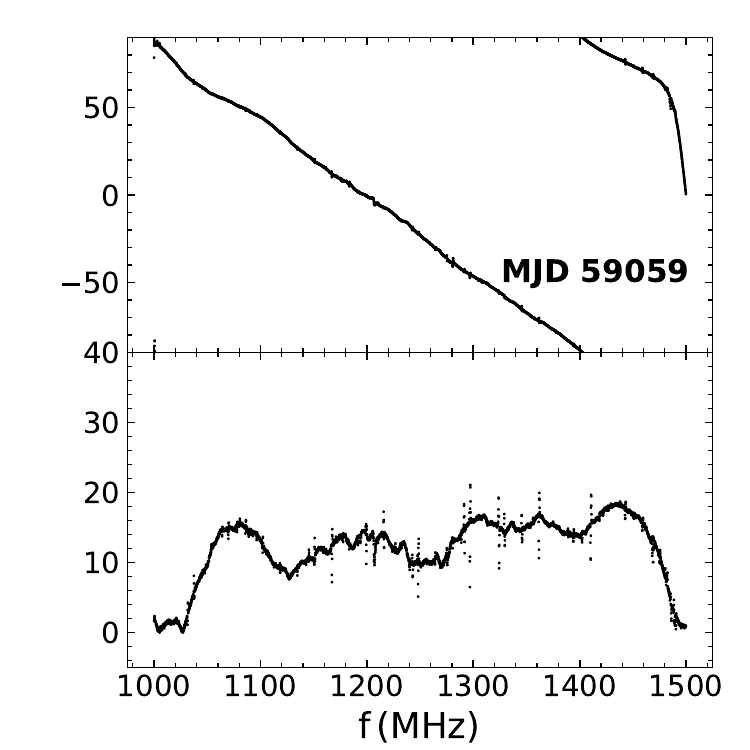}   
    \end{minipage}
    \begin{minipage}[t]{0.5\columnwidth}
    \centering
    \includegraphics[width=0.9\columnwidth]{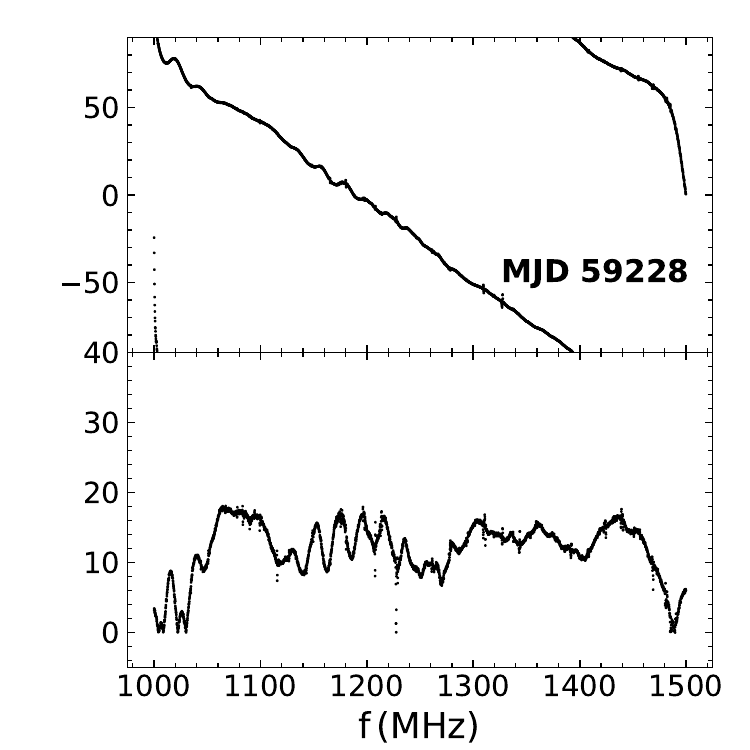}
    \end{minipage}
    \begin{minipage}[t]{0.5\columnwidth}
    \centering
    \includegraphics[width=0.9\columnwidth]{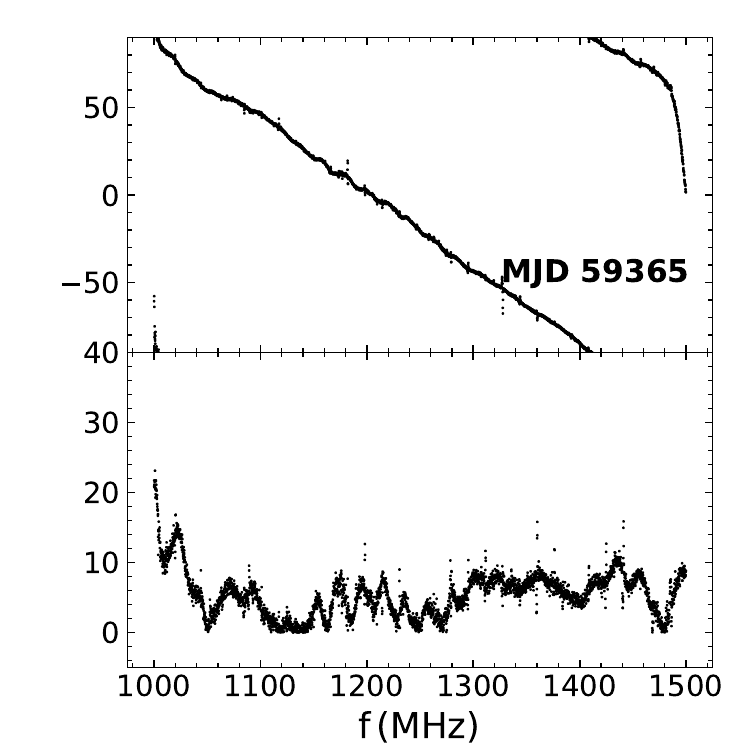}
    \end{minipage}\\
    \begin{minipage}[t]{0.5\columnwidth}
    \centering
    \includegraphics[width=0.9\columnwidth]{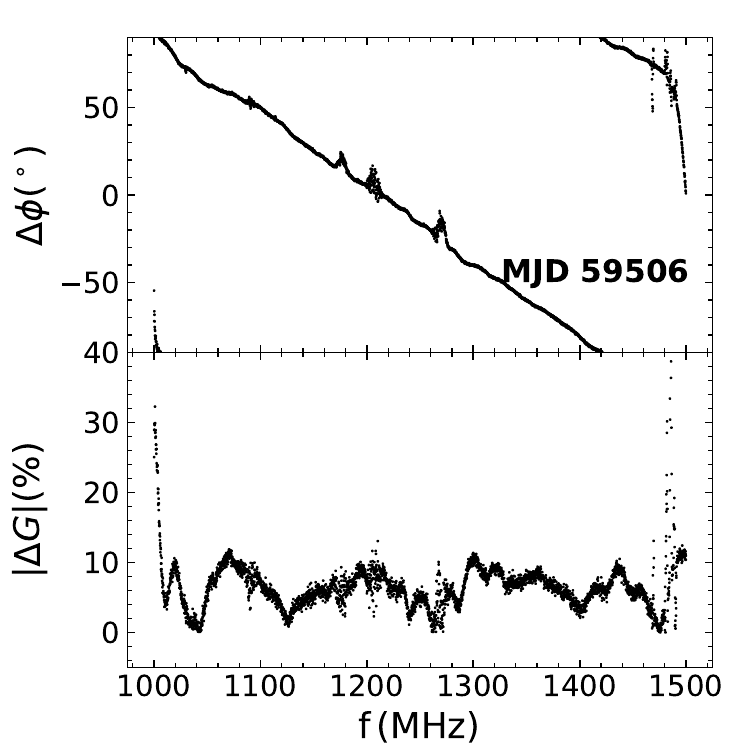}
    \end{minipage}
    \begin{minipage}[t]{0.5\columnwidth}
    \centering
    \includegraphics[width=0.9\columnwidth]{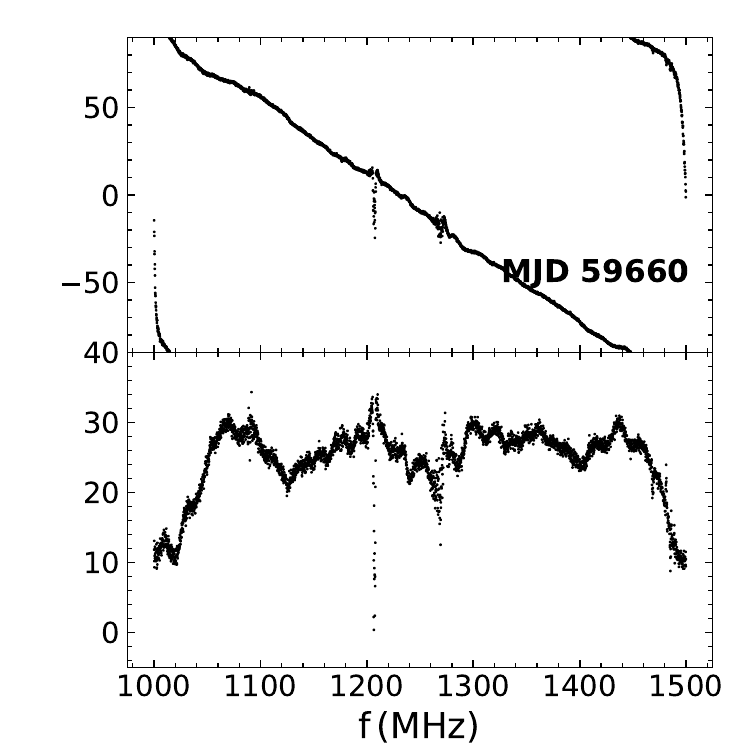}
    \end{minipage}
    \begin{minipage}[t]{0.5\columnwidth}
    \centering
    \includegraphics[width=0.9\columnwidth]{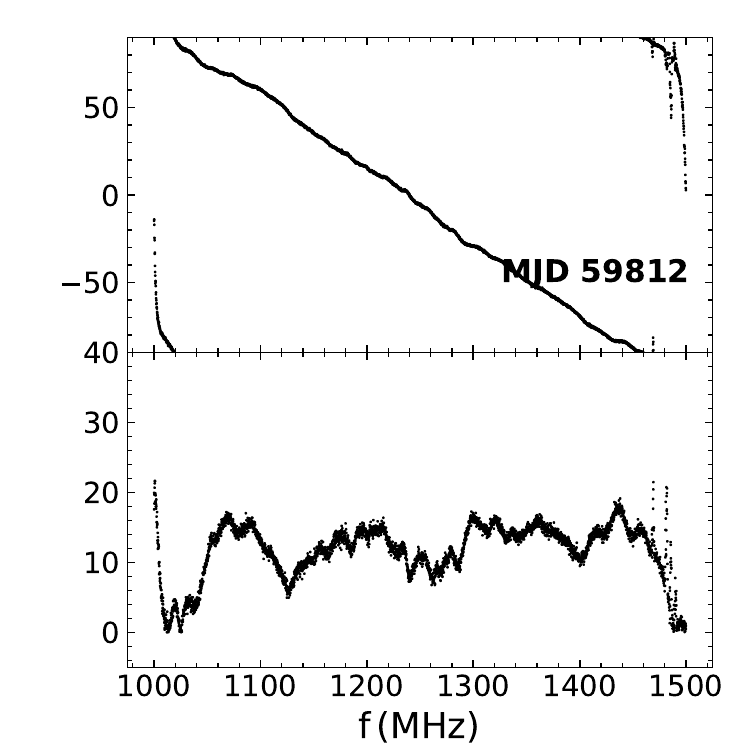}
    \end{minipage}
    \begin{minipage}[t]{0.5\columnwidth}
    \centering
    \includegraphics[width=0.9\columnwidth]{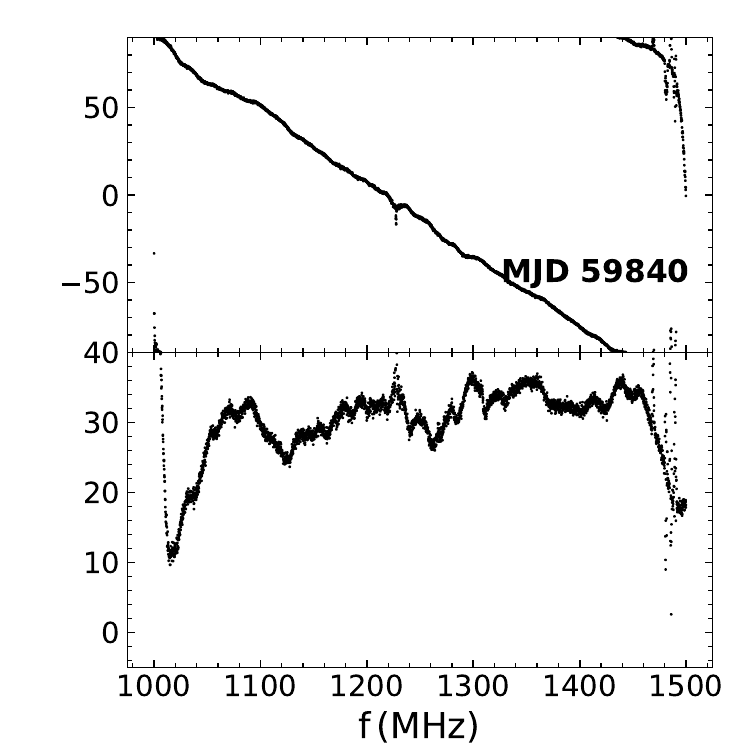}
    \end{minipage}
    \caption{Differential gains and phases as functions of observing frequencies at eight epochs for PSR J0023$+$0923. The differential phase and gains between the two amplifier chains are shown in panels (a) and (b), respectively.
    \label{fig:muller}}
\end{figure*}

As the leakage term can be neglected, we adopted the single axis model for polarization calibration using the noise signal equally injected into the two feeds~\citep{2004PASA_Hotan}, where the two feeds see the noise as of the same amplitude and phase. 
As an example, the derived M\"uller matrix elements at different epochs and frequencies through the package \textsc{PSRCHIVE} \citep{2004PASA_Hotan} are shown in \FIG{fig:muller}. We note that the differential gain of the system fluctuates by approximately 20\%, while the differential phase after July 2020 (MJD 59059) is stable up to only a few degrees. 
One can also see that the $\Delta G$ and $\Delta \phi$ change more rapidly as functions of frequency at the band edges; that is, for frequencies close to 1000 MHz or 1500 MHz. Here, the feed loses sensitivity and has a poor response. Therefore, we removed the 20 MHz band edge on each side of the band pass to obtain reliable polarization data for the subsequent analysis.

Since all pulsar observations are accompanied by calibration observations, we applied the calibration solution to the corresponding pulsar observations for correction of instrumental effects through \textsc{PSRCHIVE} \citep{2004PASA_Hotan}. The next step was to correct the Faraday rotation effects induced by the interstellar medium (ISM) and Galactic magnetic fields. 

\subsection{Faraday rotation correction}
\label{sec:rmproc}

When the radio waves propagate through the magnetoionized ISM, the linear polarization PA rotates by $\Delta\Psi(\lambda)={\rm RM}\lambda^2$, where $\lambda$ is the wavelength of radio wave and RM is the integral of parallel magnetic field strength and electron density along the line of sight: 
\begin{equation}
    {\rm RM}=\frac{e^{3} }{2 \pi m_{e}^{2} c^{4}} \int_{0}^{d} n_{e}(l) B_{\|}(l) d l\,,
\end{equation}
where $d$ is the distance from the pulsar to the earth, $n_e$ is the free electron density, and $B_{\|}$ is the magnetic field strength projected to the line of sight. Constants $e$, $m_{\rm e}$, and $c$ are electron charge, mass, and light speed, respectively. 

We used two different methods to measure the RM for cross-checking. 
The first method derived the RM value through fitting the profiles of Stokes $Q$ and $U$ across frequencies~\citep{straten2012,desvignes2019}, where the nested sampling package MULTINEST~\citep{Feroz2009} was used to infer the model parameters. The second method was the generalized RM synthesis, which searches across a range of RMs to “de-rotate” the wrapped linear polarization. The best RM was found by maximizing the total linear polarization given by~\citep{brentjens2005,2015MNRAS_Schnitzeler},
\begin{equation}
    \VEC{p}(\rm RM)= \int^{\nu_{\rm high}}_{\nu_{\rm low}}{\VEC{L}(\nu)} \VEC{v}({\rm RM}, \nu)d \nu\,,
\end{equation}
where $\nu$ is the frequency, and $\VEC{L}(\nu)$ is the complex linear polarization spectrum in the frequency range of $\nu_{\rm low}$ to $\nu_{\rm high}$, defined as $\VEC{L}(\nu)= Q(\nu)+iU(\nu)$. The de-rotation vector, $\VEC{v}({\rm RM}, \nu)$, is defined in \citet{2015MNRAS_Schnitzeler}. The differences between the RM values derived from the Bayesian $Q$-$U$ fitting and the generalized RM synthesis methods are negligible (see Appendix~\ref{sec:comprm}). In the following parts of the paper, we quote RM values from the $Q$-$U$ fitting method. 

It is necessary to clarify that, for each of the above two methods, there are two ways to perform the data analysis. One uses the phase-integrated Stokes $Q$ and $U$ at each frequency, and the other one firstly derives the RM at each phase bin and then performs averaging along the phase. The former way is less sensitive to the noise due to the phase integration, but it will be affected by the depolarization due to intrinsic PA swing in pulse profiles. The latter one avoids the depolarization effect, but it is more susceptible to noise. Since MSPs generally present violent PA swings and illusive RM evolution across the pulse phase (see \citet{Ilie2018,2015_dai} and later discussion), we used the second method to estimate the RM. 

 Step (1): We computed the Faraday spectra for each phase bin to avoid the depolarization effect and summed them up to give a preliminary RM value for each observation. 

\begin{figure}
    \centering
    \includegraphics[width=0.6\columnwidth]{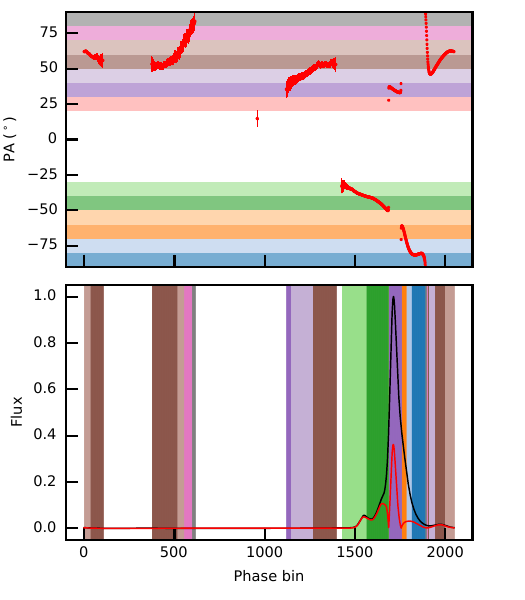}
    \caption{Illustration for the pulse phases slicing scheme for the RM measurement.  Upper panel: PA curve and the uniform interval in PA. Bottom panel: Corresponding phase interval determined by the PA interval, marked with the same color as the upper panel. The solid black and red curves are the total intensity and linear polarization of pulse profile, respectively. \label{fig:slice}}
\end{figure}

 Step (2): We corrected the Faraday rotation effect and produced the integrated polarization profile by summing up all observations. In this process, observations with a maximal phase-resolved S/N lower than 50 were discarded. In practice, the fractions of discarded observations are much less than 10 $\%$.  
In order to further reduce effects from violent PA variations, the final PA curves were sliced into intervals of $10^\circ$, which were then used to divide pulse phases for each observation. An example of such phase division is shown in \FIG{fig:slice}.

 Step (3): After the data were sliced into many phase segments, the corresponding RM value of each phase segment was derived separately through the $Q$-$U$ fitting method. We adopted the weighted average as the final RM value for each observation, where the statistical uncertainty of RM in each interval was taken as the weight. Outliers with differences larger than $30\,\mathrm{rad\cdot m^{-2}}$ were removed in the averaging process.

\begin{figure*}[ht]
\centering
\begin{minipage}[t]{0.5\columnwidth}
    \centering
    \includegraphics[width=1.0\columnwidth]{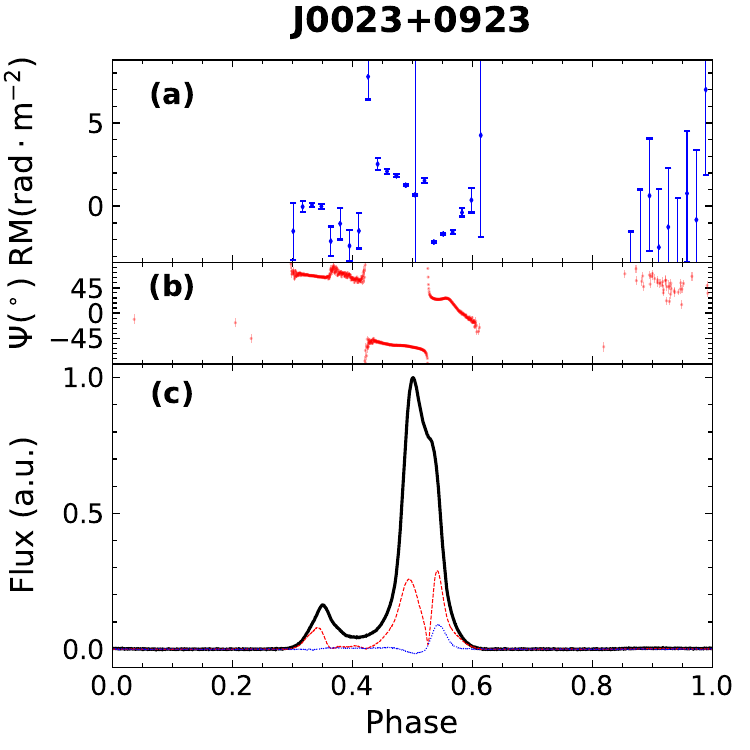}   
    \end{minipage}
    \begin{minipage}[t]{0.5\columnwidth}
    \centering
    \includegraphics[width=1.0\columnwidth]{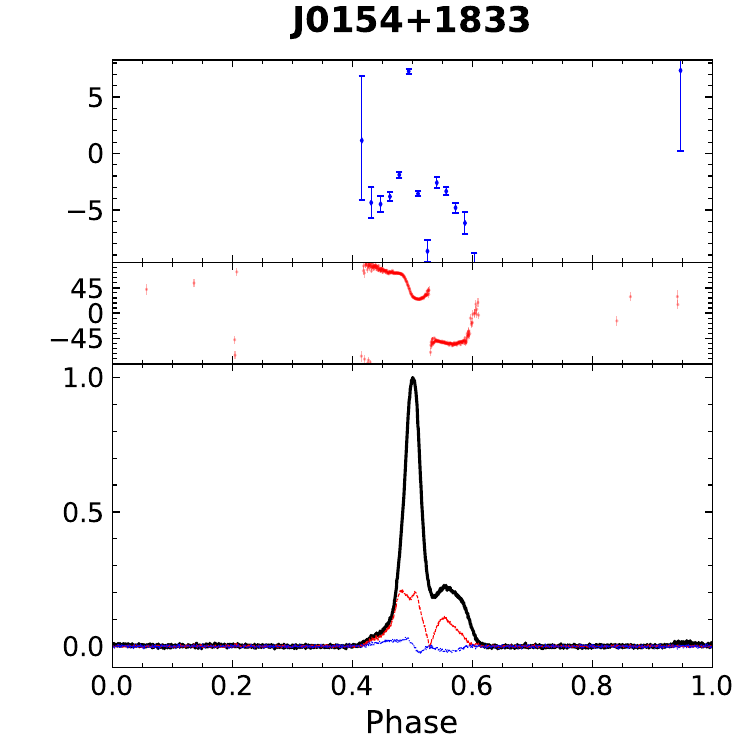}   
    \end{minipage}
    \begin{minipage}[t]{0.5\columnwidth}
    \centering
    \includegraphics[width=1.0\columnwidth]{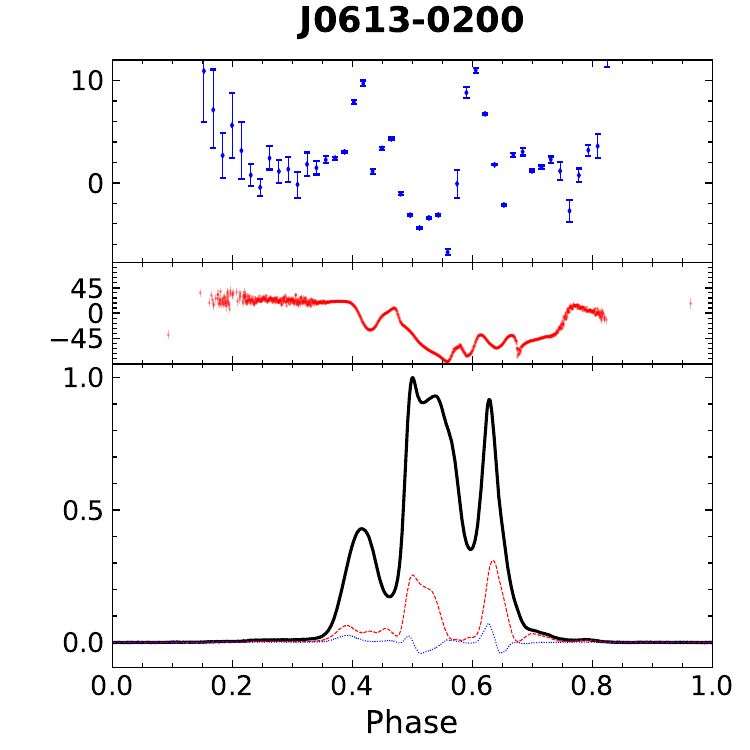}   
    \end{minipage}
    \begin{minipage}[t]{0.5\columnwidth}
    \centering
    \includegraphics[width=1.0\columnwidth]{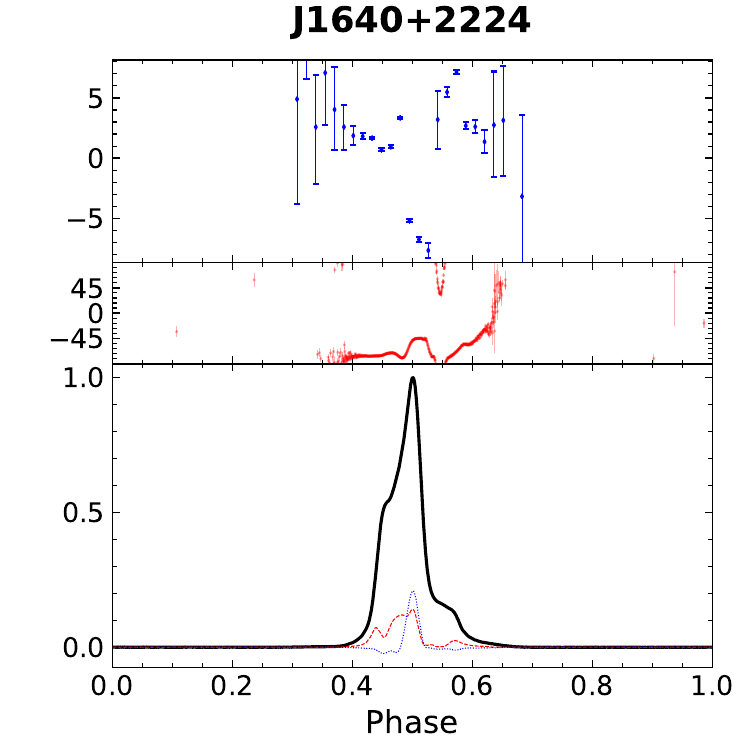}   
    \end{minipage}\\
    \begin{minipage}[t]{0.5\columnwidth}
    \centering
    \includegraphics[width=1.0\columnwidth]{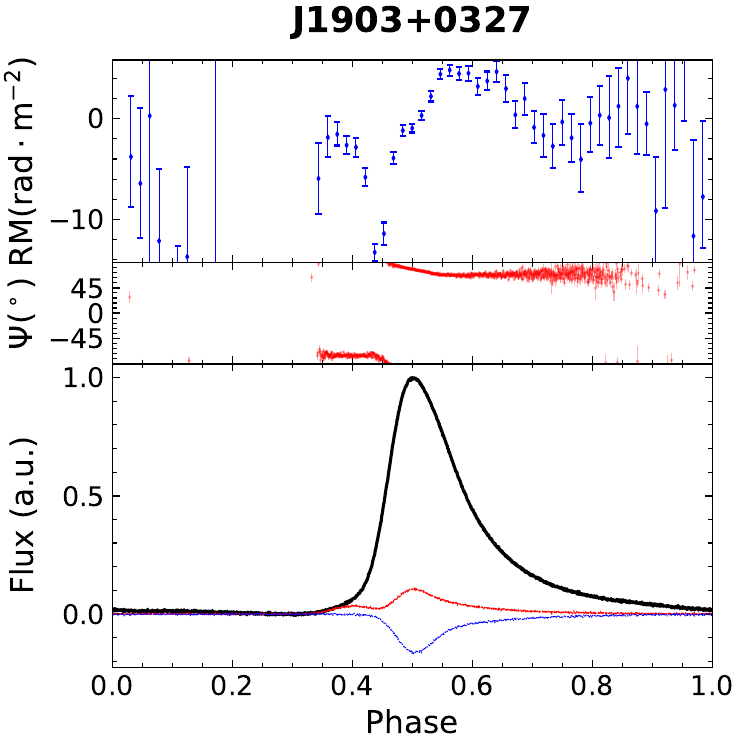}  
    \end{minipage}
    \begin{minipage}[t]{0.5\columnwidth}
    \centering
    \includegraphics[width=1.0\columnwidth]{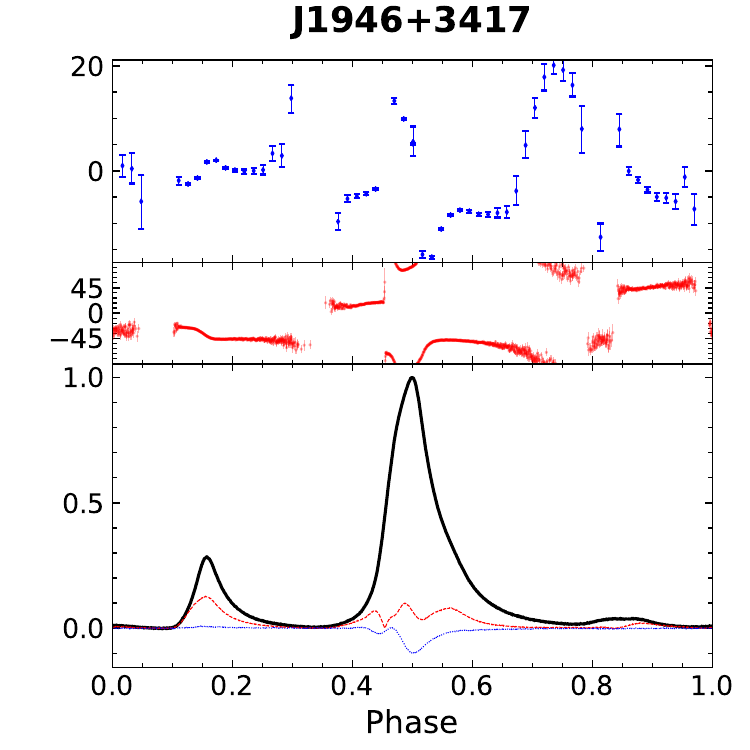}   
    \end{minipage}
    \begin{minipage}[t]{0.5\columnwidth}
    \centering
    \includegraphics[width=1.0\columnwidth]{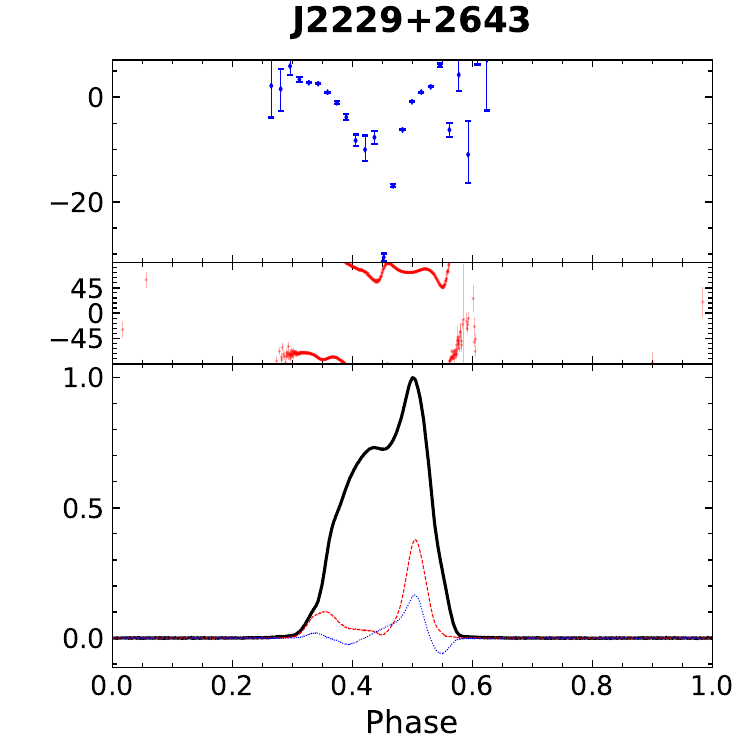}  
    \end{minipage}
    \begin{minipage}[t]{0.5\columnwidth}
    \centering
    \includegraphics[width=1.0\columnwidth]{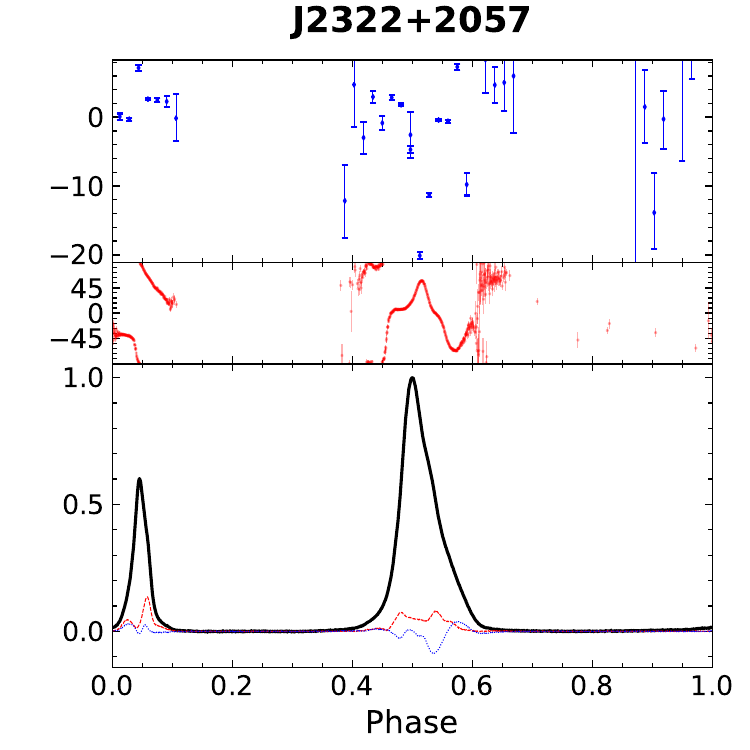}   
    \end{minipage}
    \caption{Apparent RM variations along pulse phase detected for eight pulsars. Panel (a) displays the RM variations after subtracting the average RM value. Panel (b) PA curves. Panel (c) polarization pulse profiles, where solid black, dashed red, and dotted blue curves denote the total intensity, linear intensity, and circular polarization intensity, respectively. The profiles were normalized by the maximal intensity and the phase of the intensity peak was set at 0.5.
    \label{fig:deviation}}
\end{figure*}

 Step (4): After correcting the Faraday rotation for all observations, the integrated profiles and averaged RM values were obtained for each pulsar. In order to check whether residual RM exists in the final profiles, we computed the phase-resolved RM by applying the $Q$-$U$ fitting to the data. For most pulsars, we detected few residual RM variations across the pulse phase. Some pulsars, however, show significant residual RM variation across the pulse phase, as is shown in \FIG{fig:deviation}. Such phase-resolved RM variation is likely to be illusive, as we shall discuss in Sect.~\ref{sec:phrm}.

We also checked our RM values (after correcting the Earth magnetosphere contribution) with previously published results \citep{Gentile_2018,Wahl_2022,spiewak_2022,han2023}. Our results are compatible with the published values. Here, the software package \textsc{ionFR}~\citep{Sotomayor2013} was used to compute the ionospheric corrections, where we used total electron
content CODE maps provided by GPS monitoring and interpolated to the observing epoch. The Earth magnetic field model we used is the 13th generation release of the International Geomagnetic Reference Field \citep{2021EP&S_Alken}. 

\subsection{Weisberg correction}
\label{sec:weisberg}
The linear ($L$) and total polarization ($P$) intensities are defined with the Stokes parameters:
\begin{eqnarray}
L&=&\sqrt{Q^2+U^2} \,,\\
P&=&\sqrt{Q^2+U^2+V^2}\,.
\end{eqnarray}
The $L$ and $P$ are always positive, and thus suffer from the statistical bias that their expected values are shifted due to the variance of $Q,U$, and $V$. To correct the bias, we 
performed the Weisberg correction (see \citet{Everett_2001} for the references therein and also the generalized version by \citet{Jiang_2022}); that is, $L$ and $P$ were computed with
\begin{equation}
\begin{aligned}
& L = \begin{cases}
   &\sqrt{L^2-\frac{\sum_{i,j=1,i\neq j}^2S_i^2\sigma_{S_j}^2}{L^2}},\,\forall L\geq3\,\sigma_L\,,\\
   &L,\,\textrm{otherwise}\\
\end{cases}\\
& P = \begin{cases}
   &\sqrt{P^2-\frac{\sum_{i,j=1,i\neq j}^3S_i^2\sigma_{S_j}^2}{P^2}},\,\forall P\geq3\, \sigma_P\,,\\
   &P,\,\textrm{otherwise}\\
\end{cases}\\
\end{aligned}
\end{equation}
where $S_i$ are the Stokes parameters and $S_1=Q$, $S_2=U$, and $S_3=V$. $\sigma_L$ and $\sigma_P$ are the corresponding baseline noise. We chose the threshold of 3-$\sigma$ to be compatible with later analysis. 

\section{Results}
\label{sec:result}

The final polarization profiles of CPTA pulsars are present in Fig.~\ref{fig:cpta_pol1}\footnote{The figures are available at \href{https://doi.org/10.5281/zenodo.14801349}{https://doi.org/10.5281/zenodo.14801349}}. To the authors' knowledge, the polarization profiles of three recently discovered pulsars -- namely, PSRs J0406$+$3039 \citep{2024ApJ_McEwen}, J1327$+$3423 \citep{2023ApJ_Fiore}, and J2022$+$2534 \citep{2023ApJ_Swiggum} in the L band (1-2 GHz) -- are shown for the first time. We computed the degrees of linear, circular, total polarized, and pulse width for each pulsar, which is summarized in Table~\ref{tab:tot_pol}. CPTA MSPs generally possess a moderate degree of linear polarization and a low circular polarization degree. Seven pulsars have a total fractional polarization higher than 50$\%$, where PSR~J1744$-$1134 has the highest total polarization degree ($\sim 90\%$) and PSR~J1327$+$3423 features the highest circular polarization of $\sim 27\%$. 

No correlation is found between the degree of polarization and the pulsar period, period derivative, and pulse width. Within them, the period and $|V|/I$ show the strongest correlation of 0.45 with the p value of $4.81\times10^{-4}$. However, with three long-period (>10 ms) pulsars excluded (PSRs J0621$+$1002, J1327$+$3423, and J2145$-$0750), the coefficient is reduced to 0.19. It is possible that the polarization properties of MSPs are independent of pulsar rotation parameters, or we are limited by the size of our MSP samples.

\begin{table}
        \centering
        \caption{Pulse width and polarization degree of CPTA pulsars.  The total, linear, circular, and absolute circular degree of polarization are $\Pi_{\rm P}\equiv\frac{\langle  P\rangle}{\langle I\rangle}$, $\Pi_{\rm L}\equiv\frac{\langle  L\rangle}{\langle I\rangle}$, $\Pi_{\rm V}\equiv\frac{\langle V\rangle}{\langle I\rangle}$, and $\Pi_{\rm |V|}\equiv\frac{\langle  |V|\rangle}{\langle I\rangle}$, where the phase average is denoted with $\langle \rangle$. The detectable pulse width and the effective width are $W_{\rm det}$ and $W_{\rm eff}$, respectively.
        \label{tab:tot_pol}}
        \begin{tabular}{ccccccc} 
                \hline\hline
                Pulsar& $W_{\rm det}$ & $W_{\rm eff}$&$\Pi_{\rm P}$&$\Pi_{\rm L}$&$\Pi_{\rm V}$&$\Pi_{\rm |V|}$\\&\%&\%&\%&\%&\%&\%\\
                \hline
J0023$+$0923 & 40.1 & 7.8 & 30.6  & 30.0  & 3.3  & 4.4  \\
J0030$+$0451 & 82.1 & 14.3 & 33.4  & 33.1  & 0.8  & 3.1  \\
J0034$-$0534 & 71.9 & 25.5 & 17.9  & 9.4  & -12.1  & 12.1  \\
J0154$+$1833 & 22.6 & 5.2 & 32.2  & 31.8  & 0.7  & 4.1  \\
J0340$+$4130 & 81.2 & 9.3 & 19.4  & 18.9  & -2.5  & 2.9  \\
J0406$+$3039 & 57.5 & 10.3 & 33.9  & 32.0  & -5.4  & 8.8  \\
J0509$+$0856 & 96.7 & 28.2 & 35.3  & 34.0  & 1.6  & 4.5  \\
J0605$+$3757 & 45.0 & 9.6 & 46.5  & 46.6  & -0.3  & 0.4  \\
J0613$-$0200 & 66.5 & 16.6 & 19.9  & 19.6  & 0.6  & 3.1  \\
J0621$+$1002 & 57.4 & 5.1 & 28.1  & 21.3  & -16.2  & 16.3  \\
J0636$+$5128 & 28.2 & 5.6 & 40.3  & 38.1  & -2.2  & 6.3  \\
J0645$+$5158 & 38.7 & 2.4 & 22.0  & 20.0  & 0.9  & 6.7  \\
J0732$+$2314 & 82.3 & 28.6 & 33.2  & 23.6  & 17.5  & 21.1  \\
J0751$+$1807 & 82.3 & 11.2 & 30.5  & 27.3  & -9.3  & 11.6  \\
J0824$+$0028 & 66.8 & 13.5 & 47.6  & 45.1  & 6.5  & 11.3  \\
J1012$+$5307 & 94.2 & 15.7 & 59.2  & 58.6  & 0.1  & 6.6  \\
J1024$-$0719 & 86.3 & 13.0 & 62.4  & 61.9  & 1.0  & 3.6  \\
J1327$+$3423 & 18.7 & 2.6 & 44.2  & 30.5  & 27.2  & 27.6  \\
J1453$+$1902 & 86.2 & 9.1 & 62.9  & 62.7  & -1.7  & 2.9  \\
J1630$+$3734 & 79.5 & 11.8 & 41.3  & 39.8  & 0.1  & 4.3  \\
J1640$+$2224 & 36.0 & 7.0 & 16.6  & 12.9  & 4.4  & 8.5  \\
J1643$-$1224 & 79.2 & 10.2 & 22.5  & 17.1  & -2.6  & 12.1  \\
J1710$+$4923 & 94.7 & 16.5 & 19.2  & 15.4  & -1.1  & 7.5  \\
J1713$+$0747 & 91.8 & 4.3 & 31.7  & 31.1  & -1.6  & 3.0  \\
J1738$+$0333 & 41.6 & 6.5 & 25.2  & 24.7  & -1.8  & 3.9  \\
J1741$+$1351 & 41.4 & 4.1 & 21.7  & 20.9  & 2.5  & 4.3  \\
J1744$-$1134 & 47.0 & 4.1 & 88.8  & 88.8  & -1.9  & 1.9  \\
J1745$+$1017 & 75.5 & 12.7 & 53.9  & 51.8  & -9.7  & 10.1  \\
J1832$-$0836 & 71.4 & 10.2 & 29.9  & 27.2  & -2.3  & 7.4  \\
J1843$-$1113 & 49.3 & 6.4 & 37.5  & 36.8  & -0.7  & 2.2  \\
J1853$+$1303 & 58.8 & 11.4 & 33.6  & 26.2  & 3.5  & 16.9  \\
J1857$+$0943 & 75.2 & 12.5 & 15.3  & 14.2  & -0.1  & 4.3  \\
J1903$+$0327 & 83.0 & 17.5 & 15.2  & 9.7  & -10.8  & 10.8  \\
J1910$+$1256 & 32.8 & 4.4 & 22.4  & 16.3  & -0.4  & 14.1  \\
J1911$-$1114 & 53.6 & 13.1 & 26.7  & 22.1  & -5.9  & 9.5  \\
J1911$+$1347 & 67.2 & 4.4 & 46.6  & 36.5  & 21.0  & 24.3  \\
J1918$-$0642 & 61.7 & 5.7 & 21.4  & 19.5  & -5.0  & 6.1  \\
J1923$+$2515 & 58.8 & 10.8 & 22.5  & 21.7  & -2.6  & 3.0  \\
J1944$+$0907 & 90.0 & 22.9 & 17.7  & 13.5  & 7.7  & 8.9  \\
J1946$+$3417 & 93.6 & 12.9 & 20.9  & 18.9  & -4.7  & 5.3  \\
J1955$+$2908 & 81.2 & 14.2 & 35.8  & 26.1  & -7.5  & 21.5  \\
J2010$-$1323 & 18.8 & 5.0 & 20.0  & 17.9  & 1.2  & 7.0  \\
J2017$+$0603 & 69.5 & 10.7 & 38.5  & 38.2  & -1.9  & 3.5  \\
J2019$+$2425 & 79.5 & 18.8 & 40.7  & 40.1  & 1.0  & 3.0  \\
J2022$+$2534 & 79.5 & 13.1 & 25.5  & 24.1  & 2.5  & 5.3  \\
J2033$+$1734 & 46.1 & 6.0 & 37.9  & 32.0  & -1.6  & 13.1  \\
J2043$+$1711 & 76.0 & 9.8 & 60.3  & 59.3  & 1.9  & 4.5  \\
J2145$-$0750 & 87.7 & 6.9 & 19.8  & 16.9  & 5.9  & 7.8  \\
J2150$-$0326 & 37.0 & 5.5 & 15.1  & 14.8  & -0.8  & 1.8  \\
J2214$+$3000 & 72.4 & 9.4 & 38.9  & 39.0  & 0.4  & 0.5  \\
J2229$+$2643 & 35.5 & 14.1 & 19.4  & 17.0  & 4.0  & 7.5  \\
J2234$+$0611 & 71.5 & 4.2 & 36.9  & 36.8  & 2.6  & 2.7  \\
J2234$+$0944 & 89.6 & 11.5 & 23.5  & 21.4  & 5.9  & 6.3  \\
J2302$+$4442 & 92.2 & 21.1 & 56.5  & 56.3  & -1.4  & 2.7  \\
J2317$+$1439 & 52.0 & 11.7 & 32.4  & 29.9  & 7.2  & 7.3  \\
J2322$+$2057 & 53.4 & 9.2 & 13.5  & 11.2  & -1.1  & 5.8  \\
  \hline\hline
        \end{tabular}
\end{table}

\begin{figure*}[!ht]
\begin{minipage}[t]{0.66\columnwidth}
    \centering
    \includegraphics[width=0.9\columnwidth]{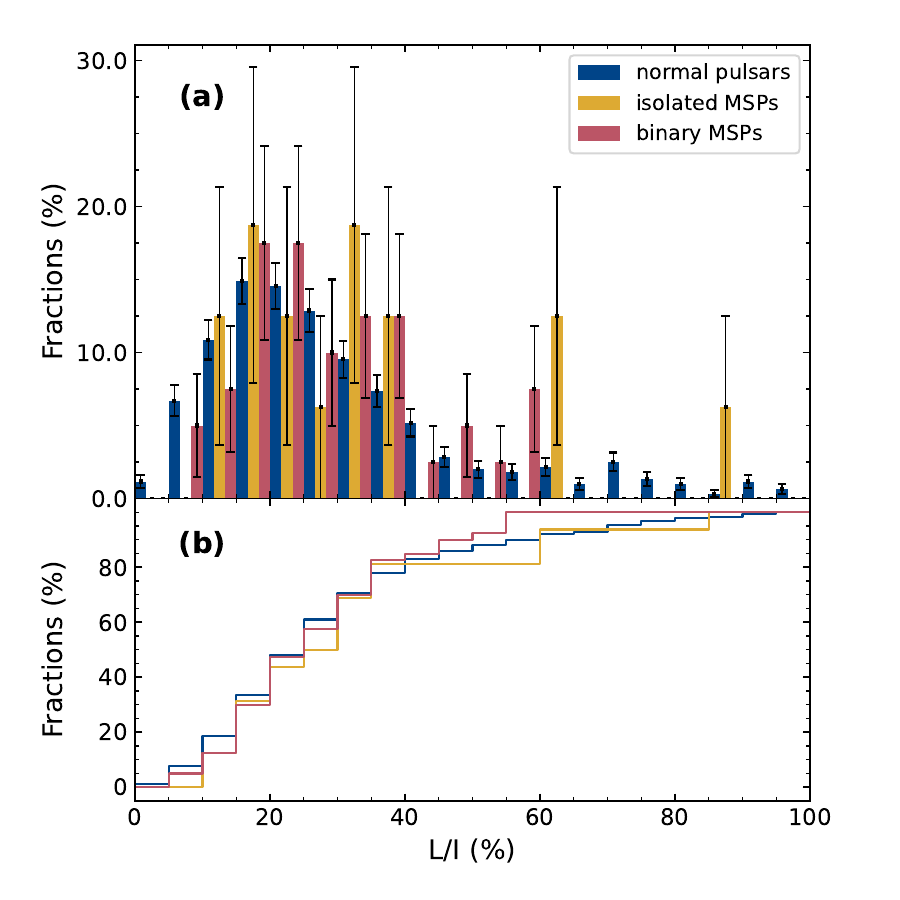}  
\end{minipage}
\begin{minipage}[t]{0.66\columnwidth}
    \centering
    \includegraphics[width=0.9\columnwidth]{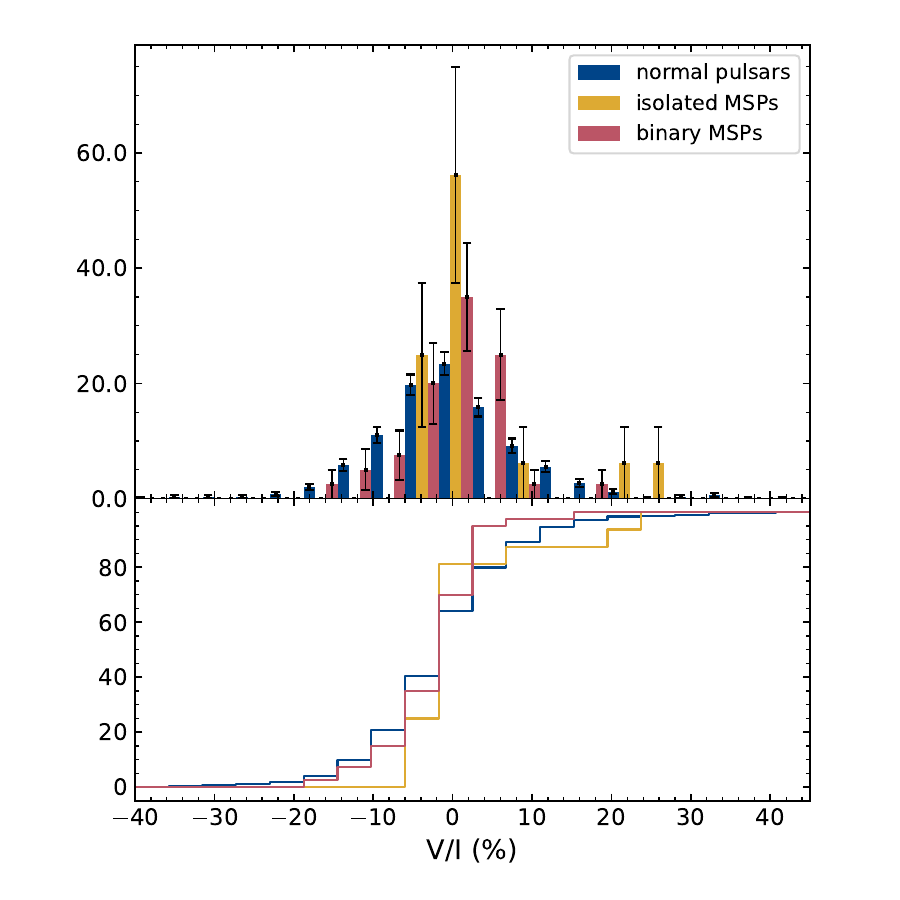}   
\end{minipage}
\begin{minipage}[t]{0.66\columnwidth}
    \centering
    \includegraphics[width=0.9\columnwidth]{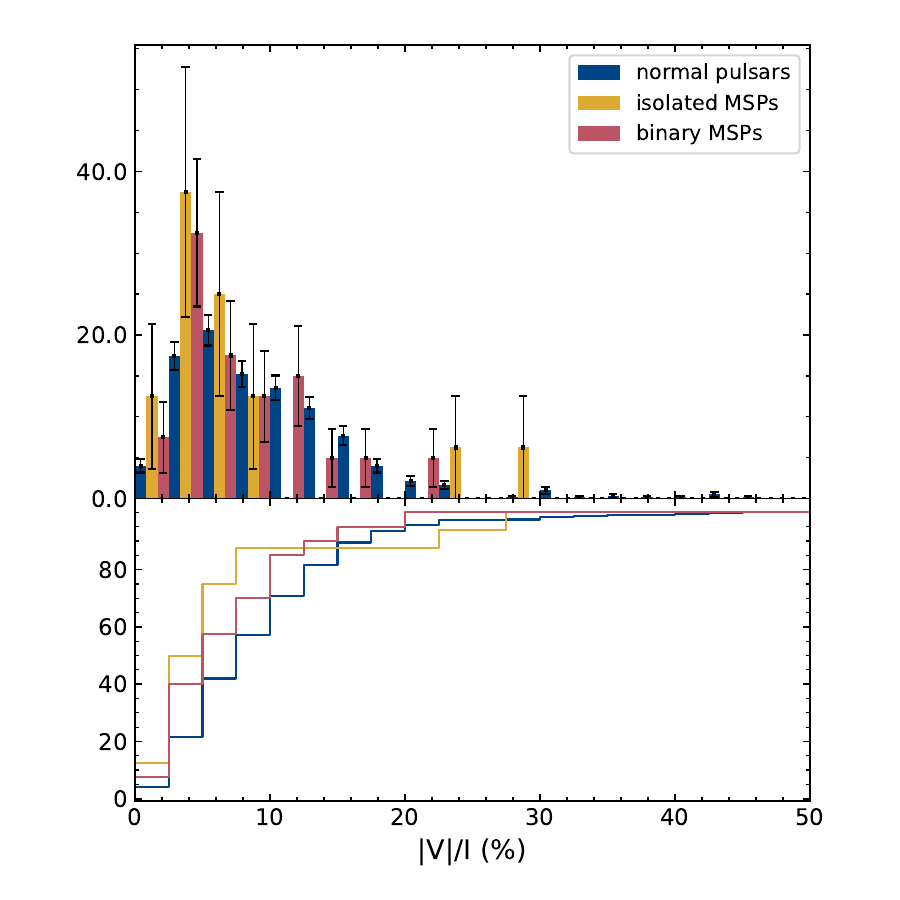}   
\end{minipage}
\caption{Distribution of polarization degree for 56 CPTA pulsars. The panels from left to right show the distribution for linear polarization, circular polarization, and the absolute value of circular polarization, respectively. The top row of panels shows the binned distribution function, while the bottom row is for the cumulative distribution function. Blue, yellow, and red colors denote the distribution for normal pulsars, isolated MSPs, and binary MSPs, respectively. Error bars are for the 68\% confidence level computed from $1/\sqrt{N}$, with $N$ being the counts per each bin. The data for normal pulsars ($P_0 > 50\,{\rm ms}$) is from \citet{han2023}. }
\label{fig:pol_dis}
\end{figure*}

The histograms of linear and circular polarization degree are present in \FIG{fig:pol_dis}. Compared to the normal pulsars \citep{han2023}, there is no difference in polarization properties with MSPs, which is the same as the conclusion of \cite{2024MNRAS_Karastergiou}. According to the Kolmogorov-Smirnov test, the corresponding p value is much higher than the 95\% confidence level, except for the distribution of the absolute circular polarization, whose value is of 0.051. Moreover, the isolated MSPs also present a similar distribution with binary MSPs. The consistencies indicate that the recycling history has no effect on polarization properties.  

We detect weak components or radiation below the level of 3\% peak flux for approximately 80\% of all 56 MSPs, except for PSRs J0509$+$0856, J0824$+$0028, J1327$+$3423, J1640$+$2224, J1643$-$1224, J1738$+$0333, J1741$+$1351, J1903$+$0327, J1910$-$1114, J2150$-$0326, J2229$+$2643, and J2322$+$2057. Six pulsars may emit radiation over the full rotation phase: PSRs J0509$+$0856, J1012$+$5307, J1710$+$4923, J1713$+$0747, J1944$+$0907, and J2302$+$4442. For these pulsars, we detected radiation ($\geq3\sigma$) within more than 90\% of the rotation phase, as is shown in Table~\ref{tab:tot_pol}. In the table, the detectable width ($W_{\rm det}$) is defined as the phase range where the radiation is three times above the noise floor ($\geq3\sigma$ ).

\begin{figure}
    \centering
    \includegraphics[width=1\columnwidth]{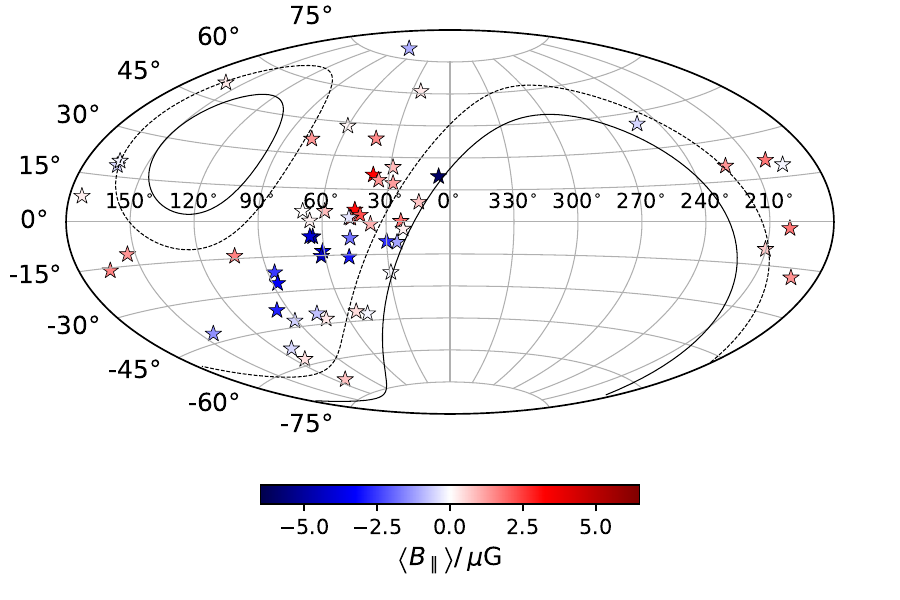}
    \caption{Averaged Galactic magnetic field parallel to the line of sight to CPTA pulsars in Galactic coordinates. Positive values indicate the Galactic magnetic field points toward the observers. The values shown here are derived from the ratios of RM and DM. The solid line denotes the sky coverage of FAST, while the dashed line presents the critical angle of $26.4^\circ$.\label{fig:gal_mag}}
\end{figure}

The RM and DM of each pulsar are listed in Table~\ref{tab:tot_pol}. The averaged Galactic magnetic field strengths, in the directions of the pulsars, were estimated from $\left\langle B_{\|}\right\rangle=1.23\,\mu\mathrm{G} \left(\frac{\mathrm{RM}}{\mathrm{rad\cdot m^{-2}}}\right)\left(\frac{\mathrm{DM}}{\mathrm{pc\cdot cm^{-3}}}\right)^{-1}$. As is shown in \FIG{fig:gal_mag}, the magnetic field distribution features reversal above and below the galactic plane,  which is consistent with the previous study~\citep{Xu_2014}.

\section{Discussion of the results}
\label{sec:disc}
\subsection{Weak emission of pulse profile}
\label{sec:weak}

From the observed pulse profiles of CPTA pulsars, a significant fraction of MSPs show weak emission outside the traditional “on-pulse window,” which was defined as the pulse phases covering the pulse peak and a dominant fraction of pulse energy \citep{1971ApJ_Taylor}. Previously, \cite{Gentile_2018} and \cite{Wahl_2022} had detected 11 pulsars with “microcomponents;” that is, pulse components with peak intensities much lower than the total pulse peak intensity\footnote{\citet{Wahl_2022} had taken the threshold as 3\% of the peak intensity.}. Such microcomponents also appear in our data (e.g., in PSRs~J1024$-$0719, J1713$+$0747, J2145$-$0750, and J2234$+$0944). We have discovered weak radiation in 44 pulsars. Furthermore, they present a very diverse phenomenology. In this way, it is hard for us to simply define the isolated components. We simply use the name “weak components” or “weak radiation” for a pulse structure with a flux lower than 3\% of the peak flux. 

\begin{figure*}
\begin{minipage}[t]{0.5\columnwidth}
    \centering
    \includegraphics[width=0.9\columnwidth]{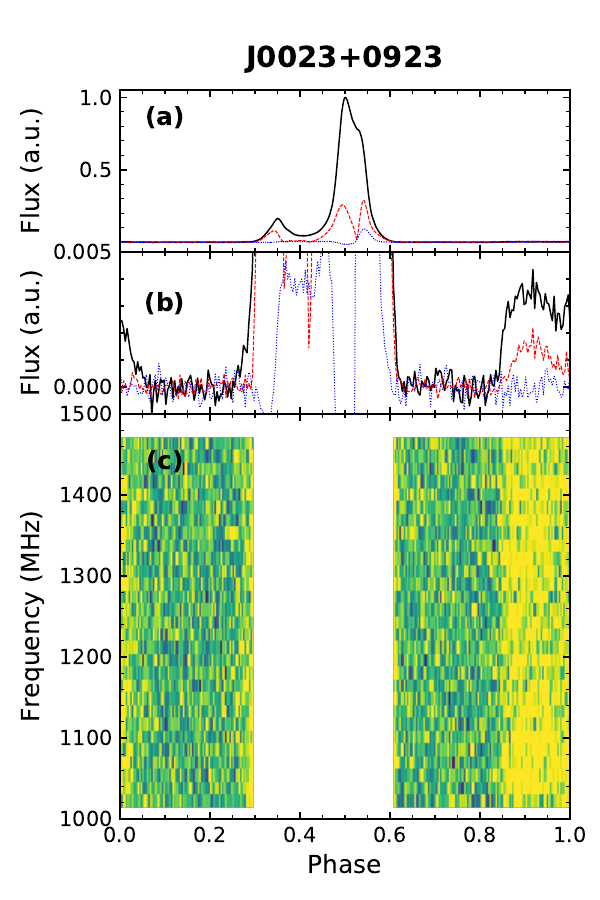}  
\end{minipage}
\begin{minipage}[t]{0.5\columnwidth}
    \centering
    \includegraphics[width=0.9\columnwidth]{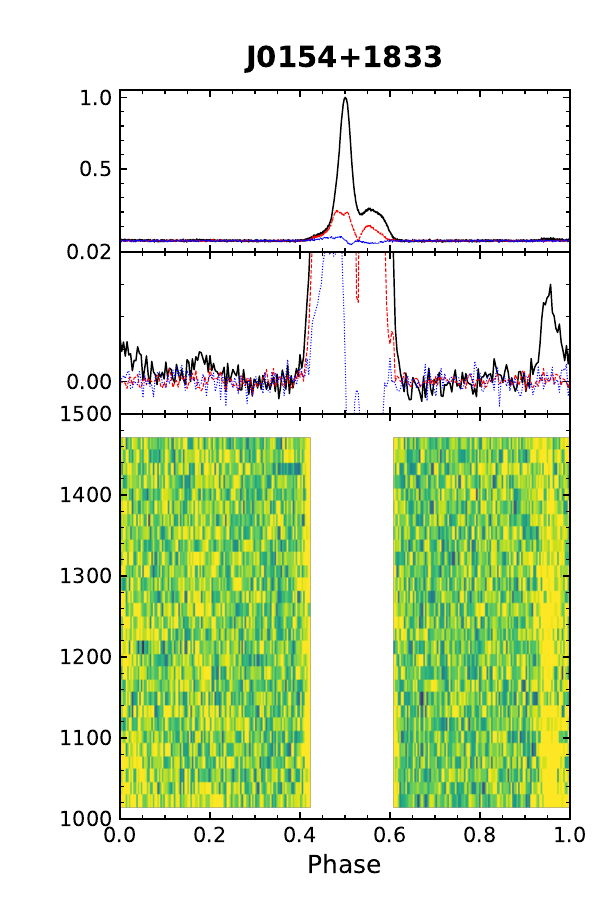}   
\end{minipage}
\begin{minipage}[t]{0.5\columnwidth}
    \centering
    \includegraphics[width=0.9\columnwidth]{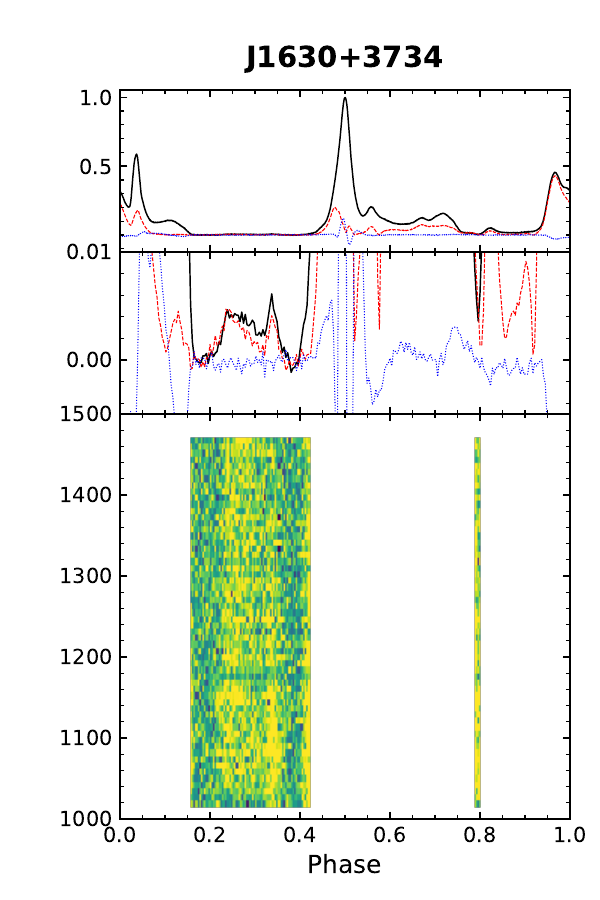}   
\end{minipage}
\begin{minipage}[t]{0.5\columnwidth}
    \centering
    \includegraphics[width=0.9\columnwidth]{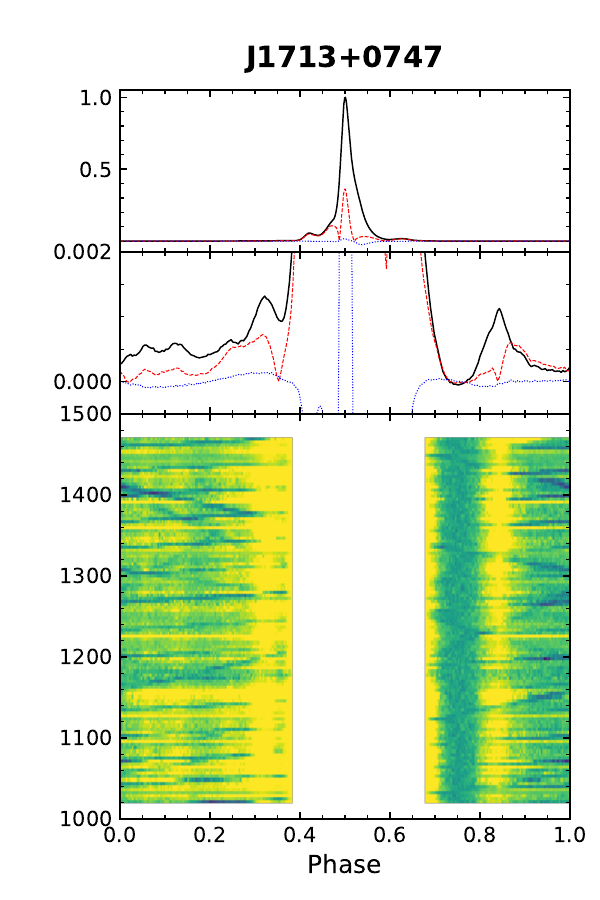}   
\end{minipage}\\
\begin{minipage}[t]{0.5\columnwidth}
    \centering
    \includegraphics[width=0.9\columnwidth]{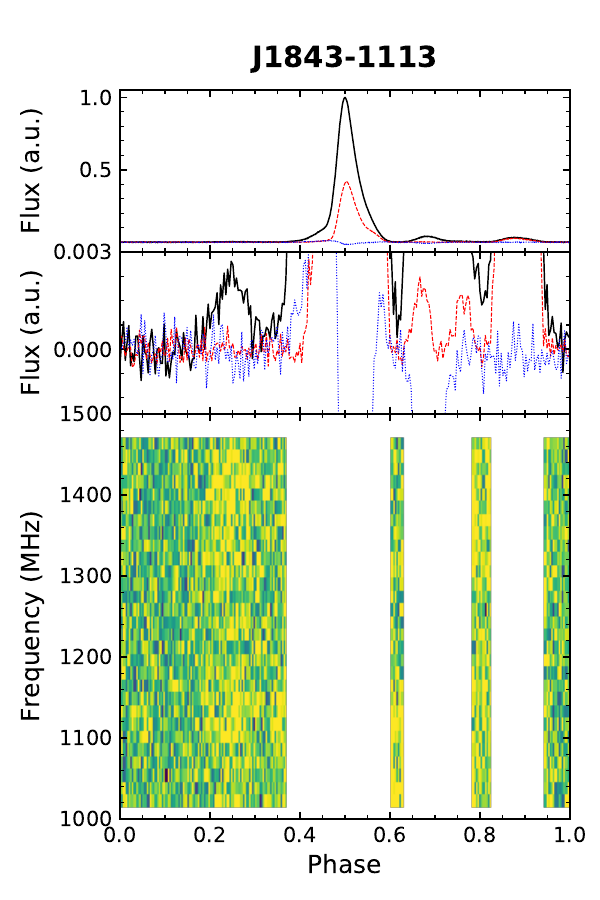}   
\end{minipage}
\begin{minipage}[t]{0.5\columnwidth}
    \centering
    \includegraphics[width=0.9\columnwidth]{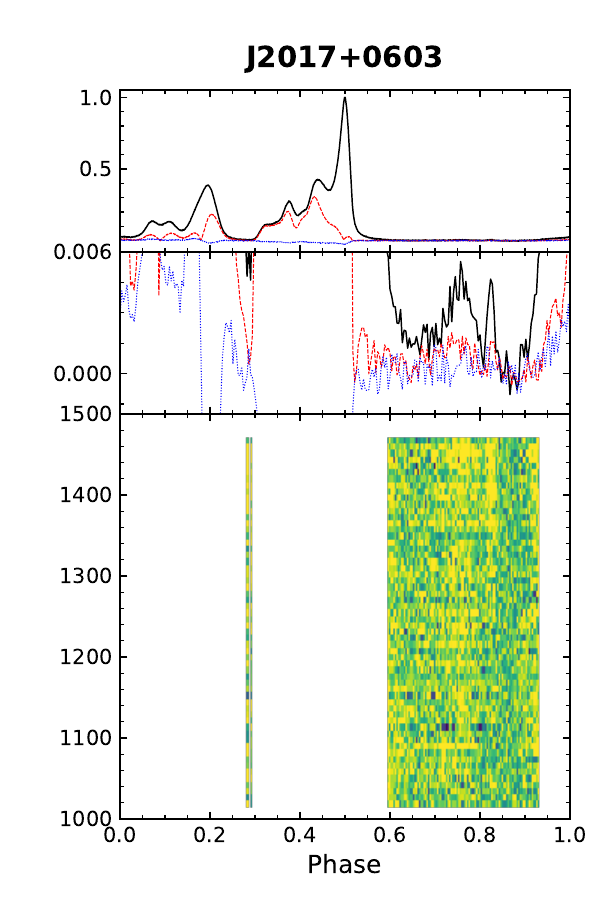}   
\end{minipage}
\begin{minipage}[t]{0.5\columnwidth}
    \centering
    \includegraphics[width=0.9\columnwidth]{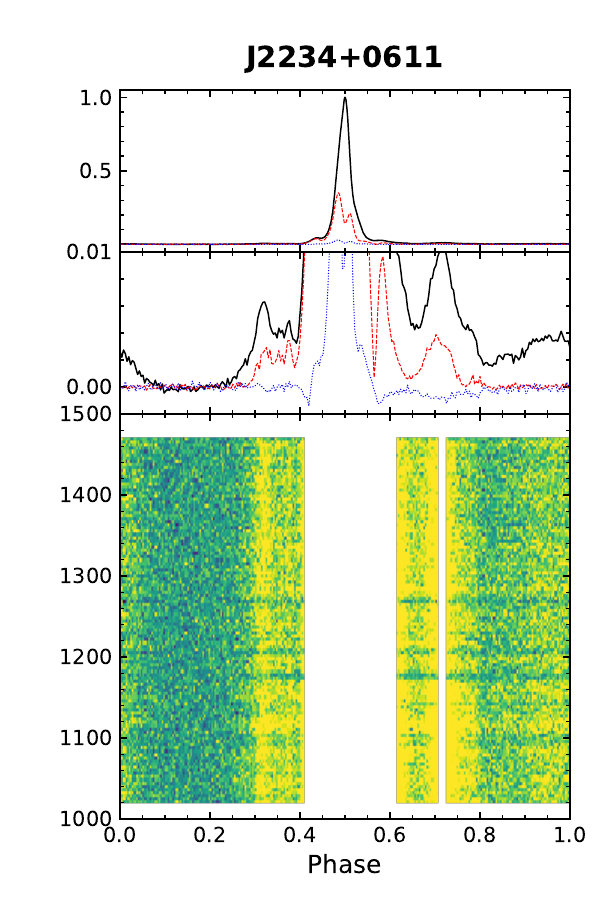}   
\end{minipage}
\begin{minipage}[t]{0.5\columnwidth}
    \centering
    \includegraphics[width=0.9\columnwidth]{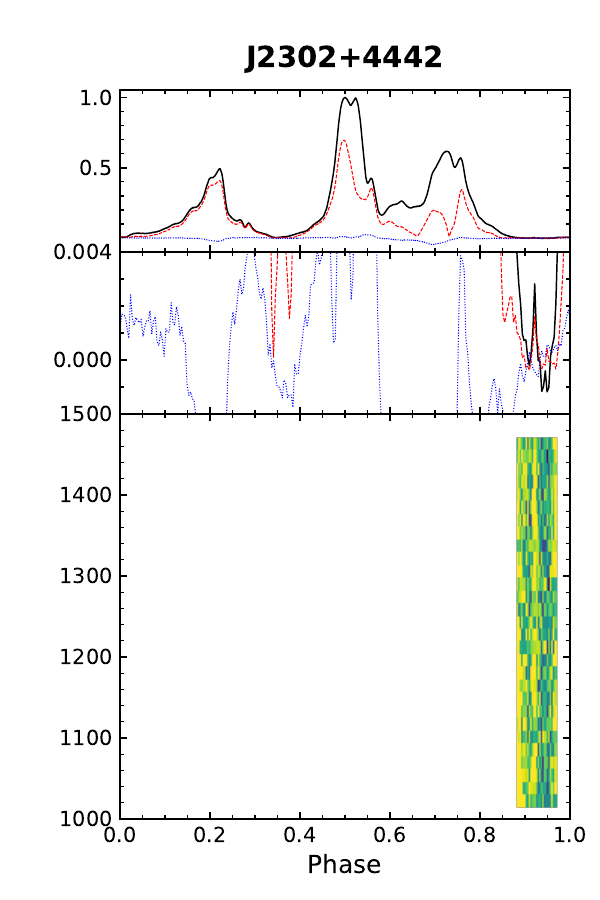}   
\end{minipage}
    \caption{Dynamic spectra of weak components for eight CPTA MSPs. Panel (a) shows the full polarization profile, panel (b) zooms into the low-flux region to show the weak radiation or weak components, and panel (c) shows the dynamic spectra, where total fluxes larger than the range displayed in panel (b) are masked out (in white).}
\label{fig:weak}
\end{figure*}

It is well known that spectral leakage at the band edges and polyphase filter response functions introduce low-amplitude signal artifacts. In fact, any band-limited process will introduce artificial structures into the time domain. However, the weak components we found here do not belong to the instrumental artifacts mentioned above. We show the dedispersed dynamic spectra of eight pulsars in \FIG{fig:weak}. There is no apparent “reflection” at the band edges (i.e., frequencies close to 1500 MHz or 1000 MHz). Moreover, the weak components possess the same DM as the main pulse of the pulsars, since they also present as vertical strips after dedispersion.

The weak components are not narrowband features. As is shown in \FIG{fig:weak}, they span the entire bandwidth, from approximately 1000 MHz to 1500 MHz. In the time domain, these weak components are well separated from the high-flux pulse structures. For certain pulsars, such as PSR~J0023$+$0923, there are distinct flux gaps between the weak components and the main pulse. The polarization properties of the weak components, as is demonstrated in \FIG{fig:weak}, differ significantly from those of main pulses in most pulsars. These three facts make it unlikely that the weak components arise from polyphase filter leakage.

However, we observe a new kind of artifact in CPTA pulsars that is different from previous studies. Examples of such artifacts are shown in \FIG{fig:weak} within the dynamic spectra of PSR~J1713$+$0747. They present as weak inclined strips with different slopes. Such artifacts frequently appear in two bright pulsars, PSRs~J1713$+$0747 and J2145$-$0750, but they also occur in weak pulsars such as PSR~J0636$+$5128. Fortunately, these artifacts have a much lower amplitude than the weak components, and even noise level, so the integrated pulse profile is minimally affected. Further discussion on these artifacts can be found in Appendix~\ref{sec:immrfi}.

The weak radiation, besides forming extra pulse components, can also be a tail-like or prelude-like structure flanking the main pulse, such as in PSR J0340$+$4130, or bridge emission between pulse components, such as in PSR J1857$+$0943, or interpulse-like components, such as in PSR J0751$+$1807. In at least five pulsars (PSRs J0023$+$0923, J0154$+$1833, J0406$+$3039, J1946$+$3417, and J2234$+$0611), the weak emission components are spaced approximately 180$^\circ$ from the main pulse, suggesting that those components are possibly from the other magnetic pole. These emerging components offer us new information about the radiation geometry, which will be discussed in a subsequent paper.

Most of the weak components show linear polarization, and we detect circular polarization in weak components for PSR~J0406$+$3039, J0751$+$1807, J1713$+$0747, and J2145$-$0750. Some non-polarized weak components are also observed in PSR~J0154$+$1833, J1843$-$1113, and J2234$+$0611. However, the current limited sample size prevents us from investigating whether the polarization properties of the weak components are significantly different from normal pulse components.

In terms of pulse width, the weak components last for about $10\%-20\%$ of full rotation. PSR~J2017$+$0603, J2022$+$2534, and J2302$+$4442 show very narrow, spiky, weak components at the scale of less than 5\% rotation. The timescales for the narrow weak components are around 50 $\mu s$. Such a small timescale indicates that the plasma beam powering such radiation has a rather limited angular diameter and implies an extremely nonuniform environment for the radiation regions.

\subsection{Pulse width}
\label{sec:pulwid}
As a common consensus, the pulse widths of MSPs are generally wider than the ones of normal pulsars. In the case of CPTA pulsars,
as is listed in Table~\ref{tab:tot_pol}, the average value of the effective width is on the level of 10\%\footnote{Here, we quoted the effective width due to the complex shape of the pulse profile. For Gaussian profiles, the widths under different definition can be converted that $W_{\rm eff}=1.06W_{50}=0.58W_{90}$.}, which is defined as the ratio of integrated flux divided by the peak flux (i.e., $W_{\rm eff}\equiv\int_0^{1} I(\phi) d\phi/I_{\rm max}$). 
The average and median for $W_{\rm eff}$ are  11\% and 10\%, respectively. The values are comparable to previous published results  \citep{2011_yan,Gentile_2018}.
Such a pulse width exceeds that of normal pulsars, which is generally lower than $10\%$~\citep{1998_Gould,2018_Johnston,han2023}. The effective pulse width measurements indicate that the MSP radiation energy is still beamed, although it is more extended than that of a normal pulsar. 

Instead of asking which phase range the most pulse energy is concentrated in, we can pose a different question: whether it is possible to detect pulsed radiation outside the pulse window. Our observations show that the detectable pulse width is much larger than the $W_{\rm eff}$. Here, the detectable pulse width ($W_{\rm det}$) is the total phase range where the flux is above three times the noise floor. We compare the effective pulse width ($W_{\rm eff}$) and the detectable pulse width ($W_{\rm det}$) in Table~\ref{tab:tot_pol}. The average and median of $W_{\rm det}$ for 56 MSPs are 65\% and 70\%, respectively. In this way, a small fraction of the radiation energy from an MSP spreads to a much larger solid angle, and nearly covers the full rotation phase.

The higher value of $W_{\rm det}$ is not caused by the interstellar scattering effect, with the exception of PSRs~J1903$+$0327 and J1946$+$3417. Firstly, the pulse profiles show no obvious scattering tails for most pulsars in our list. Secondly, ISM models, such as the NE2001 and YMW models \citep{NE2001, YMW17}, predict that the scattering timescale of CPTA pulsars is of only a few microseconds. This is naturally expected, as all the 56 MSPs here are observed for timing purposes. Their distance, DM, and scattering measure are all expected to be small compared to other MSPs. 

If we extrapolate the radiation beam size -- period relation derived for normal pulsars to MSPs \citep{1970ApJ_Gunn,1988_lyne,1993A&A_Gil,2011MNRAS.414.1314M}, we would expect a correlation between $W_{\rm det}$ or $W_{\rm eff}$ and the pulsars' spin periods and/or period derivatives. However, no significant correlations can be found, possibly because our sample size is very limited where the range of pulsar period is only approximately 1 dex. It is also possible that the different magnetosphere conditions for MSPs and normal pulsars erase such a correlation. For example, one argument is that the magnetospheres of MSPs are smaller than the ones of normal pulsars, so the multipolar magnetic field becomes important and radiation is less beamed \citep{1991ApJ_Krolik, 1998ApJ_Kramer}. 

A large value for $W_{\rm det}$ may indicate that the MSPs radiate over the full rotation phase. As is mentioned in Sect.~\ref{sec:result}, six CPTA pulsars are promising candidates. If we lower the threshold of identifying the emission, in other words, identifying a signal above the 1-$\sigma$ noise level, another 13 pulsars are potentially 360$^\circ$ radiators: PSRs~J0030$+$0451, J0751$+$1807, J1024$-$0719, J1453$+$1902, J1630$+$3734, J1745$+$1017, J1832$-$0836, J1955$+$2908, J2017$+$0603, J2145$-$0750, J2214$+$3000, J2234$+$0944, and J2302$+$4442. For normal pulsars, a similar phenomenon had been reported before \citep{1981_hankins,1997JApA_Rankin,WangZL2022}.
Future monitoring of CPTA pulsars will further increase the integration time and S/N, which will help us to identify and confirm these 360$^\circ$ candidates. 

A large value of $W_{\rm det}$ and the weak radiation of MSPs may affect our predictions for future MSP searches using upcoming high-gain telescopes such as the Square Kilometre Array (SKA). In the near future, SKA will become operational, with pulsar searches being one of the key scientific objectives\footnote{See document SKA-TEL-SKO-0000015 at \url{https://www.skao.int}.}. As the telescope's sensitivity increases, the combination of a large detectable pulse width and the weak radiation components observed in many MSPs will improve our chance of discovering more MSPs, because the weak radiation extends the size of the pulsar beam and increases the probability of detection.

\subsection{Polarization properties}
\label{sec:pol_property}
In this paper, we focus on investigating the frequency-integrated
polarization pulse profiles of MSPs. The investigation of the frequency
evolution of polarization properties will be published in the future.
There are no significant differences between our polarization profiles and
published results \citep{2011_yan,Gentile_2018,spiewak_2022}, except for PSR J0154$+$1833. For this source,
we had detected the presence of both linear and circular polarization,
which was absent in the profile obtained by MeerKAT~\citep{spiewak_2022}.

The polarization profiles of MSPs are more complex than ones of normal pulsars. We
find that they may all belong to the “complex type” of Rankin's
classification scheme \citep{rankin83}, and the detection of weak
components makes the pulse profiles even more complex.  The
interpulse-like components show up in 25\% of the 56 pulsars, including
PSR~J0023$+$0923, J0030$+$0451, J0154$+$1833, J0406$+$3039, J1453$+$1902,
J1630$+$3734, J1857$+$0943, J1946$+$3417, J2017$+$0603, J2043$+$1711,
J2150$-$0326, J2214$+$3000,  J2234$+$0611, and J2322$+$2057. The more frequent appearance of interpulses in CPTA pulsars may be due to the larger radiation beam sizes, which makes it easier to catch up with the interpulse emission.

Across the pulse phase, most pulsars exhibit a higher degree of
linear polarization at the edges of pulse components, consistent with
the findings of \citet{2015_dai}. This suggests an anticorrelation
between the degree of linear polarization and flux intensity. This is consistent with the wave propagation model (see Fig. 9 of
\citet{2010MNRAS_Wang}), which suggests that the wings of the pulse,
due to their higher emission altitude, are less affected by propagation
effects, and therefore exhibit a higher degree of linear polarization. On
the other hand, such limb enhancement of linear polarization may indicate
that all MSP pulse profiles are affected by the propagation effects.

We note that the sign of circular polarization can reverse multiple
times across the pulse phase. Similarly, the sense of the PA sweep
can also reverse. For instance, PSR J0613$-$0200 exhibits oscillatory
patterns in both PA and $V$ within the main pulse window. Previous
studies \citep{1990radhak, 2006ChJAA_You} have suggested a
correlation between the sense of circular polarization and the PA sweep
in conal-double pulsars. However, for MSPs, it is difficult to define
a global sense of the PA sweep and circular polarization, as multiple
reversals of both occur quite frequently. The correlation may break
down in MSPs. For example,  both PSR~J1744$-$1134 and J1745$+$1017 have
negative $V$, but opposite senses for the PA sweep. We suspect that the
complex magnetic field configuration introduces an extra ingredient into the
propagation model.  However, the local sense of circular polarization
(sign of $V$) of a given pulsar may still correlate with the slope
of the PA sweep.  For example, in PSR~J0824$+$0028, the first two peaks have
$V<0$ and the third peak shows $V>0$, while the PA sweep sign is reversed
for the third peak. A similar feature is also seen in PSR~J1453$+$1902
and J0154$+$1833.

The PA jumps are common in CPTA pulsars, which are believed to be induced
by the orthogonal polarization modes. We note that PA jump can be away
from exactly 90$^\circ$, as is shown in PSR J1630$+$3734. Also, the PA
jumps seem to fall into two categories, I) the PA jumps when the polarization
intensity goes to zero (e.g., in PSR~J0154$+$1833), and II) the PA jumps
around a local circular polarization peak (e.g., in PSR~J0023$+$0923).
The two types of PA jump can occur in the same pulsar. For example,
the first PA jump of J0340$+$4130 is type I, while the second PA jump is
type II. It is worth mentioning that the two types of jump show very
different paths on the generalized Poincar\`e sphere:\footnote{Strictly
speaking, the Poincar\`e sphere cannot be used to describe the unpolarized
components, so it is not applicable to the pulsar problem. However, if we
normalize the Stokes parameters $Q$, $U$, and $V$ using the intensity, $I$,
we get a generalized Poincar\'e sphere, which is then valid for the
partial polarization.} the type I PA jump happens when the polarization
vector shrinks back to the origin (e.g., see \citep{1998ApJ_McKinnon}),
while the type II PA jump happens when the polarization vector goes through
the north or south poles \citep{Dyks2020}. The PA curves of many CPTA pulsars evidently deviate from the RVM model. Only five pulsars -- PSRs J0605$+$3757, J0636$+$5158, J1903$+$0327, J2214$+$3000, and J2234$+$0611 -- are found to be well described by the RVM model.

In the low-flux regime, we note that, sometimes, linear polarization
has a flux that appears to be barely larger than the total intensity, as happened in
PSRs~J0509$+$0856, J1453$+$1902, J1630$+$3734, J1710$+$4923, J1713$+$0747,
J1745$+$1017, and J1955$+$2908. This excess is not caused by our baseline
selection. We note that the excesses remain, even if we
reduce the baseline phase range. In the cases of PSRs J1713$+$0747 and
J1955$+$2908, high-S/N data from the Arecibo telescope~\citep{Gentile_2018}
seem to show similar excesses at the same pulse phases; that is, phase 0.9
and phase 0.3 for PSR J1713$+$0747 and J1955$+$2908, respectively.
Thus, the problem is more likely caused by small DC offsets in the
polarization profiles. For example, if one adds a tiny DC offset to
the total intensity pulse profiles, there will be no linear polarization
excess. This indicates that those pulsars may also be 360$^\circ$ radiator
candidates. Indeed, the pulsar list here overlaps with the pulsar list
showing a detectable width ($W_{\rm det}$) larger than 0.9 rotation.
The interferometric observations can be used to determine the true
baseline of the pulsar signal \citep{Navarro1995,Marcote2019}. We expect
that future high-sensitivity arrays such as SKA will be able to provide
better baseline estimation.

\subsection{Apparent RM variation across the pulse phase}
\label{sec:phrm}

As is discussed in Sect.~\ref{sec:rmproc}, the RMs of most pulsars
are constant across the pulse phase. However, we detected residual
“RM variations” across the pulse phase within some pulsars, as is shown in
\FIG{fig:deviation} for some examples. Such behavior should not
be regarded as the intrinsic RM variation due to the Faraday rotation
in the pulsar magnetosphere, as the relativistic plasma contribute
only negligible RM \citep{2011MNRAS.417.1183W}. The apparent RM
variation may be a consequence of pulse profile evolution in the frequency
domain. We postpone this analysis to a future work focusing on the
frequency evolution. As one can
see, the apparent RM variation is more prominent at the phases where
the PA jumps. In this case, the two orthogonal polarization modes with different spectral indexes will induce frequency-dependent variations in the PA~\citep{Ramachandran_2004,Noutsos2009,Ilie2018}. For
PSRs~J1903$+$0327 and J1946$+$3417, the RM variation may be a result
of the interstellar scattering effect, which flattens the PA curves \citep{karastergious2009,Noutsos2009,Noutsos2015}.

\subsection{Description of each pulsar}
\label{sec:des_sin}

PSR J0023$+$0923: We detected a solitary weak component whose intensity is only $\sim0.3$\% of its highest peak. This component exhibits a higher degree of linear polarization than the main peak and has a flat PA. It is roughly separated from its main pulse region for 0.5 rotations. The PA jumps at phase 0.52 coincides with the reversal of circular polarization, while another jumps at phase 0.37 corresponds to the local peak.

PSR J0030$+$0451: The pulse width is approximately a full rotation phase. If we consider the weak components at the trailing edge of the interpulse (phase 0.15), the PA has a global linear trend across the full phase, although it shows complex structure around the main peak. The main pulse presents a higher polarization degree at its edges.

PSR J0034$-$0534: The pulse profile shows a dominant double-peak structure with a trailing pulse component, whose fractional linear polarization is much higher than the main pulse. There is a weak bridge emission at phase 0.7 at the level of 2\% peak flux. The pulse signal covers more than 50\% rotation. Similar to PSR J0030$+$0451, the PA shows a linear trend, outside the main peak. 

PSR J0154$+$1833: The profile contains two weak components that form the interpulse (phase 0.2 and 0.95). They are separated from the main pulse by approximately half a rotation, which is similar to the case of PSR J0023$+$0923. No polarization is detected for them. There is a 90$^\circ$ PA jump around the main pulse peak (phase 0.53), where both linear polarization and circular polarization decrease to zero. 

PSR J0340$+$4130: We detected two weak components with peaks at the leading edge and the trailing edge (phase 0.1 and 0.8), respectively. Both of these components exhibit moderate degrees of polarization. The PA curve exhibits four 90$^\circ$ jumps in the pulse window (phase 0.4-0.6), while the circular polarization shows no corresponding variation at these phases. The PA curve presents smooth variation outside the pulse window. 

PSR J0406$+$3039: The pulse profile shows a weak component roughly 50\% rotation away from the main pulse (phase 0.05). It shows a 90$^\circ$ PA jump in the main pulse (phase 0.54) where the circular polarization reaches a local maximum. Similarly, the PA curve presents a monotonic decreasing trend across the whole pulse phase.

PSR J0509$+$0856: The pulsar shows radiation for nearly the full rotation. The PA curve is rather flat, except at the leading edge of the peak with a lower flux. The 90$^\circ$ PA jump occurs at phase 0.04 and 0.9.

PSR J0605$+$3757: There is a weak bridge emission  between the main pulse and the trailing weak pulse (phase 0.7). Unlike the main pulse, the weak trailing component presents extremely low linear polarization. The PA curve seems to follow the S-shape curve in the pulse window, although there are 90$^\circ$ PA jumps at the center of the pulse peak (phase 0.5). The derived magnetic inclination angle, $\alpha$, is $0.9^\circ\pm9.7^\circ$, and the sight angle, $\zeta$, between the line of sight and the spin axis is $0.7^\circ\pm7.6^\circ$. 

PSR J0613$-$0200: There are weak pulse components at the leading and trailing edges (phase 0.2 and 0.8). We are inclined to treat them as separated components, as PA variations in the weak components are smoother than the one in the main pulse. The PA jumps occur frequently across the pulse phase, and some of them coincide with the sense reversal or the local extremum of circular polarization. At phase 0.67, both linear and circular polarization decrease to zero. 

PSR J0621$+$1001: The pulse profile has a double-peaked shape, and the two peaks are connected by bridge emission (phase 0.3). There is a weak component at the leading edges of the peak with a lower flux (phase 0.1). There is a roughly 90$^\circ$ PA jump between the leading weak component and the main pulse (phase 0.1). In addition, PA jumps are frequent within the main pulse, one of which occurs at the local maximum of circular polarization around phase 0.48.

PSR J0636$+$5128:  There is a weak component at the leading edges of the pulse (phase 0.3). A nearly 90$^\circ$ PA jump occurs near the center of the pulse peak (phase 0.46). From the PA curve, we can derive the radiation geometry of $\alpha=6.5^\circ\pm28.5^\circ$ and $\zeta=11.9^\circ\pm49.0^\circ$. We should caution that the pulse width of this pulsar is narrow, which causes large uncertainty in the fitting result.  

PSR J0645$+$5158: It has a bridge-like weak component between the main pulse and interpulse (phase 0.7). Although this weak component is connected to the interpulse, it may not be the weak leading edge of the interpulse, due to its long duration and low degree of polarization compared to that of the interpulse. A 90$^\circ$ PA jump occurs near the center of the pulse peak (phase 0.5) where the circular polarization changes its sign. Compared with the main pulse, the interpulse shows a smoother PA curve.

PSR J0732$+$2314: This pulsar has an S-shaped PA swing in the main pulse window. However, the PA becomes very complex for the leading pulse (phase 0.1). There are bridge-like emissions (phase 0.2) between the two pulse peaks separated by approximately 180$^\circ$. Two PA jumps occur at phase 0.1 and 0.42, which coincides with the sense reversal of $V$. 

PSR J0751$+$1807: We detect a weak component preceding the leading edge of the main pulse (phase 0.1), and a wide weak component at phase 0.8 spanning for $\sim0.3$ rotations. Both of them exhibit a low degree of polarization. PA jumps occur multiple times over the whole phase, two of which, at phase 0.12 and 0.42, are orthogonal. The circular polarization reaches the local extremum at jump phase 0.42 and changes sign at jump phase 0.52.

PSR J0824$+$0028: There seem to be no weak components for this pulsar. Unlike the main pulse, the interpulse at phase 0.1 is nearly 100\% linearly polarized. The former two pulse components present a different sense of circular polarization than the main pulse.

PSR J1012+5307: There are three peaks on the pulsar's pulse profile. The weaker peaks from phase 0.8 to 1 may be an interpulse. However, we detected bridge emission at a level of 1\% of the peak flux connecting to the main pulse (phase 0.7). A PA jump occurs at the leading edge of the interpulse around phase 0.75. The PA curve of the main pulse has a good S-shape.

PSR J1024$-$0719: We find that there is weak bridge emission between the leading weak component and the main pulse (phase 0.1-0.4). There is also a trailing weak component (phase 0.9) with an intensity of $\sim0.5\%$ of the highest peak and nearly 100\% degree of linear polarization at the spin phase of $\sim0.9$. 90$^\circ$ PA jumps occur at the leading and trailing edges of the main pulse (phase 0.12, 0.35, and 0.64). 

PSR J1327$+$3423: No weak component is detected. A 90$^\circ$ PA jumps occurs at the leading edges of the pulse (phase 0.45) and the jump at phase 0.52 is non-orthogonal.  

PSR J1453$+$1902: This is a rather weak pulsar, even for FAST. The pulse component at phase 0.2 has a much higher linear polarization degree than the main pulse. A weak component is present at the phase of $\sim0.82$ in the profile. 
In addition, we identity a rather weak interpulse with an intensity of 2\% of the main pulse (phase 1.0). There are possible 90$^\circ$ PA jumps at the trailing edge of the main pulse (phase 0.67). We note that there is a non-orthogonal PA jump accompanied by a small spike in the degree of circular polarization around the main peak (phase 0.53).

PSR J1630$+$3734: A very weak component with a nearly 100\% degree of linear polarization is located between the interpulse and the main pulse (phase $\sim0.3$). The intensity of this component is only about 0.4\% of that of the highest peak. The PA curve presents multiple jumps, which are nearly  90$^\circ$ except for those around the main pulse. Some of them are accompanied by a sign change in circular polarization at phase 0.1, 0.50, 0.52, and 0.8.

PSR J1640$+$2224: Four PA jumps happen within this pulsar, one of which coincides with the sense reversal of $V$ at phase 0.53.

PSR J1643$-$1224: The pulse width seems to be rather wide, with the leading edge extending for more than half the rotation. Both orthogonal and non-orthogonal PA jumps are visible, some of which occur at the phase of the sense reversal of $V$ (phase 0.36, 0.49, and 0.53). 

PSR J1710$+$4923: This pulsar radiates over the full rotation phase, and a weak component is observed around phase 0.1, which contains a low polarization degree. The PA jumps are common in this pulsar, three of which are nearly $90^\circ$, at phase 0.21, 0.28, and 0.34, respectively. The circular polarization profile reaches the local maximum at jump phase 0.45 and 0.64.

PSR J1713$+$0747: This is a well-known pulsar; however, our data show that the pulsar radiates over nearly the whole period. We note that this pulsar has an interpulse-like structure at phase 0.85. The leading component spans roughly 40\% of the rotation phase. There are several PA jumps for this pulsar, either in the main pulse peak or in the weak components. Except for the one at phase 0.59, they are all orthogonal. Two jumps around the flux peak are accompanied by a sign change in the circular polarization profile. After correcting the PA jumps, the PA curve has an S-shape. The RVM model gives the geometry of $\alpha=48.0^\circ\pm5.6^\circ$ and $\zeta=74.7^\circ\pm6.8^\circ$.

PSR J1738$+$0333: No weak component is detected. The PA curve swings rapidly at the trailing edge of the main pulse.

PSR J1741$+$1351: No weak component is detected. Several PA jumps happen within the main pulse. At phase 0.52, both linear and circular polarization reduce to zero.

PSR J1744$-$1134: A weak component is observed at phase 0.15. The main pulse is nearly 100\% linearly polarized, while the weak components have a lower linear polarization degree. In addition, the PA curve of the main pulse is smoother. The 90$^\circ$ PA jumps occur around the pulse peak (phase 0.45 and 0.6). 

PSR J1745$+$1017: The profile exhibits weak components at phase 0.05 and 0.9, both of which are highly linearly polarized. This pulsar is also a potential $360^\circ$ radiator, since the linear polarization flux is higher than the intensity flux at phase 0.35. 

PSR J1832$-$0836: This pulsar has a very complex pulse profile that spans nearly the whole pulse phase. A weak component with a high degree of linear polarization is situated between the two highest peaks. Additionally, a bump-like weak component precedes the low-intensity peak (at phase 0.05 in this study). PA jumps are common in this pulsar. At jump phase 0.77, the linear polarization decreases to zero, while the circular polarization shows as a peak.

PSR J1843$-$1113: The profile shows a weak component preceding the main pulse (phase 0.25), with an intensity that is only 0.2\% of the highest peak. There is no detectable polarized emission in this weak component. Additionally, we have detected weak bridge emission between the three prominent peaks. No PA jump is detected.

PSR J1853$+$1303: The profile shows a weak bridge emission (phase 0.7) connecting the main pulse and the interpulse. PA jumps occur frequently in this pulsar. There are four jumps accompanied by sign changes in $V$ at phase 0.44, 0.51, 0.56, and 0.58, respectively. 

PSR J1857$+$0943: The profile shows a weak bridge emission (phase 0.3) connecting the main pulse and the interpulse. The PA curve has a complex shape, and presents a rapid sweep between phase 0.4 and 0.6.

PSR J1903$+$0327: The profile shows a scattering-like pulse tail. Neither a PA jump nor a weak component is detected. The RVM model gives the radiation geometry of $\alpha=0.4^\circ\pm3.6^\circ$ and $\zeta=0.8^\circ\pm8.2^\circ$. However, we should caution that such a result may be influenced by the scattering effect.

PSR J1910$+$1256: No weak component is detected. Nearly $90^\circ$ PA jumps occur around the peak, one of which corresponds to the sense reversal of $V$ (phase 0.5).

PSR J1911$-$1114: No weak component is detected. A $90^\circ$ PA jump is observed at phase 0.6. The leading component at phase 0.35 has a much higher linear polarization degree and a smoother PA curve.

PSR J1911$+$1347: There is weak radiation at the trailing edge of the main pulse (phase 0.8-0.9). Weak bridge emission is also detected around phase 0.6. The PA jump at phase 0.51 coincides with the sign change of $V$.

PSR J1918$-$0642: The pulse profile has two peaks separated for a roughly 0.5 rotation. No clear weak pulse component is detected. There is also no bridge emission between the main pulse and the interpulse down to the level of a few thousandths of the peak flux. This indicates that the interpulse comes from the other magnetic pole. The PA curve has a complex shape in which PA jumps are frequent within the main pulse. One of them occurs at phase 0.56 which is of little offset to the local peak of $V$. Two jumps at phase 0.44 and 0.48 coincide with the sense reversal of V.

PSR J1923$+$2515: A weak component at phase 0.95 is detected that is connected to the interpulse at phase 0.15. A bridge emission of 1\% peak flux around phase 0.4 is also observed between the main pulse and the prevailing component. We notice that they have different fractional polarizations. Two PA jumps at phases 0.49 and 0.61 are a little offset from the local peak of $V$. 

PSR J1944$+$0907: This pulsar is a candidate radiating over the full rotation. We observe a weak component at the leading edge of the main pulse (phase 0.3). The PA curve apparently deviates from the S-shape. The profile V reaches local maximum at jump phase 0.6. 

PSR J1946$+$3417: There is a weak component separating from the brightest peak for half a rotation (phase 0), and it shows roughly a 70\% degree of linear polarization. We observe two weak bridge emission regions (phase 0.3 and 0.7). Similar to J1903$+$0327, the interstellar scattering effect is also obvious. PA jumps happen around the main pulse peak (phase 0.5) and around the bridge emission region (phase 0.34, 0.78, and 0.83).

PSR J1955$+$2908: There is weak bridge emission (phase 0.3) between the weak interpulse and main pulse. As one can see, the linear polarization is apparently higher than the total intensity. One possible explanation is that the baselines of the polarization profiles have not been subtracted correctly, indicating that it could be a full-phase pulsar. PA jumps happen in the main pulse phase (phase 0.44 and 0.52). The PA curve presents a rapid swing after phase 0.52.

PSR J2010$-$1323: A weak component with a long leading tail is observed at phase 0.4. It exhibits a higher degree of linear polarization than the main pulse. The circular polarization changes its sign at jump phase 0.49 and reduces to zero with linear polarization at phase 0.56.

PSR J2017$+$0603: Two distinct weak components are detected, following the trailing edge of the brightest pulse (phase 0.7-0.8). Both of them show a moderate degree of linear polarization. Their fluxes are only 0.4\% of the peak flux. We also notice a very weak bridge emission between the main pulse and the interpulse at phase 0.3. Four PA jumps happen at phase 0.09, 0.18, 0.3, and 0.5. At phase 0.18, the circular polarization changes its sign. It reaches the local maximum when linear polarization reduces to zero at phase 0.5. Although the PA curve is smooth, the RVM model cannot give a reasonable result, which may be due to the non-orthogonal jump at phase 0.3.

PSR J2019$+$2425: A weak bridge emission at a level of 2\% of the peak intensity between the main pulse and the leading component is detected (phase 0.3). The PA curve shows a sudden dip at the phase around 0.43, which may be due to an OPM transition. We notice that the directions of the sense reversal of circular polarization are opposite for the main pulse and the trailing component. A $90^\circ$ PA jump is detected in the trailing components (phase 0.82), where the V changes its sign.

PSR J2022$+$2534: There may be a weak component at phase 0.9. PA jumps occur around the main pulse peak, which are both accompanied by a sense reversal in V. In addition, the leading pulse component presents much steeper PA slopes than the main pulse. Sense reversal of circular polarization happens multiple times in this pulsar. 

PSR J2033$+$1734: The extended trailing component covers roughly 30\% of the spin period. Unlike the main pulse, it presents violent PA variations. There may be a weak component around phase 0.85. Sense reversal of circular polarization occurs several times. Both a PA jump and sense reversal emerge at the main pulse peak (phase 0.5).

PSR J2043$+$1711: A weak component of 2\% $I_{\rm max}$ at phase 0.1 and low-level bridge emission around phase 0.8 are detected. The PA jumps are observed at its trailing component (phase 0.7).

PSR J2145$-$0750: A weak component of 0.4\% $I_{\rm max}$ at phase 0.9 and the bridge emission of 0.2\% $I_{\rm max}$ at phase 0.4 are observed. The new detected weak component and the interpulse (at phase 0.3) contain different fractional linear polarizations from the main pulse. The main pulse and interpulse have different senses of circular polarization. The PA curve is complex and the PA jumps are common.

PSR J2150$-$0326: No weak components are detected in this pulsar. The PA jump happens within the interpulse at phase 0.1, where V changes sign. The sense reversals of circular polarization occur at phases near the main pulse peak. At phase 0.5, before the main pulse peak, we note that a circular polarization peak coincides with a local PA sweep.

PSR J2214$+$3000: Two weak components are observed, one at the leading edge of the main pulse (phase 0.3) and another one between the main pulse and the interpulse (phase 0.7-0.8). We also found two bridge emission regions, one connecting the main pulse and the second weak components (phase 0.3) and another one between the second weak component and interpulse (phase 0.7). The intensities of both weak bridge emission are at the level of 0.2\% of the peak flux. All those weak components appear to be highly linearly polarized. There is a PA jump around phase 0.35, shortly before the main pulse peak. The PA curve has an S-shape, although it presents little deviation at phase 0.05. From the RVM model, we derive $\alpha=31.6^\circ\pm3.7^\circ$ and $\zeta=30.3^\circ\pm4.5^\circ$. This contrasts with the presence of the interpulse, which supposed to be from another magnetic pole. Similar cases are also discussed in PSR B0950$+$08 \citep{1981_hankins} and PSR B1929$+$10 \citep{1997JApA_Rankin}. We leave this to future work.  

PSR J2229$+$2643: No weak components are observed. The sense reversals of circular polarization occur around phases 0.4 and 0.55.

PSR J2234$+$0611: We found three weak components, two of which flank the main pulse (phase 0.3 and 0.7). The third one at phase 0 is very weak and wide (it extends to the second weak component at the trailing edge of the main pulse). The third weak component separates from the main peak for about 0.5 rotation, indicating that it is possibly the interpulse of this pulsar. A PA jump is seen at the trailing edge of the main peak (phase 0.57), accompanied by the sense reversal of V. Similar to J2214$+$3000, the fitting result on the PA curve gives $\alpha=8.1^\circ\pm7.1^\circ$ and $\zeta=14.7^\circ\pm12.7^\circ$, which suggests a nearly aligned rotator.

PSR J2234$+$0944:  The pulsar is potentially radiating for the full rotation. We found a weak component at phase 0.2, whose intensity is only 0.2\% of that of the highest peak. There is bridge emission of less than 1\% between the two pulse peaks (phase 0.7). Both orthogonal and non-orthogonal PA jumps occur in this pulsar. In addition, the PA curve of the main pulse precursor seems to follow an S-shaped swing.

PSR J2302$+$4442: We identify a weak, distinct, and narrow component around phase 0.9 and weak bridge emission between the main pulse and the interpulse around phase 0.35. These components make this pulsar a potential $360^\circ$ radiator. A $90^\circ$ PA jump occurs at phase 0.34, where V changes its sign. Although the PA curve cannot be described by the RVM model, the curves of the main pulse and the interpulse have separate S-shapes. 

PSR J2317$+$1439: We found a weak component with a plateau at the trailing edge of the main pulse (phase 0.9-1.0). The flux of this weak component is roughly 0.1\% of that of the highest peak. The PA curve has a complex shape. The degree of circular polarization peaks around the second pulse peak and coincides with a dip in the PA curve. Both linear and circular polarization reach the local minimum at jump phase 0.53. Circular polarization arrives at local maximum at phase 0.65.

PSR J2322$+$2057: No weak component is detected. The pulse profile shows two peaks with approximately 0.4 rotation separation. At phase 0.04 and 0.6, the PA curve presents jumps coinciding with the sense reversal of V.

\section{Conclusions}
\label{sec:conclu}

As a summary, the major conclusions in this paper are: 1) The polarization 
profiles for 56 CPTA pulsars are presented. Most of them are compatible with 
previous publications, but have higher S/Ns to shed light on the detail of 
structures. The polarization profiles of 
three pulsars -- PSR J0406$+$3039, J1327$+$3423, and J2022$+$2534 -- are published for the first time. 2) We find that there is no difference 
between the distribution functions of the polarization percentage between MSPs and 
normal pulsars. 3) Radiation below 3\% of the pulsar peak flux are detected for 
80\% of pulsars, which implies that the microcomponents may be a common feature 
among MSPs. In addition, some pulsars may sustain radiation over the whole rotation period. 4) PA jumps are 
detected in the majority of MSPs. Two types of PA jumps can coexist in the pulse profile, 
in which the type I PA jump happens when the polarization flux drops to 0 and the type II PA jump is 
accompanied by nonzero circular polarization. 5) Polarization properties 
imply that the wave propagation effects in the MSP magnetosphere are important for 
the shape of the polarization pulse profile.  

\section*{Data availability}
The data of polarization pulse profiles are available at \url{https://psr.pku.edu.cn/publications/CPTADR1/}, and the corresponding figures are published at \url{https://doi.org/10.5281/zenodo.14801349}.

\begin{acknowledgements}  
Observation of CPTA is supported by the FAST Key project. FAST is a Chinese national mega-science facility, operated by National Astronomical Observatories, Chinese Academy of Sciences. This work is supported by the National SKA Program of China (2020SKA0120100), the National Key Research and Development Program of China No.2022YFC2205203, the National Natural Science Foundation of China grant no. 12041303, 12250410246, 12173087 and 12063003, China Postdoctoral Science Foundation No. 2023M743518 and 2023M743516, Major Science and Technology Program of Xinjiang Uygur Autonomous Region No. 2022A03013-4, the CAS-MPG LEGACY project, and funding from the Max-Planck Partner Group. KJL acknowledges support from the XPLORER PRIZE.
\end{acknowledgements}

\bibliographystyle{aa}
\bibliography{ref} 




\begin{appendix}
\section{Comparison of polarization profiles at small and large zenith angles}
\label{sec:compara}

Although the backward illumination strategy facilitates FAST to observe sources with zenith angles larger than 26.4$^\circ$, the main beam axis is not aligned with principal axis of the illumination area in this case. The misalignment causes angular separations between the left and right circular beams, known as beam squint and elliptical illumination patterns called as beam squash~\citep{chu1973,Heiles_2001,2002heiles,2021Robishaw}. Consequently, the polarization profiles derived at large zenith angles (>26.4$^\circ$) may be different from those obtained at small zenith angles ($\leqslant$26.4$^\circ$). 

Nearly half of the CPTA pulsars have both observations taken at small and large zenith angles. For each pulsar, after combining all observations at both kinds of zenith angles separately, a rough estimation on the effect of backward illumination can be made by comparing the derived integrated polarization profiles. Because of the three reasons, 1) such effect may be frequency dependent, 2) pulse profile evolves along the frequency, and 3) ISM effects, such as scintillation and scattering, are frequency dependent, we need to compare polarization profiles per frequency channel. The observation band is  thus divided into several sub-bands. For each channel, we compute the difference in polarization profiles between the large and small zenith angle observation. 

To obtain reliable result, we align, rescale and rebaseline the two pulse profiles 
and then subtract them to get the difference. The profiles 
alignment is performed in the Fourier domain \citep{Taylor92}. Then, since the two profiles 
have different S/N and baselines, they need to be matched before the subtraction 
as explained by \citet{Men2019,Jiang_2022}. The profile matching is obtained 
through minimizing $\chi^2$ defined as,
\begin{equation}
\chi^2_{\nu} = \left(\mathbfcal{P}_{\nu}-\alpha_\nu\mathbfcal{T}_{\nu}-\beta_\nu\right)^2,
\end{equation}
where $\mathbfcal{T}_{\nu}$ and $\mathbfcal{P}_{\nu}$ are polarization profile vectors normalised by respective noise levels at central frequency $\nu$. The integrated profiles at small zenith angle is taken as template $\mathbfcal{T}_{\nu}$. The scalar parameters $\alpha_\nu$ and $\beta_\nu$ are  S/N scaled and profile baseline offset for the large zenith observations. The minimization of $\chi^2$ leads to 
\begin{equation}
    \begin{aligned}
        &\alpha_\nu = \frac{\mathbfcal{P}_{\nu}\cdot\mathbfcal{T}_{\nu}-\frac{1}{N}\sum_i\mathcal{P}_{\nu,i}\sum_i\mathcal{T}_{\nu,i}}{\sum_i\mathcal{T}_{\nu,i}^2-\frac{1}{N}\sum_i\mathcal{T}_{\nu,i}\sum_i\mathcal{T}_{\nu,i}},\\
        &\beta_\nu = \frac{1}{N}\sum_i(\mathcal{P}_{\nu,i}-\alpha\mathcal{T}_{\nu,i}),
    \end{aligned}
\end{equation}
where $N$ is the number of bins of the profiles. The residual profiles are then accumulated across the frequency to obtain the final profile differences that
\begin{equation}
    \delta \mathbfcal{P} = \int\left(\mathbfcal{P}_{\nu}-\alpha_\nu\mathbfcal{T}_{\nu}-\beta_\nu\right) d\nu\,.
\end{equation}

\begin{figure}[!ht]
\begin{minipage}[t]{0.49\columnwidth}
    \centering
    \includegraphics[width=1.0\columnwidth]{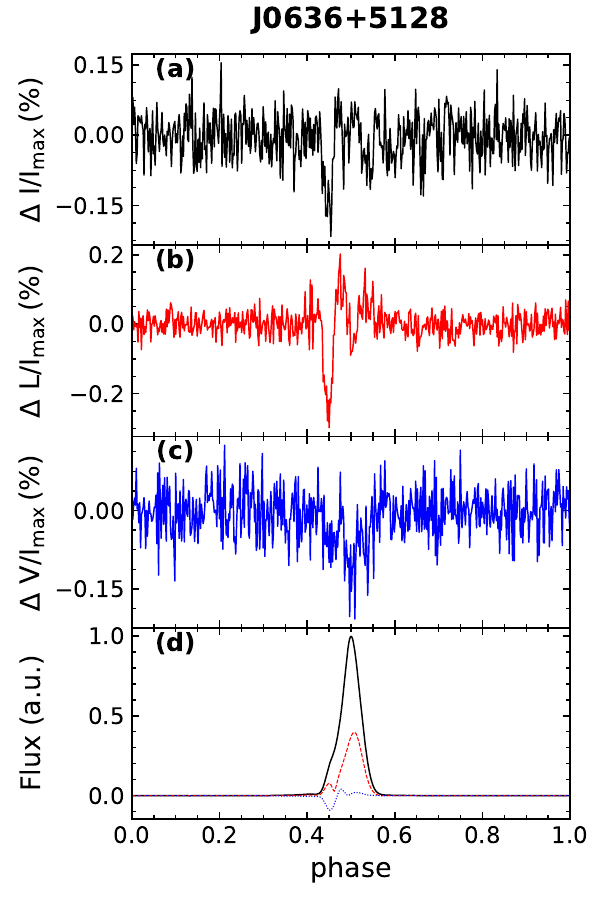}  
\end{minipage}
\begin{minipage}[t]{0.49\columnwidth}
    \centering
    \includegraphics[width=1.0\columnwidth]{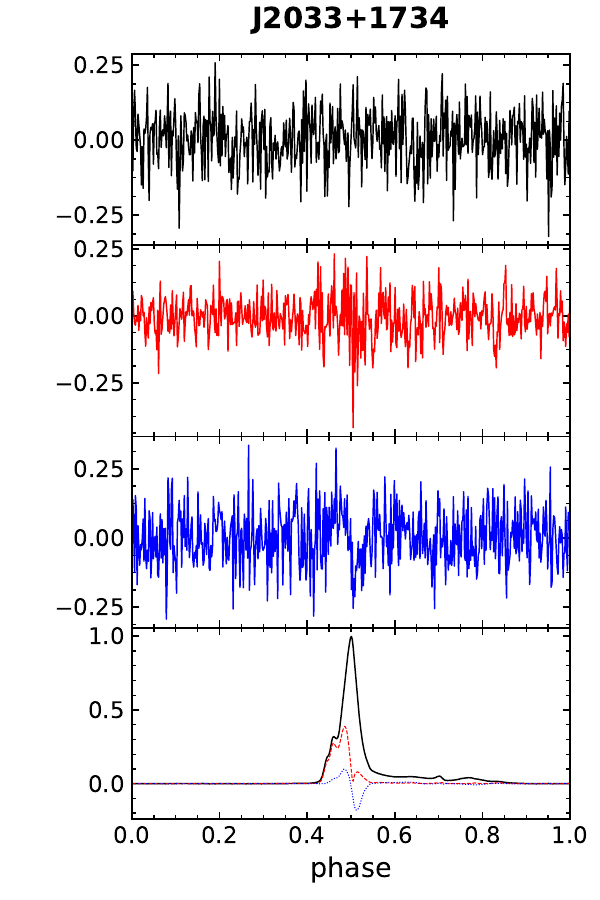}   
\end{minipage}

\caption{Polarization profile differences between small and large zenith angle observations for PSRs~J0636$+$5128 and J2033$+$1734, which contain enough number of observations. Panel (a), (b), (c) are the profile residuals for total intensity, linear and circular polarization, respectively. Panel (d) is the polarization profile from small zenith angle observations. The black, red, and blue represents total intensity, linear, and circular polarization, respectively. 
We normalize all the curves, such that the peak total intensity is 1. \label{fig:diffa}}
\end{figure}

Furthermore, to compare the difference in a reliable fashion, the integrated polarization profiles should be of high S/N for both small zenith and large zenith angles, which requires a long observation time. Only for PSRs~J0636$+$5128 and J2033$+$1734, we have enough observations for both conditions. As shown in \FIG{fig:diffa}, the maximal differences in polarization profiles between large and small zenith angle is at the level of 0.2\%, barely above noise floor to be visible. 
These findings are consistent with previous studies\citep{Jiang_2020,Jiang_2022}.

\section{Comparison of RM derived with the Bayesian and RM synthesis }
\label{sec:comprm}
\begin{figure*}
    \centering
    \includegraphics[width=0.9\textwidth]{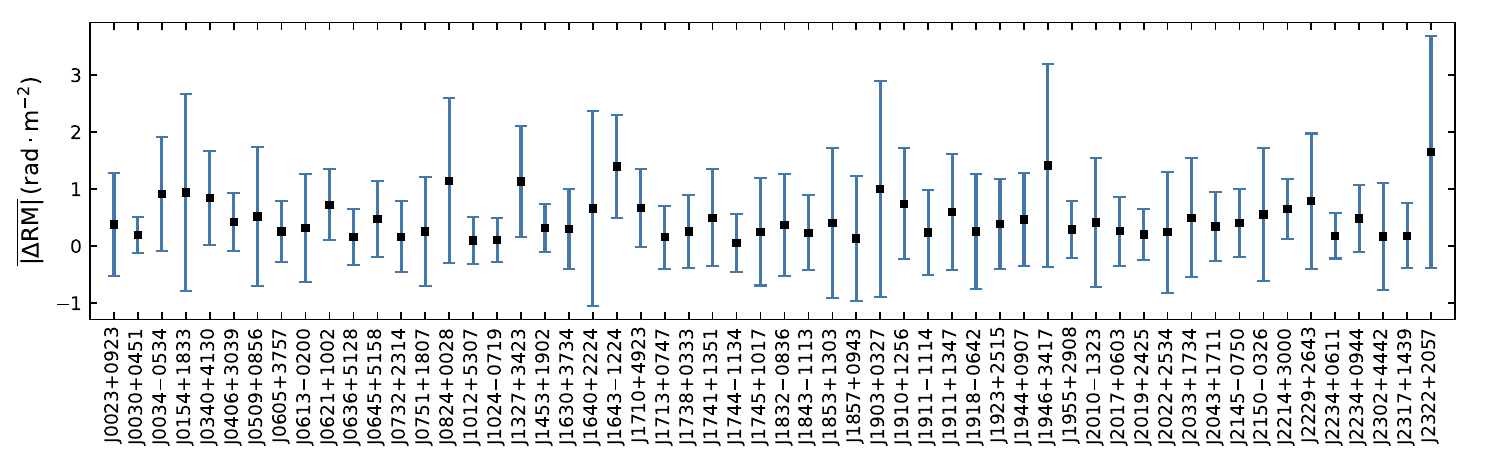}
    \caption{Averaged absolute differences between RM values derived with the Bayesian $Q$-$U$ fitting and the generalized RM synthesis. The error bars indicate the 68\% confidence level (1-$\sigma$).
    \label{fig:rmcompa}}
\end{figure*}

We had computed the RM value from two methods, the Bayesian $Q$-$U$ fitting and the generalized RM synthesis.
The averaged absolute differences of all observations between these two methods are given in \FIG{fig:rmcompa}. As one can see, the difference between the two method is close to zero for most of the CPTA pulsars, and the difference is within statistical fluctuation.
\FloatBarrier
\FloatBarrier

\section{Weak artifacts in dynamic spectra}
\label{sec:immrfi}
\begin{figure}
    \centering
    \includegraphics[width=1.05\linewidth]{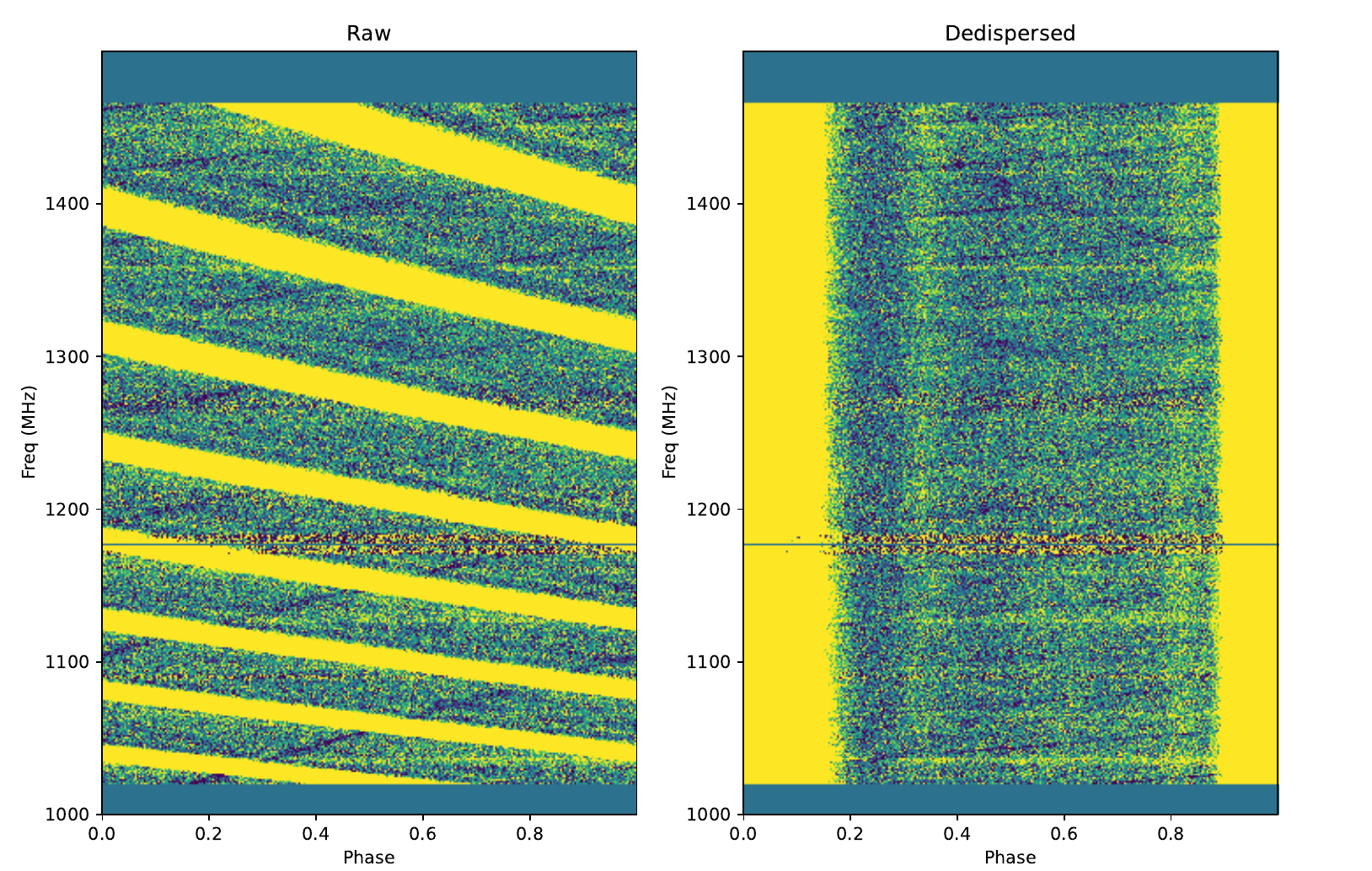}
    \caption{Dynamic spectrum of weak artifacts in observation for PSR~J1713$+$0747, i.e. signal flux as function of frequency (y-axis) and pulse phase (x-axis). Left: raw dynamic spectrum without dedispersion. Right: dedispersed dynamic spectrum. The dynamic spectrum is intentionally saturated to enhance visual inspection of the weak artifacts. 
    \label{fig:weak2}}
\end{figure}

We have detected weak artifacts in pulsar dynamic spectra shown in \FIG{fig:weak2}.
These dark inclined strip-like artifacts are detected mostly in bright pulsars, 
i.e. PSR~J1713$+$0747, J1744$-$1134 and J2145$-$0750.  For the three pulsars, the 
occurrence rates are $80\%$, $28\%$, and $31\%$ respectively.  We note that the 
amplitudes of these artifacts correlate to signal strengths of pulsar signals, 
as they are only found in observations with high SNRs. Additionally, 
PSRs~J1713$+$0747 and J2145$-$0750, which exhibit these artifacts most often, are 
the strongest pulsars in CPTA sample. In addition to the three pulsars mentioned 
above, we also detected such weak artifacts for a few occurrences ($<20\%$) in 
PSR~J0030$+$0451, J0621$+$1002, J0636$+$5128, J0645$+$5158, J1327$+$3423, J1640$+$2224, 
J1857$+$0943 and J2229$+$2643. 

Similar weak artifacts were also observed in other works \citep{Yuan_2023}, where it 
was claimed to be weak interference. We believe that the artifacts are not the 
radio frequency interferences (RFIs) themselves. RFIs are concentrated in narrow 
bands and are much brighter than pulsar signal, while the weak artifacts exhibit a 
lower flux density than the noise floor and occupy the full bandwidth.  
They look similar to a previous case \citep{Alam_2021}, where one can see 
artifacts as the mirror of dispersed signal about the central frequency.
In contrast to the 
artifact reported by \citet{Alam_2021}, the artifacts we found are significantly 
weaker, which is below the noise floor.
In addition, \citet{Alam_2021} concludes that the artifacts were induced by mismatch between the interleaved 
analog-to-digital converters \citep[ADC;][]{Kurosawa_2001}. It is unlikely that ADC alone causes artifacts we detected. The FAST digital backends adopt the KatADC board, which is mounted with the ADC 
chip ADC08D1520
(from \textsc{Texas Instruments} \url{https://casper.berkeley.edu/wiki/KatADC}). The two ADC cores are not used 
in the interleaved mode, so we do not expect the artifact is due to the same reason as in the case of \citet{Alam_2021}.

The artifact is probably not caused by the leakage of polyphase filter banks alone, which, instead of creating a `dip', introduces low amplitude power excess. 
We note that the artifacts follows a band-flipped DM signature as shown in Fig. \ref{fig:weak2}. The dip appears at the lowest frequency at the time pulsar signal enters the highest frequency channel. As the dispersed pulsar signal drifts towards lower frequency, the dip turns to higher frequency. Other artifacts are tracks of dips being parallel shifted of the dispersed signal in frequency for $1/4$ and $1/2$ of the full band. As the FAST backend design used a four-tap polyphaser filter bank, it is likely that the artifact is caused by a combination of polyphaser filter leakage and digitizing of strong signal.

\begin{figure}
\centering
    \includegraphics[width=0.8\columnwidth]{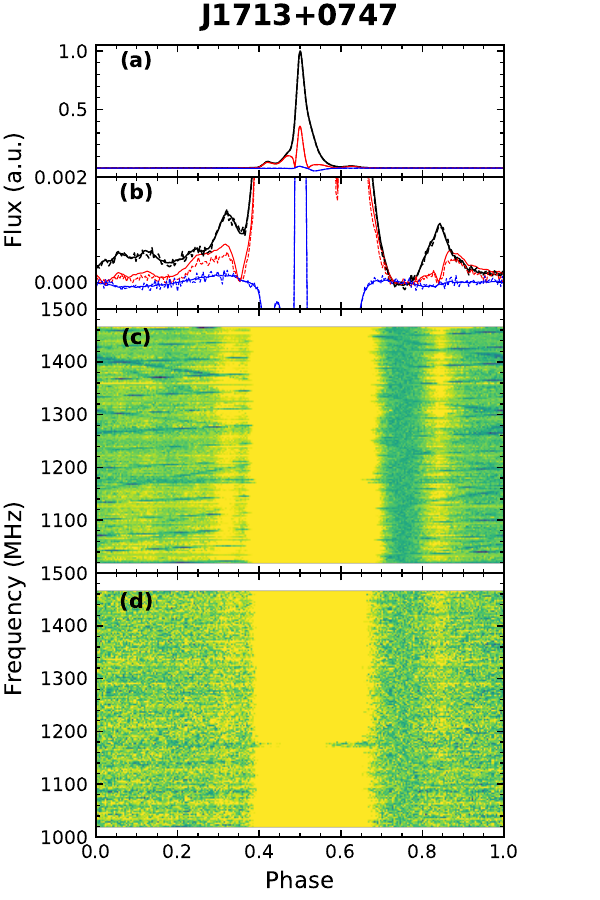}
    \caption{Comparison of profiles with and without weak artifacts. Panel a), polarization pulse profile of PSR J1713$+$0747 using the same color code as other plots in the paper. Panel b), pulse profiles from data with (solid curves) and without (dashed curves) artifact. Panel c) and d), dynamic spectra for data with and without artifact, respectively.  \label{fig:weak3}}
\end{figure}

We could not track down the exact reason for the artifacts, which is left for future detailed studies. Instead, we evaluate the impact of the artifacts on the data quality. We combine observations with and without weak artifacts separately to form and compare integrated polarization profiles as shown in \FIG{fig:weak3}. Luckily, we found that the major difference is only a slight offset of linear polarization intensity at the level of 0.1\%, well below any effects we are trying to address in the current work.

\FloatBarrier
\FloatBarrier

\section{Collection of plots and tables}
\subsection{Table of basic parameters for CPTA pulsars}
\begin{table*}
        \centering
        \caption{Parameters of 56 CPTA pulsars. All error bars are of 68\% confidence level. Systematics $\rm \Delta DM_{\rm sys}$ and $\rm \Delta RM_{\rm sys}$ are estimated from the standard deviation for all observations, i.e. the values are for the DM/RM fluctuation or evolution over time. \label{tab:small_summary}}
        \begin{tabular}{ccccccccc}
                \hline\hline
                Pulsar&Obs.dura.&N$_{\rm obs}$&Length   & DM&$\Delta{\rm DM_{sys}}$ & RM &$\Delta{\rm RM_{sys}}$&B$_{\parallel}$\\&MJD&& min  & $\mathrm{pc\cdot cm^{-3}}$& $10^{-4}\mathrm{pc\cdot cm^{-3}}$&$\mathrm{rad\cdot m^{-2}}$ &$\mathrm{rad\cdot m^{-2}}$&${\rm \mu G}$\\
                \hline
J0023$+$0923 & 58907$-$59840 & 56 & 1281.5 & 14.33201 (16) & 2.3 & -5.6 (8) & 0.6 & -0.50 $\pm$ 0.07 \\
J0030$+$0451 & 58883$-$59851 & 57 & 1290.0 & 4.33235 (4) & 2.0 & 1.55 (23) & 0.5 & 0.40 $\pm$ 0.13 \\
J0034$-$0534 & 58956$-$59827 & 13 & 358.6 & 13.7764 (11) & 6.0 & 9.2 (6) & 1.4 & 0.78 $\pm$ 0.11 \\
J0154$+$1833 & 58907$-$59697 & 23 & 450.0 & 19.79691 (10) & 1.2 & -19.6 (16) & 2.1 & -1.42 $\pm$ 0.07 \\
J0340$+$4130 & 58684$-$59840 & 66 & 1693.0 & 49.58564 (24) & 4.0 & 52.8 (5) & 0.8 & 1.332 $\pm$ 0.016 \\
J0406$+$3039 & 59403$-$59847 & 30 & 609.9 & 49.37599 (35) & 2.5 & 64.00 (26) & 0.4 & 1.588 $\pm$ 0.009 \\
J0509$+$0856 & 58962$-$59848 & 35 & 681.9 & 38.32843 (18) & 3.1 & 43.4 (11) & 0.8 & 1.410 $\pm$ 0.024 \\
J0605$+$3757 & 58962$-$59830 & 14 & 268.9 & 20.9433 (4) & 5.0 & 4.34 (24) & 0.4 & 0.247 $\pm$ 0.019 \\
J0613$-$0200 & 58686$-$59849 & 67 & 1526.5 & 38.78188 (11) & 2.5 & 19.5 (6) & 0.9 & 0.687 $\pm$ 0.027 \\
J0621$+$1002 & 58666$-$59849 & 66 & 1682.5 & 36.56186 (25) & 27.0 & 52.0 (5) & 0.7 & 1.770 $\pm$ 0.017 \\
J0636$+$5128 & 58687$-$59823 & 34 & 807.0 & 11.10857 (6) & 0.6 & -4.22 (28) & 1.1 & -0.49 $\pm$ 0.06 \\
J0645$+$5158 & 58687$-$59850 & 60 & 1572.2 & 18.25136 (12) & 1.0 & -1.92 (30) & 1.2 & -0.13 $\pm$ 0.05 \\   
J0732$+$2314 & 58961$-$59830 & 22 & 409.8 & 44.66919 (25) & 3.2 & -4.3 (4) & 0.4 & -0.122 $\pm$ 0.013 \\
J0751$+$1807 & 58668$-$59847 & 69 & 1719.1 & 30.241064 (31) & 2.9 & 42.83 (24) & 0.6 & 1.740 $\pm$ 0.021 \\
J0824$+$0028 & 58961$-$59830 & 16 & 278.7 & 34.5469 (4) & 5.0 & 38.4 (11) & 0.4 & 1.349 $\pm$ 0.032 \\
J1012$+$5307 & 58684$-$59842 & 67 & 1574.5 & 9.02113 (5) & 1.1 & 2.28 (22) & 0.8 & 0.30 $\pm$ 0.04 \\
J1024$-$0719 & 58947$-$59842 & 63 & 1459.1 & 6.48699 (6) & 1.2 & -2.77 (20) & 0.6 & -0.53 $\pm$ 0.06 \\
J1327$+$3423 & 59408$-$59855 & 28 & 598.0 & 4.1863 (5) & 5.0 & -3.9 (9) & 1.0 & -1.06 $\pm$ 0.35 \\
J1453$+$1902 & 58686$-$59841 & 46 & 1222.0 & 14.05479 (24) & 2.4 & 3.23 (27) & 0.7 & 0.300 $\pm$ 0.031 \\
J1630$+$3734 & 58887$-$59843 & 41 & 795.8 & 14.12885 (9) & 1.4 & 1.0 (6) & 0.6 & 0.11 $\pm$ 0.06 \\
J1640$+$2224 & 58709$-$59848 & 66 & 1709.9 & 18.42912 (4) & 1.6 & 20.7 (16) & 1.1 & 1.51 $\pm$ 0.07 \\
J1643$-$1224 & 58883$-$59846 & 53 & 1241.9 & 62.39833 (6) & 15.0 & -305.6 (8) & 1.0 & -6.040 $\pm$ 0.018 \\
J1710$+$4923 & 58967$-$59839 & 20 & 369.9 & 7.08647 (13) & 1.6 & 7.5 (6) & 0.9 & 1.34 $\pm$ 0.18 \\
J1713$+$0747 & 58707$-$59318 & 59 & 3096.8 & 15.987098 (26) & 0.4 & 11.36 (21) & 0.4 & 0.876 $\pm$ 0.017 \\
J1738$+$0333 & 58709$-$59847 & 62 & 1467.5 & 33.76716 (4) & 6.0 & 34.12 (31) & 0.6 & 1.227 $\pm$ 0.015 \\
J1741$+$1351 & 58686$-$59845 & 66 & 1670.1 & 24.195740 (34) & 4.0 & 63.1 (6) & 0.7 & 3.20 $\pm$ 0.04 \\
J1744$-$1134 & 58884$-$59844 & 57 & 1315.3 & 3.138334 (16) & 0.9 & 1.78 (17) & 0.7 & 0.72 $\pm$ 0.05 \\
J1745$+$1017 & 58967$-$59839 & 20 & 368.5 & 23.97061 (12) & 5.0 & 26.0 (4) & 0.7 & 1.371 $\pm$ 0.032 \\
J1832$-$0836 & 58889$-$59850 & 52 & 1206.0 & 28.19076 (4) & 4.0 & 41.8 (6) & 1.0 & 1.82 $\pm$ 0.04 \\
J1843$-$1113 & 58919$-$59850 & 47 & 1059.4 & 59.96115 (14) & 3.4 & 9.32 (21) & 0.7 & 0.193 $\pm$ 0.006 \\
J1853$+$1303 & 58710$-$59838 & 51 & 1201.2 & 30.57140 (8) & 0.8 & 77.8 (10) & 0.8 & 3.13 $\pm$ 0.04 \\
J1857$+$0943 & 58706$-$59844 & 54 & 1245.5 & 13.29843 (4) & 3.4 & 24.0 (4) & 0.4 & 2.25 $\pm$ 0.04 \\
J1903$+$0327 & 58906$-$59844 & 47 & 1049.0 & 297.5275 (9) & 40.0 & 233.4 (18) & 1.8 & 0.968 $\pm$ 0.009 \\
J1910$+$1256 & 58709$-$59850 & 49 & 1115.9 & 38.06725 (5) & 5.0 & 53.5 (6) & 0.8 & 1.718 $\pm$ 0.031 \\
J1911$-$1114 & 58890$-$59844 & 51 & 1176.1 & 30.96652 (13) & 7.0 & -28.8 (5) & 0.7 & -1.144 $\pm$ 0.020 \\
J1911$+$1347 & 58709$-$59844 & 56 & 1288.9 & 30.978352 (21) & 1.9 & -7.18 (35) & 0.4 & -0.350 $\pm$ 0.018 \\
J1918$-$0642 & 58906$-$59850 & 50 & 1117.0 & 26.58877 (4) & 1.2 & -58.2 (4) & 0.9 & -2.69 $\pm$ 0.04 \\
J1923$+$2515 & 58709$-$59844 & 49 & 1140.4 & 18.86027 (12) & 2.8 & 12.6 (5) & 0.6 & 0.84 $\pm$ 0.04 \\
J1944$+$0907 & 58984$-$59850 & 48 & 1088.0 & 24.35836 (34) & 4.0 & -36.1 (5) & 0.6 & -1.81 $\pm$ 0.04 \\
J1946$+$3417 & 58969$-$59844 & 48 & 1088.5 & 110.20096 (12) & 40.0 & 3.8 (18) & 1.0 & 0.029 $\pm$ 0.022 \\
J1955$+$2908 & 58884$-$59850 & 54 & 1266.1 & 104.49092 (17) & 10.0 & 14.45 (34) & 0.6 & 0.172 $\pm$ 0.008 \\
J2010$-$1323 & 58907$-$59844 & 48 & 1056.1 & 22.16240 (6) & 1.5 & -1.56 (31) & 0.6 & -0.085 $\pm$ 0.026 \\
J2017$+$0603 & 58906$-$59821 & 49 & 1137.6 & 23.92264 (9) & 2.0 & -58.29 (28) & 0.6 & -2.981 $\pm$ 0.018 \\
J2019$+$2425 & 58890$-$59844 & 52 & 1229.5 & 17.19986 (21) & 2.2 & -67.60 (24) & 0.4 & -4.838 $\pm$ 0.025 \\
J2022$+$2534 & 59409$-$59850 & 27 & 539.2 & 53.66169 (9) & 1.5 & -173.3 (5) & 0.4 & -3.979 $\pm$ 0.011 \\
J2033$+$1734 & 58886$-$59457 & 33 & 858.0 & 25.06888 (34) & 3.1 & -71.8 (6) & 0.4 & -3.525 $\pm$ 0.032 \\
J2043$+$1711 & 58884$-$59814 & 35 & 817.1 & 20.71497 (12) & 2.7 & -72.75 (28) & 0.5 & -4.336 $\pm$ 0.028 \\
J2145$-$0750 & 58885$-$59852 & 64 & 1477.2 & 9.00785 (14) & 1.9 & -0.4 (4) & 0.8 & -0.12 $\pm$ 0.06 \\
J2150$-$0326 & 59392$-$59840 & 31 & 618.5 & 20.67409 (6) & 2.0 & 7.3 (7) & 0.8 & 0.43 $\pm$ 0.04 \\
J2214$+$3000 & 58883$-$59851 & 62 & 1441.6 & 22.55644 (8) & 3.2 & -45.41 (31) & 0.5 & -2.519 $\pm$ 0.028 \\
J2229$+$2643 & 58883$-$59850 & 61 & 1438.9 & 22.72851 (10) & 2.5 & -62.3 (10) & 1.2 & -3.36 $\pm$ 0.07 \\
J2234$+$0611 & 58885$-$59844 & 65 & 1485.2 & 10.76582 (5) & 4.0 & 2.60 (17) & 0.5 & 0.291 $\pm$ 0.021 \\
J2234$+$0944 & 58922$-$59846 & 64 & 1499.0 & 17.83440 (6) & 7.0 & -10.8 (4) & 0.5 & -0.772 $\pm$ 0.031 \\
J2302$+$4442 & 58922$-$59851 & 62 & 1465.3 & 13.71841 (9) & 4.0 & 17.68 (19) & 0.6 & 1.594 $\pm$ 0.030 \\
J2317$+$1439 & 58885$-$59840 & 69 & 1606.4 & 21.89851 (6) & 2.6 & -9.69 (25) & 0.4 & -0.559 $\pm$ 0.023 \\
J2322$+$2057 & 58885$-$59847 & 58 & 1337.5 & 13.38275 (9) & 1.6 & -32.2 (19) & 1.0 & -2.82 $\pm$ 0.14 \\
                \hline\hline
        \end{tabular}
\end{table*}
\subsection{Polarization pulse profiles of all 56 MSPs}

\begin{figure*}
\centering
\begin{minipage}[t]{0.66\columnwidth}
    \centering
    \includegraphics[width=1\columnwidth]{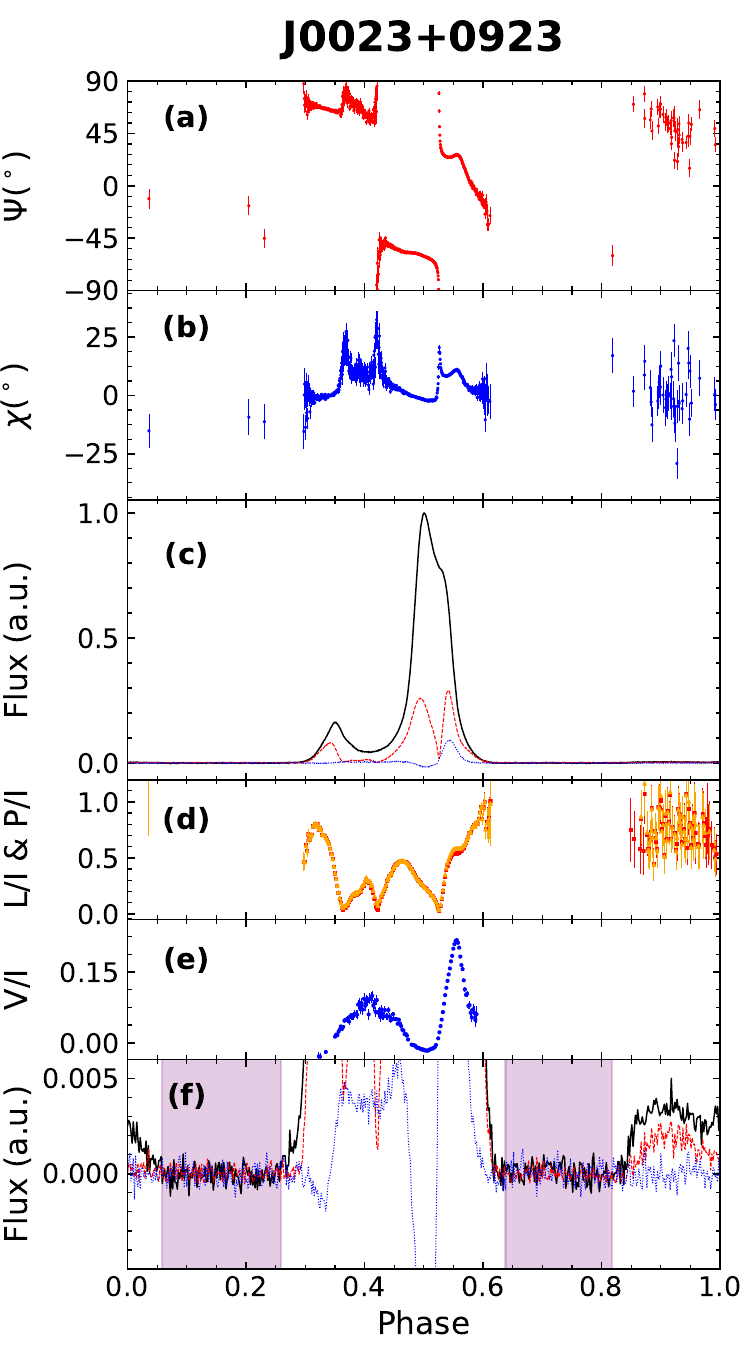}  
\end{minipage}
\begin{minipage}[t]{0.66\columnwidth}
    \centering
    \includegraphics[width=1\columnwidth]{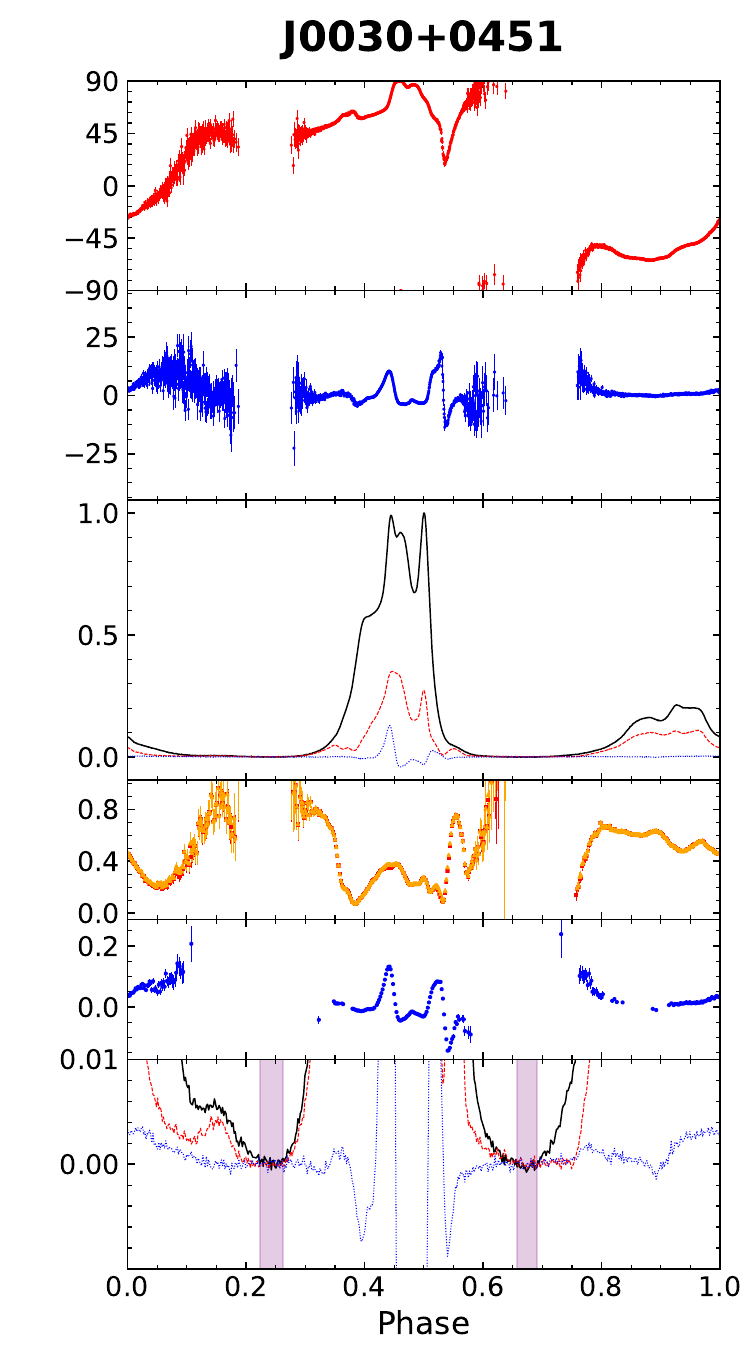}   
\end{minipage}
\begin{minipage}[t]{0.66\columnwidth}
    \centering
    \includegraphics[width=1\columnwidth]{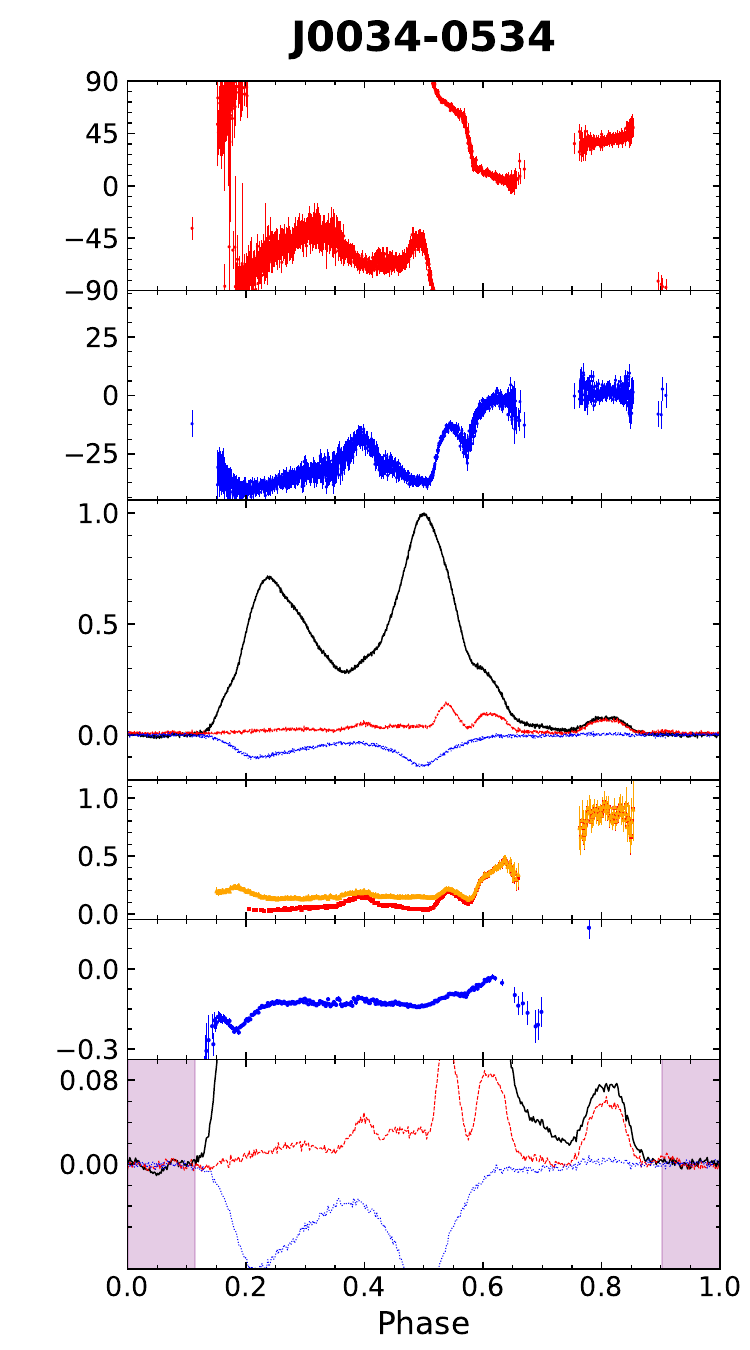}   
\end{minipage}\\
\begin{minipage}[t]{0.66\columnwidth}
    \centering
    \includegraphics[width=1\columnwidth]{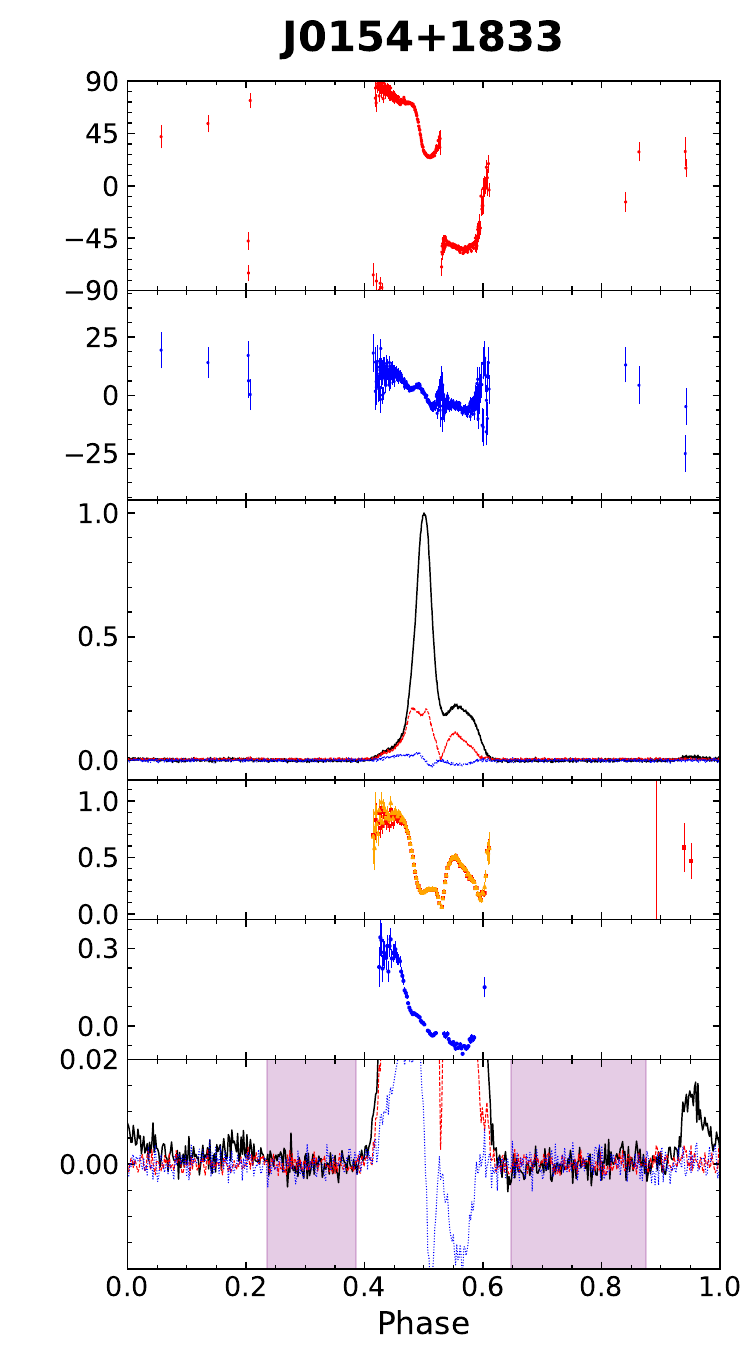}   
\end{minipage}
\begin{minipage}[t]{0.66\columnwidth}
    \centering
    \includegraphics[width=1\columnwidth]{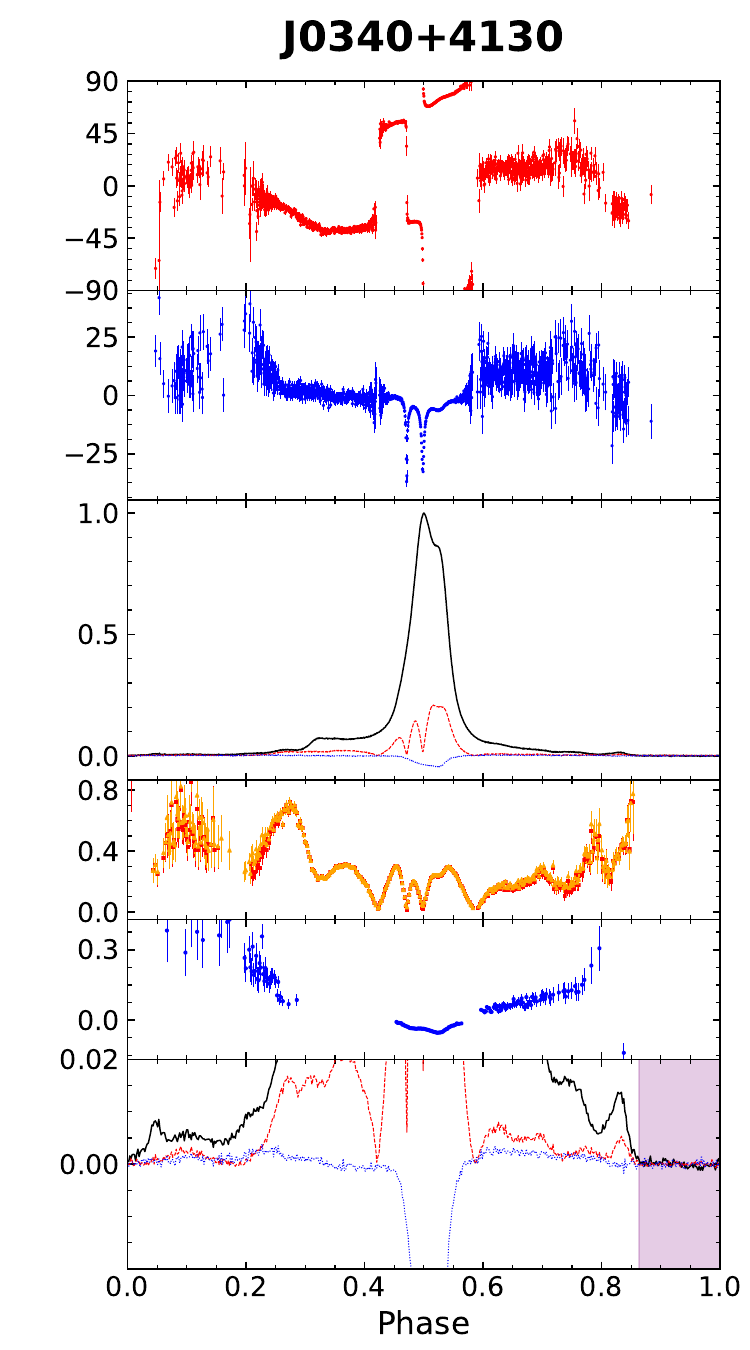}   
\end{minipage}
\begin{minipage}[t]{0.66\columnwidth}
    \centering
    \includegraphics[width=1\columnwidth]{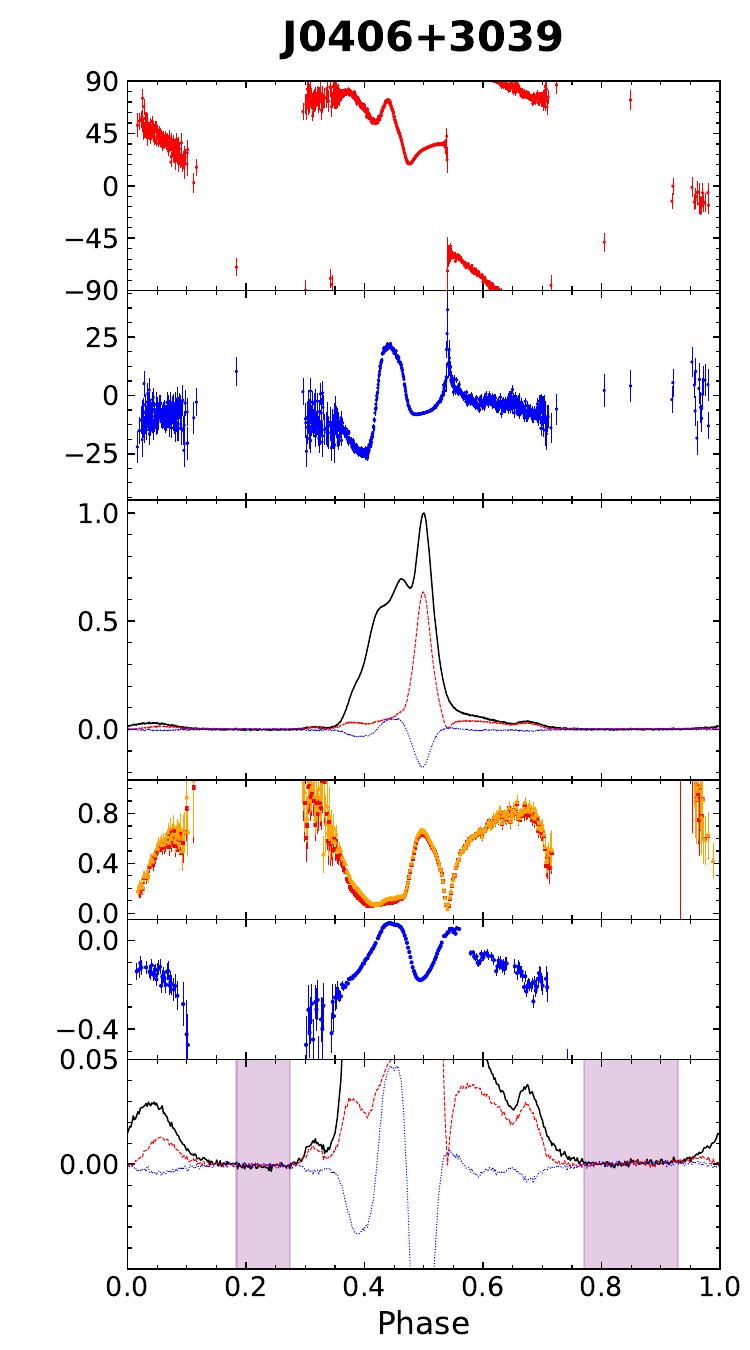}  
\end{minipage}
\caption{Integrated polarization profiles of CPTA pulsars.  Panels from (a) to (f) are linear polarization position angle ($\Psi$), 
ellipticity angles ($\chi$), pulse profiles (black curves for $I$, red dashed 
curves for $L$, and blue dotted curves for $V$), degree of total ($P/I$; orange dots), linear ($L/I$; red dots), and circular ($V/I$; blue dots) polarization, and zoom-in pulse profile to see 
the polarization of lower fluxes. Here, we only show $\Psi$, $\chi$, $P/I$, $L/I$ and 
$V/I$ with significant statistics, i.e. when total polarization intensity ($P$) 
is larger than $3\sigma_P$. We mark the phases with no significant polarization 
intensity in the zoomed-in pulse profile panels with the purple shades. Visual 
inspection is applied to make sure the pulse signal is minimal in the red 
shades.  Data of those region is used to infer the baseline of pulse profiles.  
\label{fig:cpta_pol1}}
\end{figure*}

\begin{figure*}
\ContinuedFloat
\centering
\FIGSWITCH{
\begin{minipage}[t]{0.66\columnwidth}
    \centering
    \includegraphics[width=1\columnwidth]{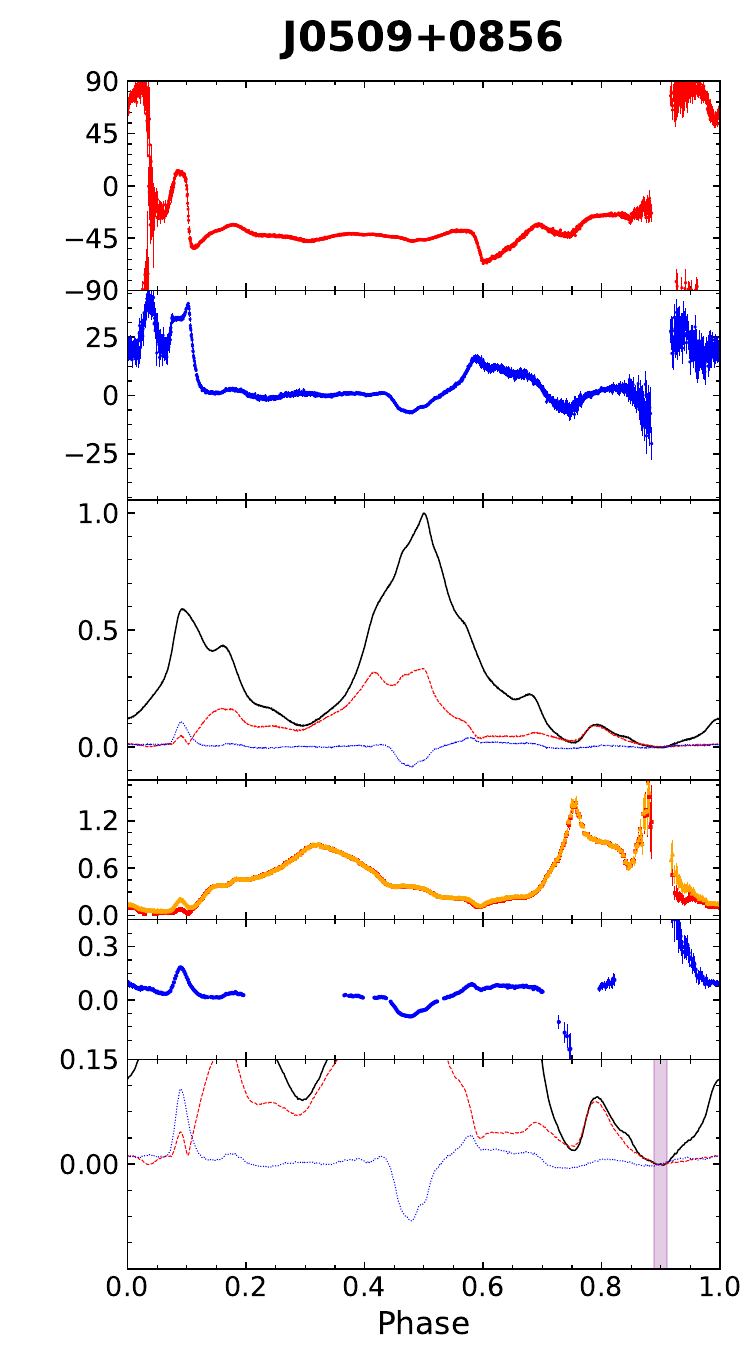}  
\end{minipage}
\begin{minipage}[t]{0.66\columnwidth}
    \centering
    \includegraphics[width=1\columnwidth]{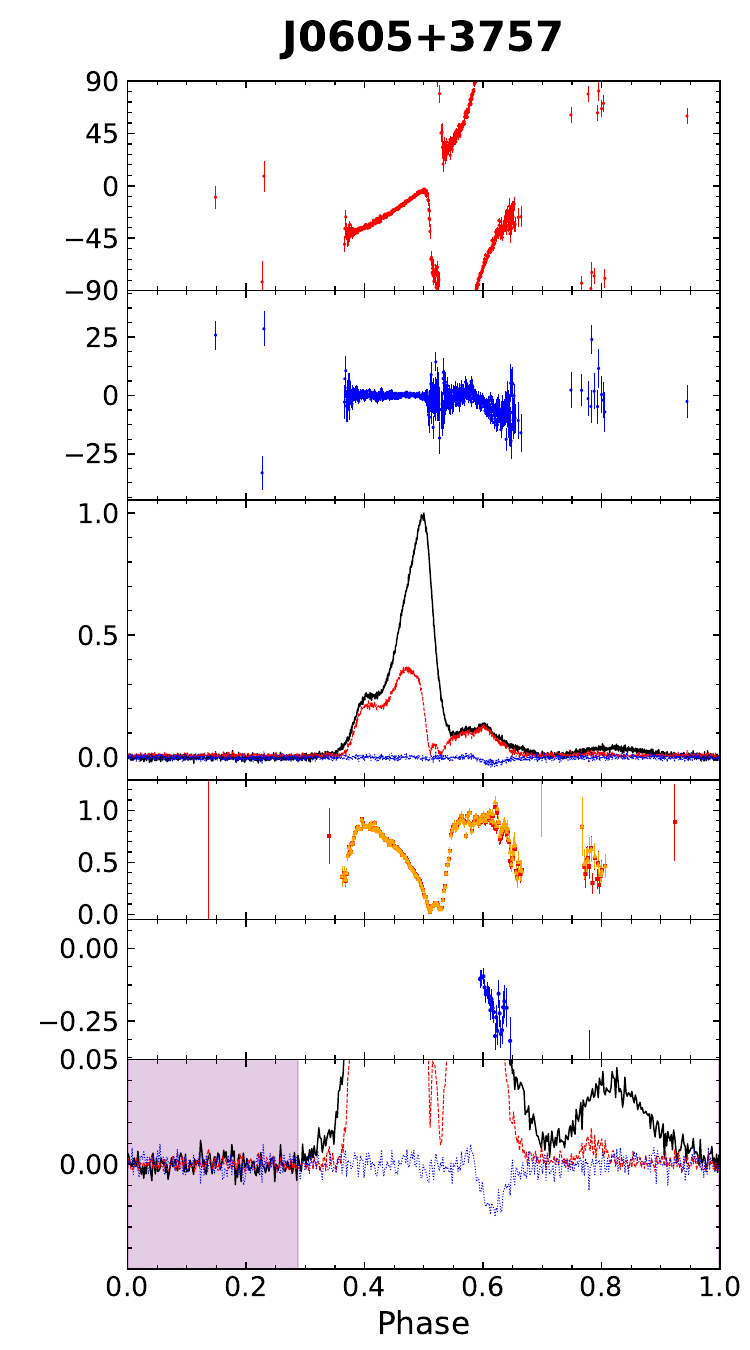}   
\end{minipage}
\begin{minipage}[t]{0.66\columnwidth}
    \centering
    \includegraphics[width=1\columnwidth]{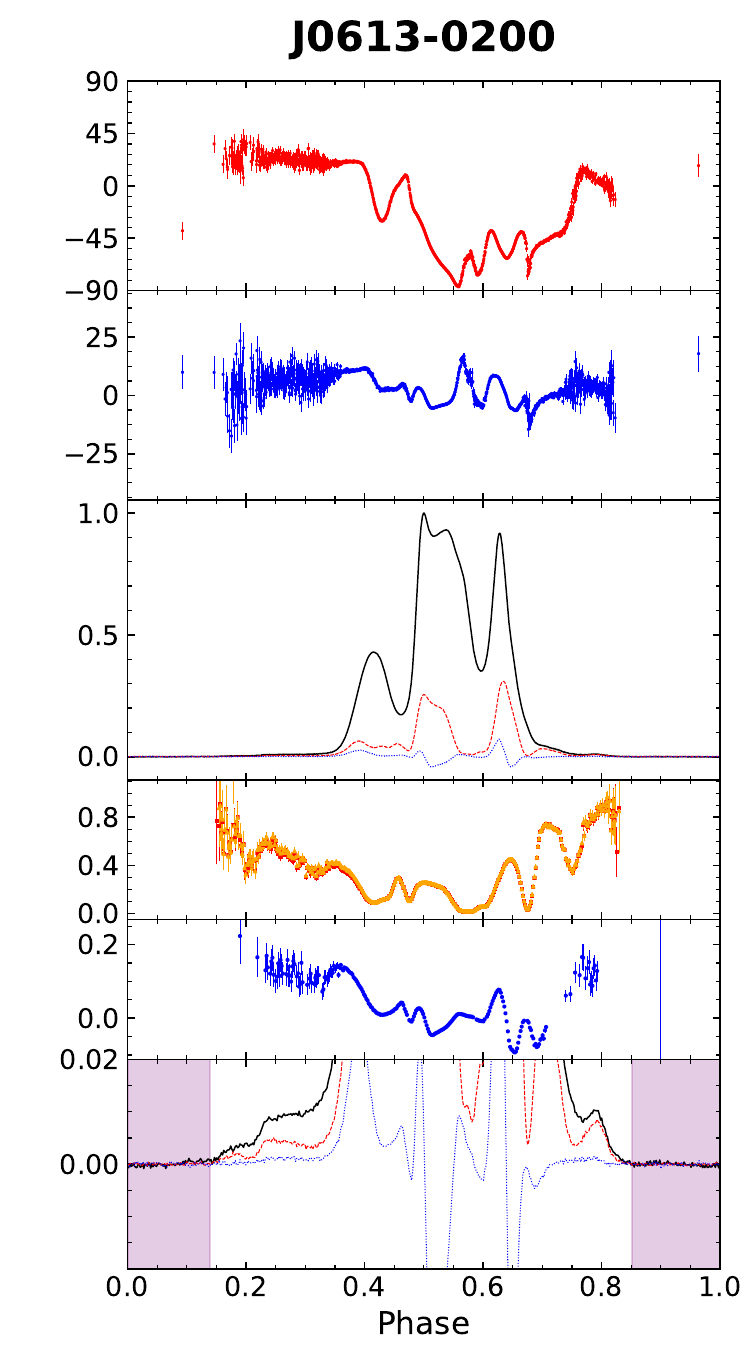}   
\end{minipage}\\
\hspace{1mm}
\begin{minipage}[t]{0.66\columnwidth}
    \centering
    \includegraphics[width=1\columnwidth]{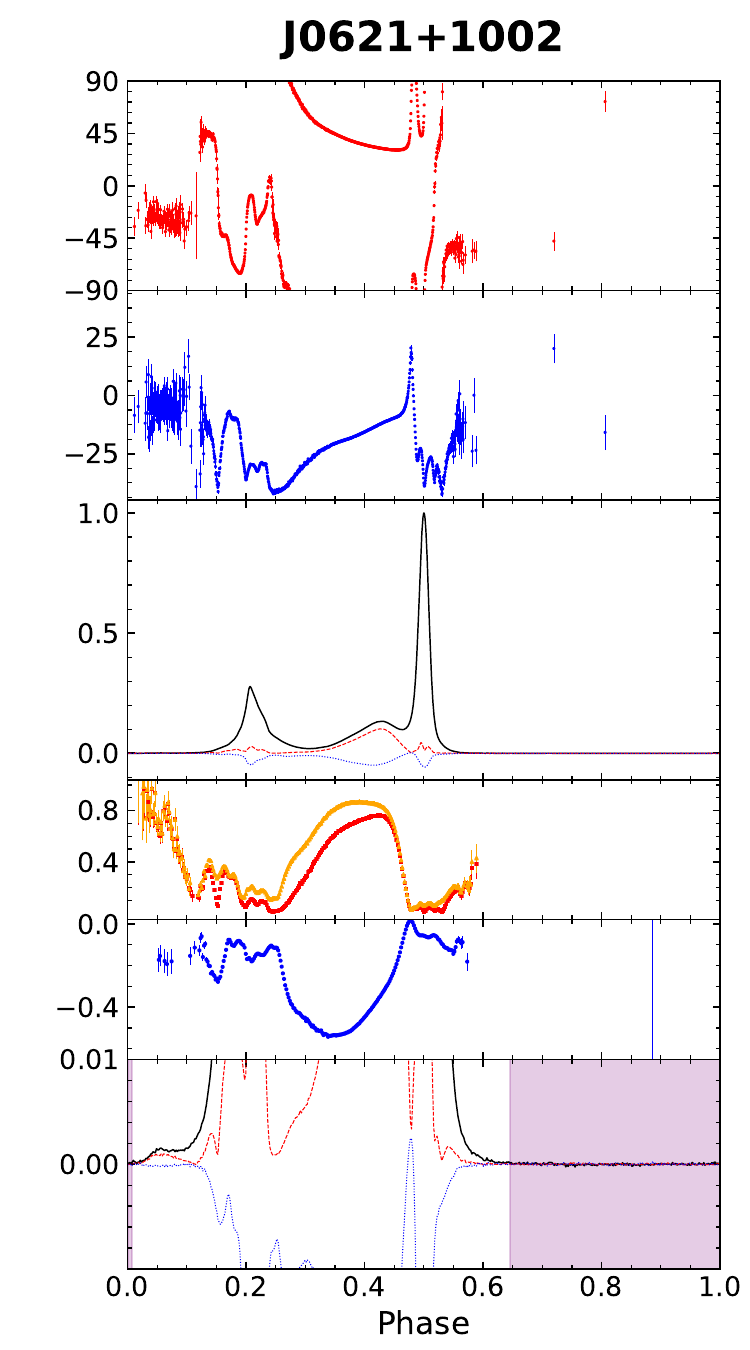}   
\end{minipage}
\begin{minipage}[t]{0.66\columnwidth}
    \centering
    \includegraphics[width=1\columnwidth]{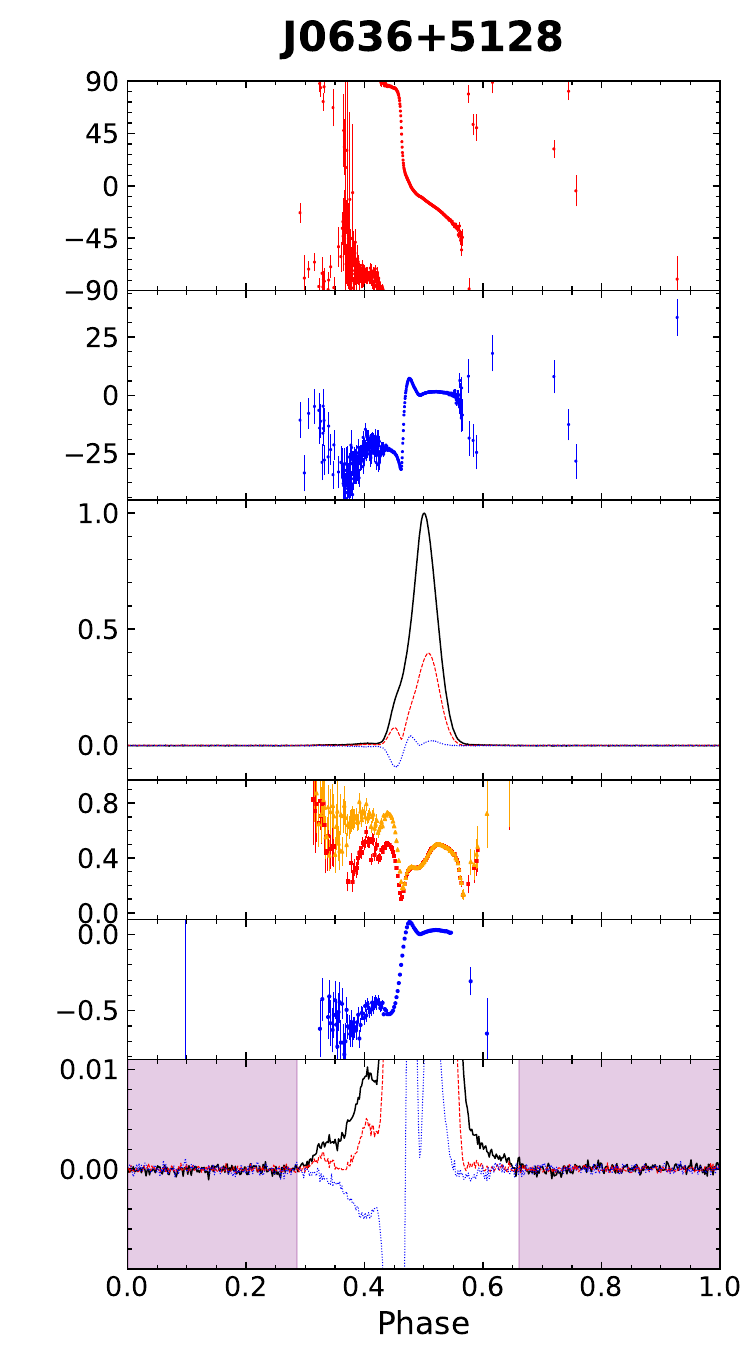}   
\end{minipage}
\begin{minipage}[t]{0.66\columnwidth}
    \centering
    \includegraphics[width=1\columnwidth]{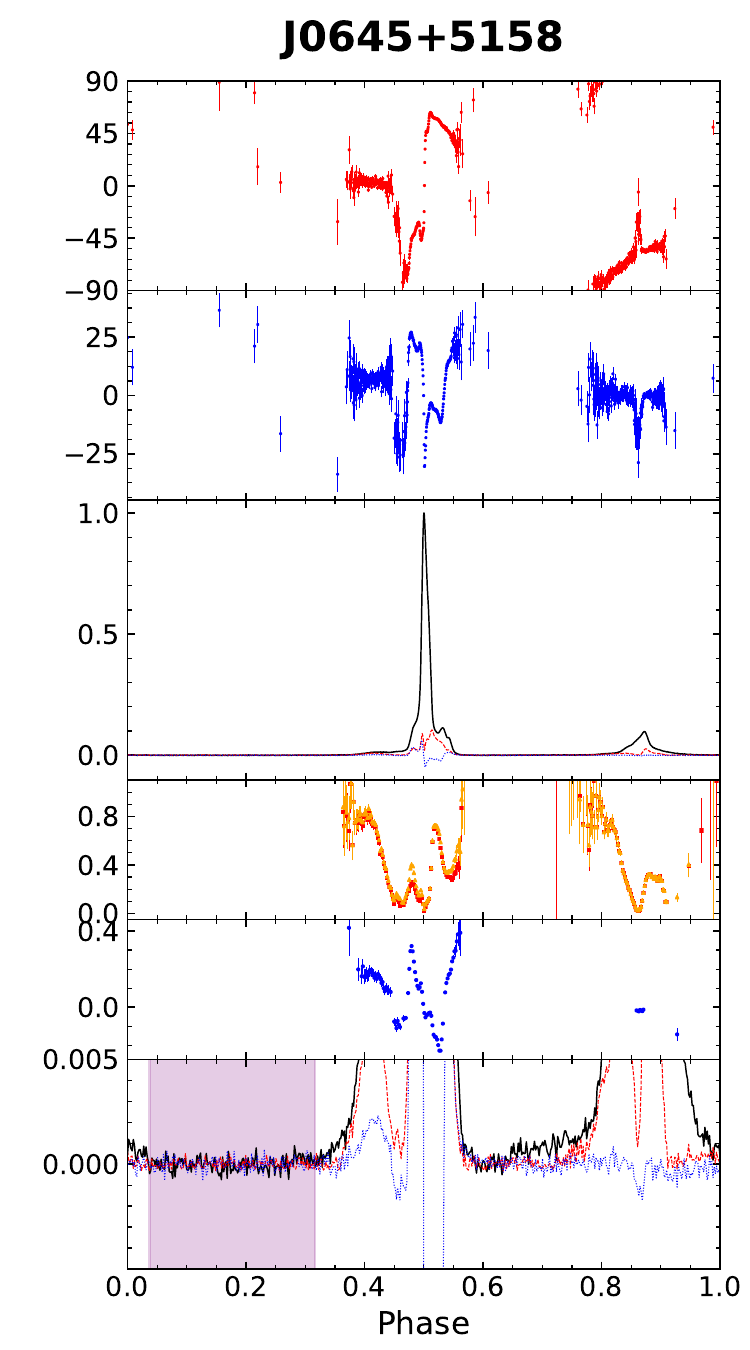}   
\end{minipage}
}
\caption{Continued.}
\end{figure*}

\begin{figure*}
\ContinuedFloat
\centering
\begin{minipage}[t]{0.66\columnwidth}
    \centering
    \includegraphics[width=1\columnwidth]{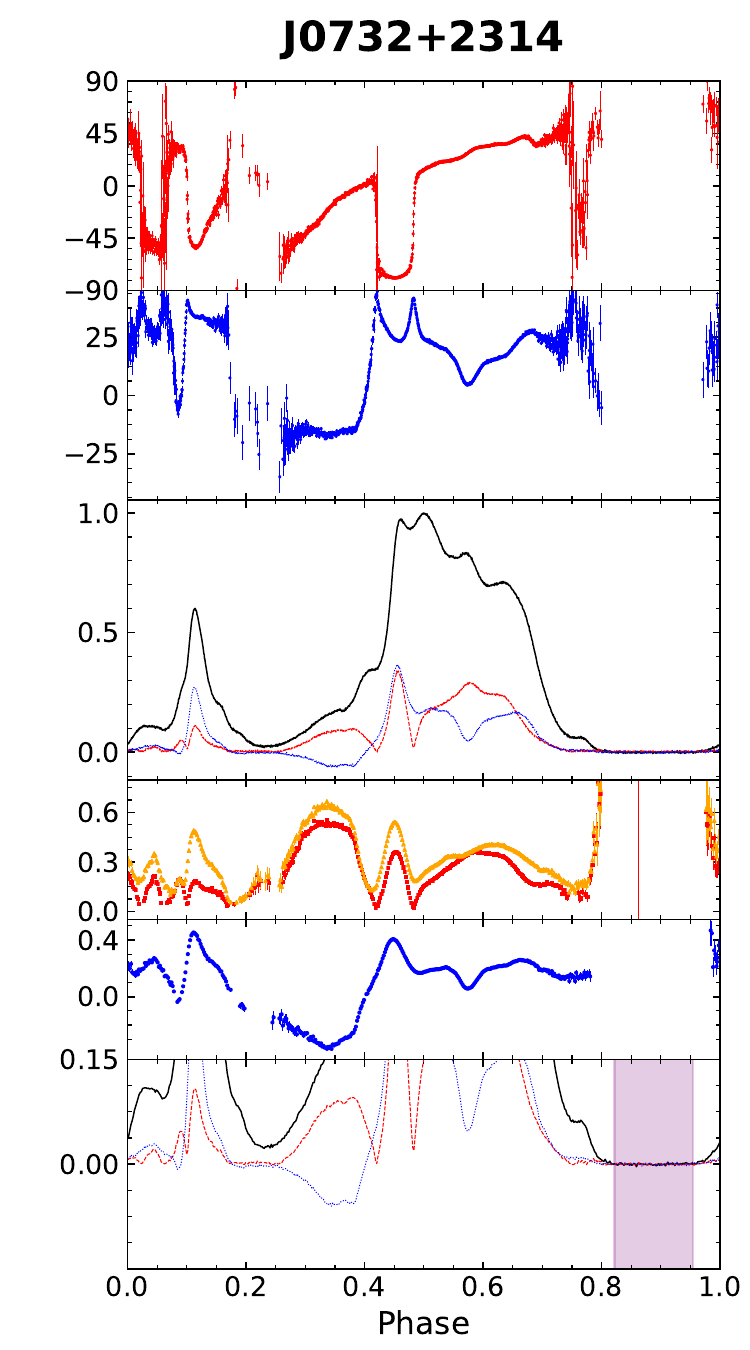}   
\end{minipage}
\begin{minipage}[t]{0.66\columnwidth}
    \centering
    \includegraphics[width=1\columnwidth]{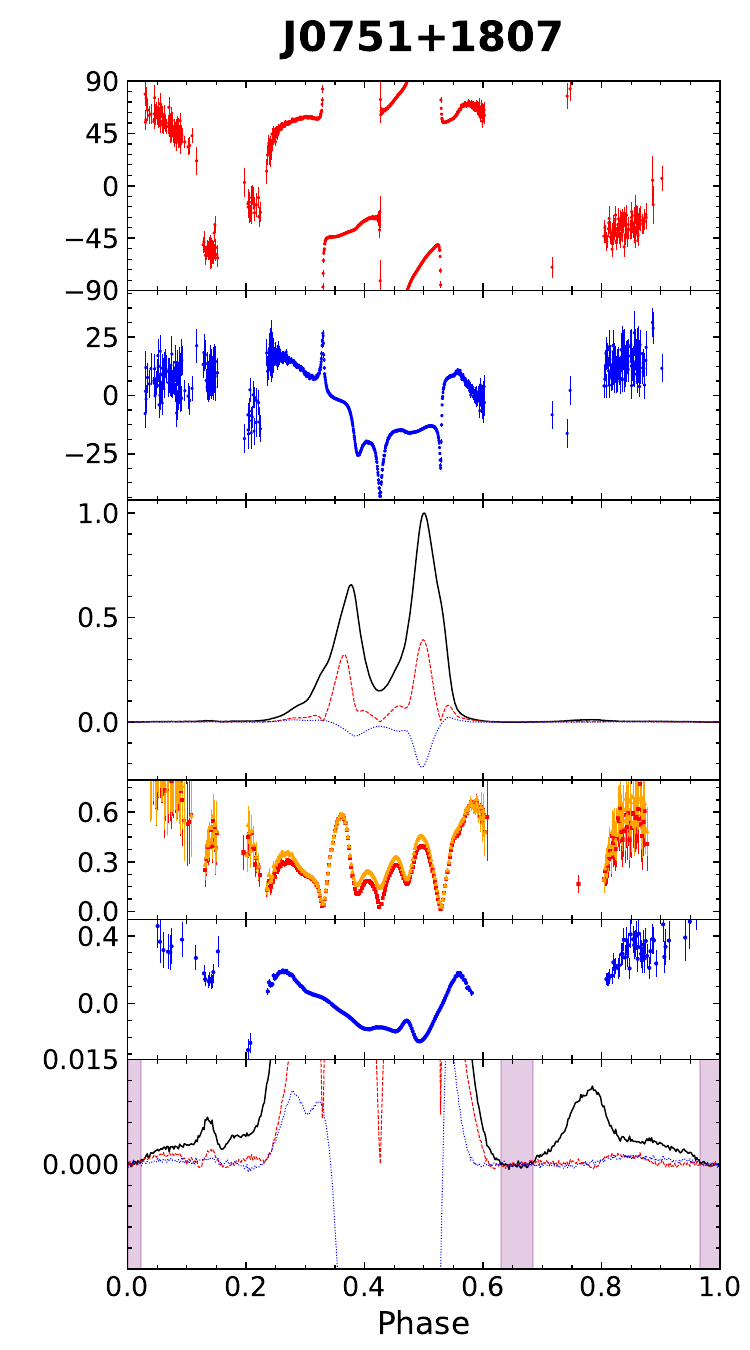}   
\end{minipage}
\begin{minipage}[t]{0.66\columnwidth}
    \centering
    \includegraphics[width=1\columnwidth]{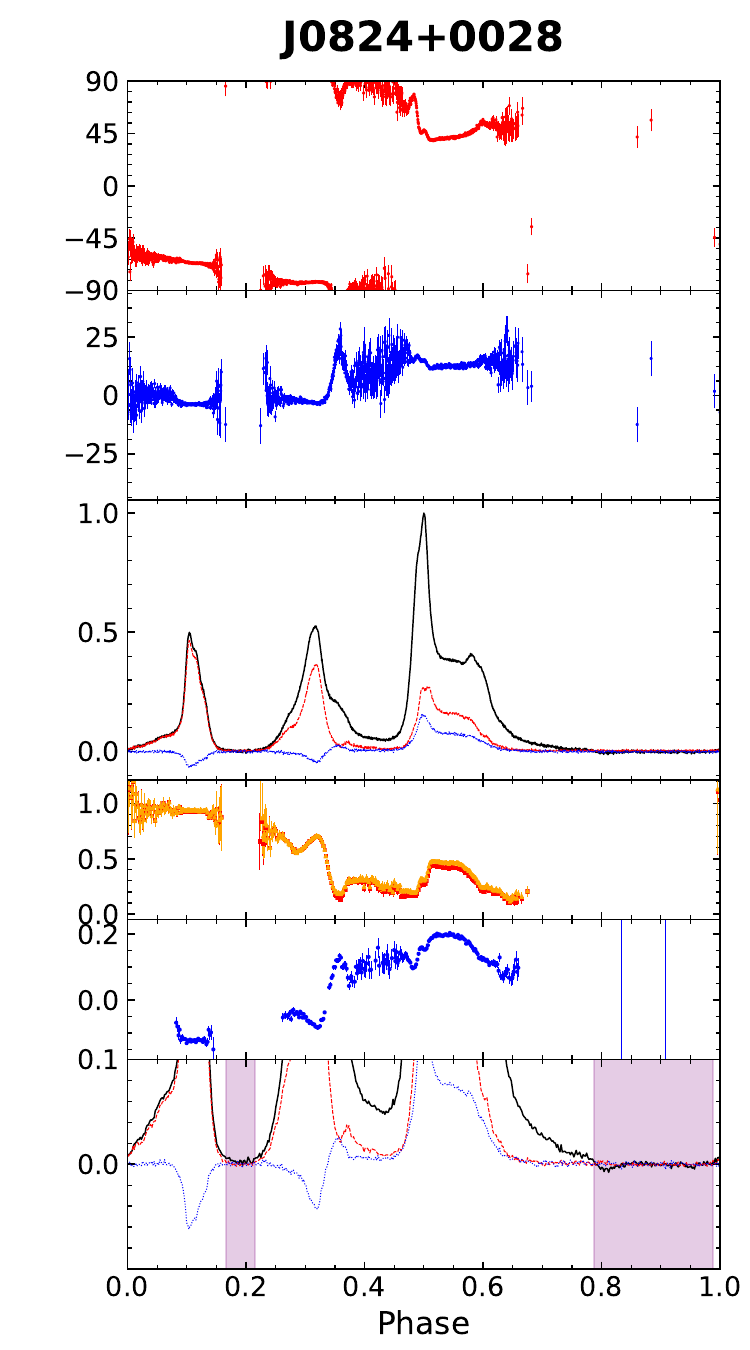}   
\end{minipage}\\
\begin{minipage}[t]{0.66\columnwidth}
    \centering
    \includegraphics[width=1\columnwidth]{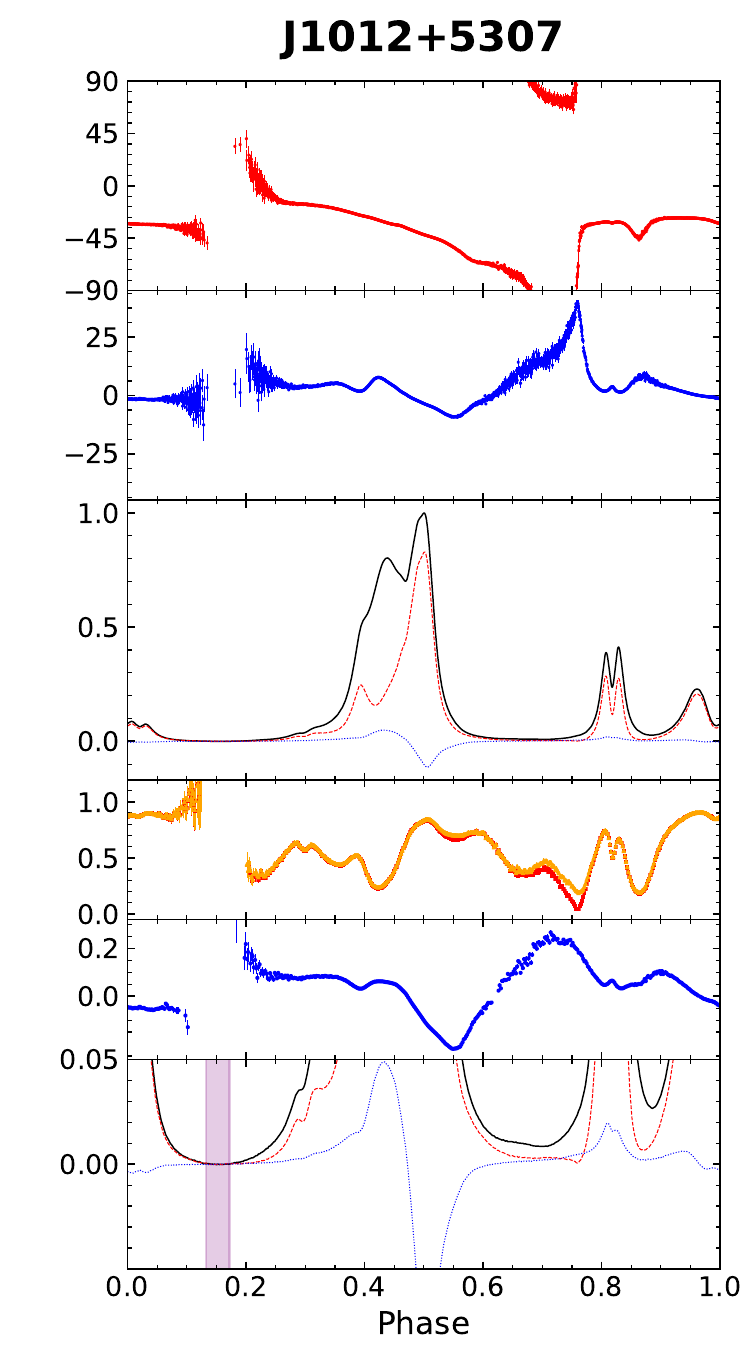}   
\end{minipage}
\begin{minipage}[t]{0.66\columnwidth}
    \centering
    \includegraphics[width=1\columnwidth]{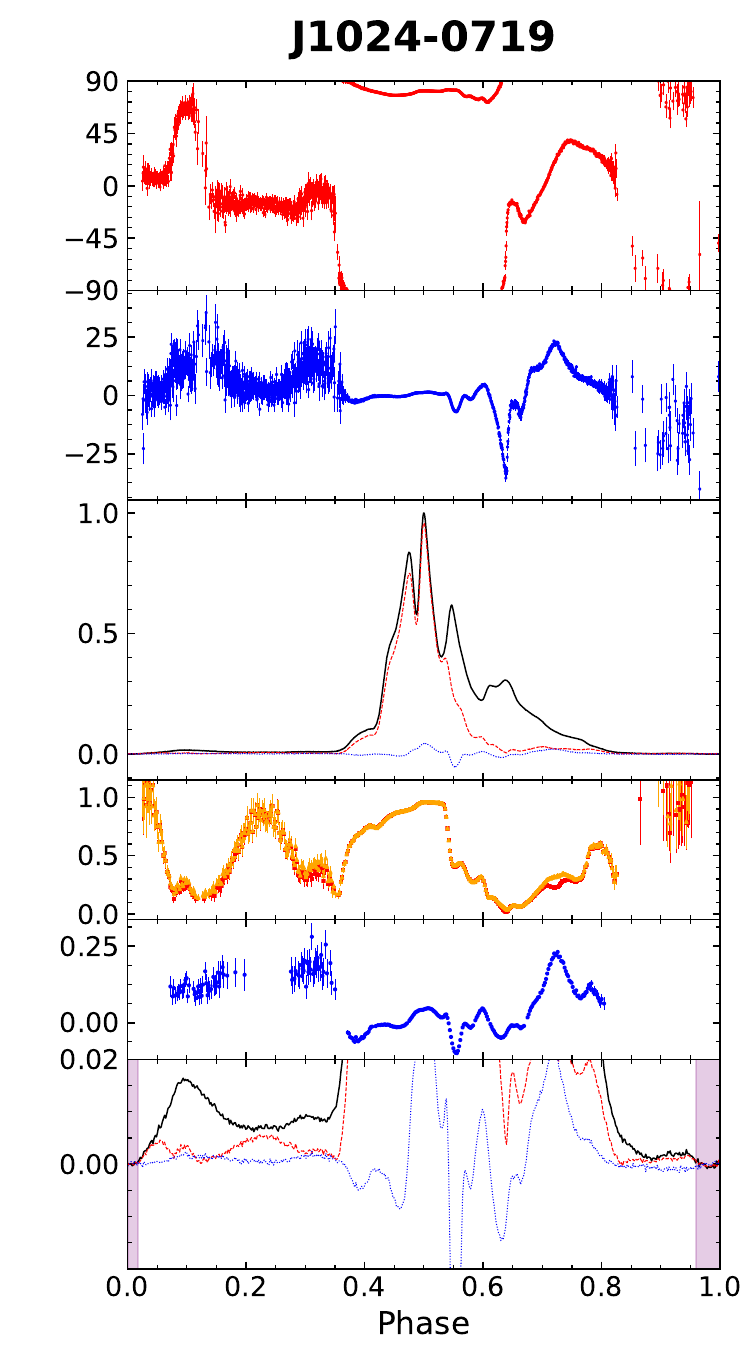}   
\end{minipage}
\begin{minipage}[t]{0.66\columnwidth}
    \centering
    \includegraphics[width=1\columnwidth]{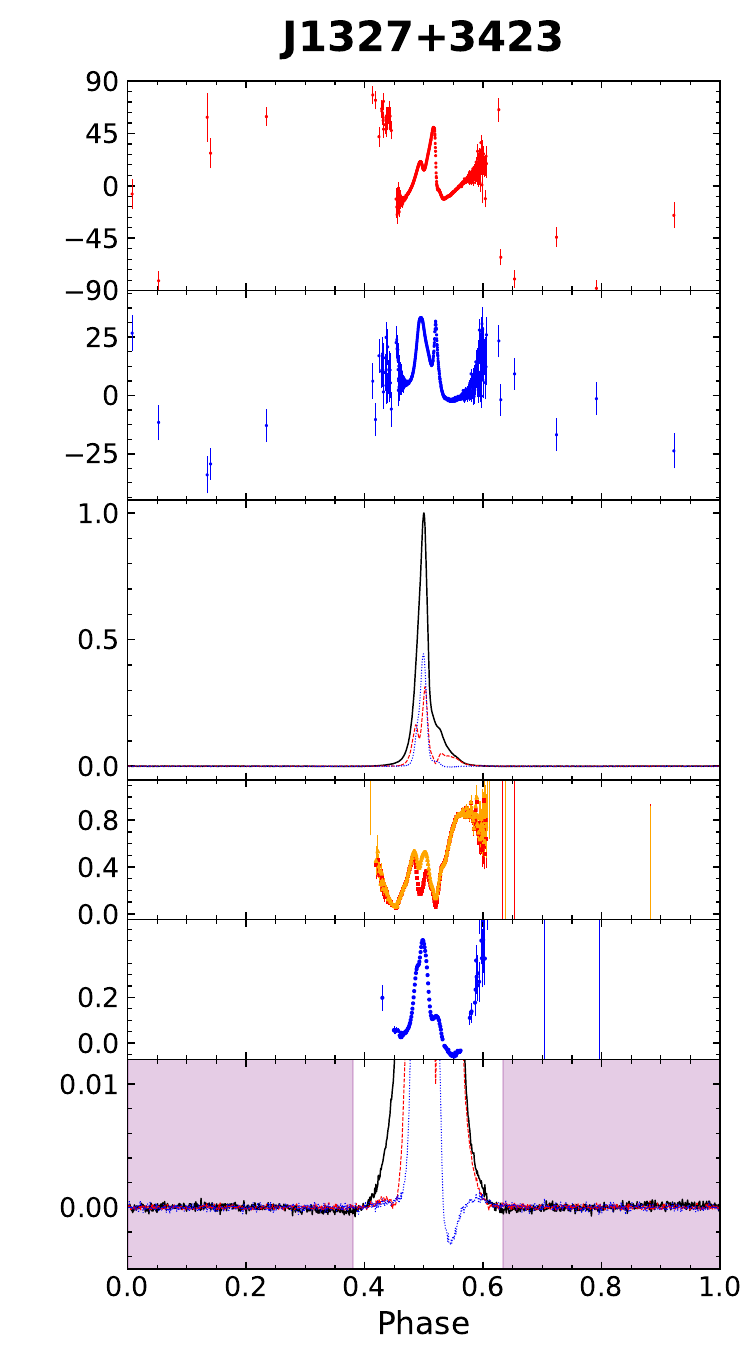}   
\end{minipage}
\caption{Continued.}
\end{figure*}

\begin{figure*}
\ContinuedFloat
\centering
\FIGSWITCH{
\begin{minipage}[t]{0.66\columnwidth}
    \centering
    \includegraphics[width=1\columnwidth]{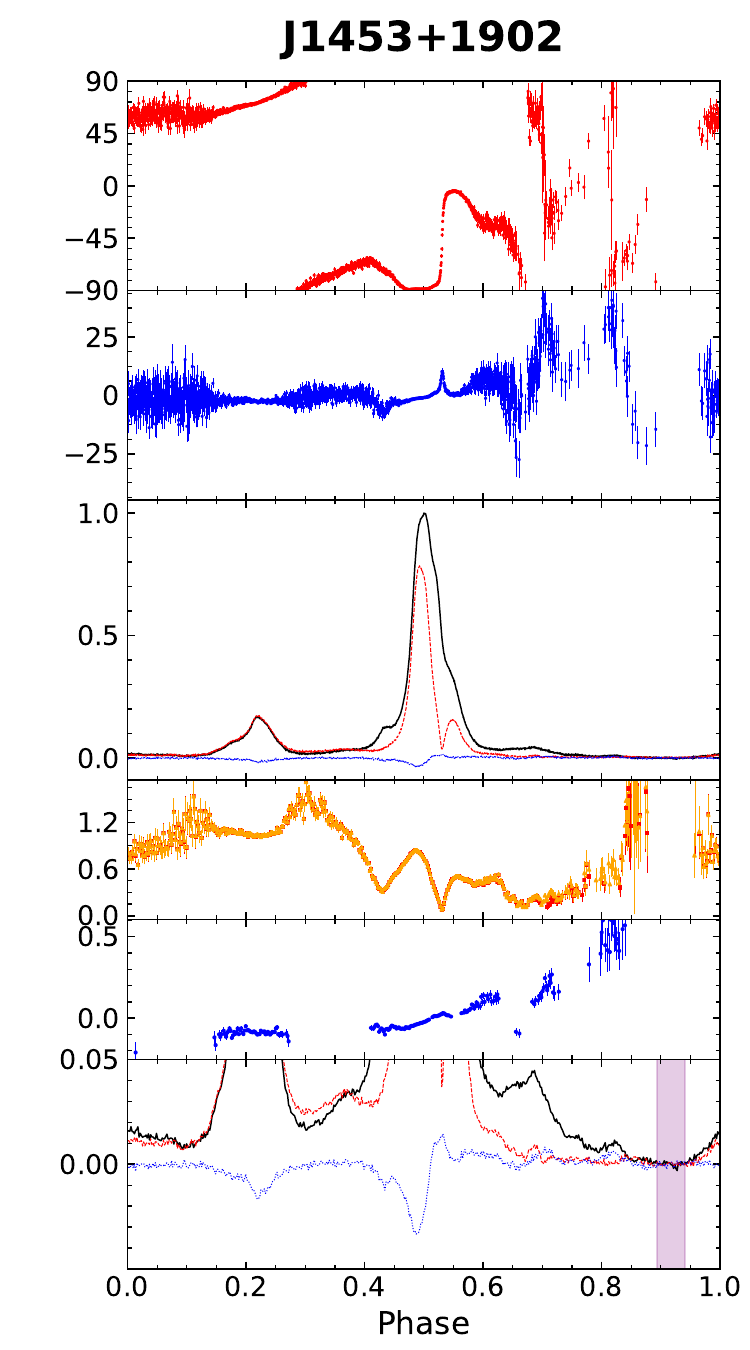}   
\end{minipage}
\begin{minipage}[t]{0.66\columnwidth}
    \centering
    \includegraphics[width=1\columnwidth]{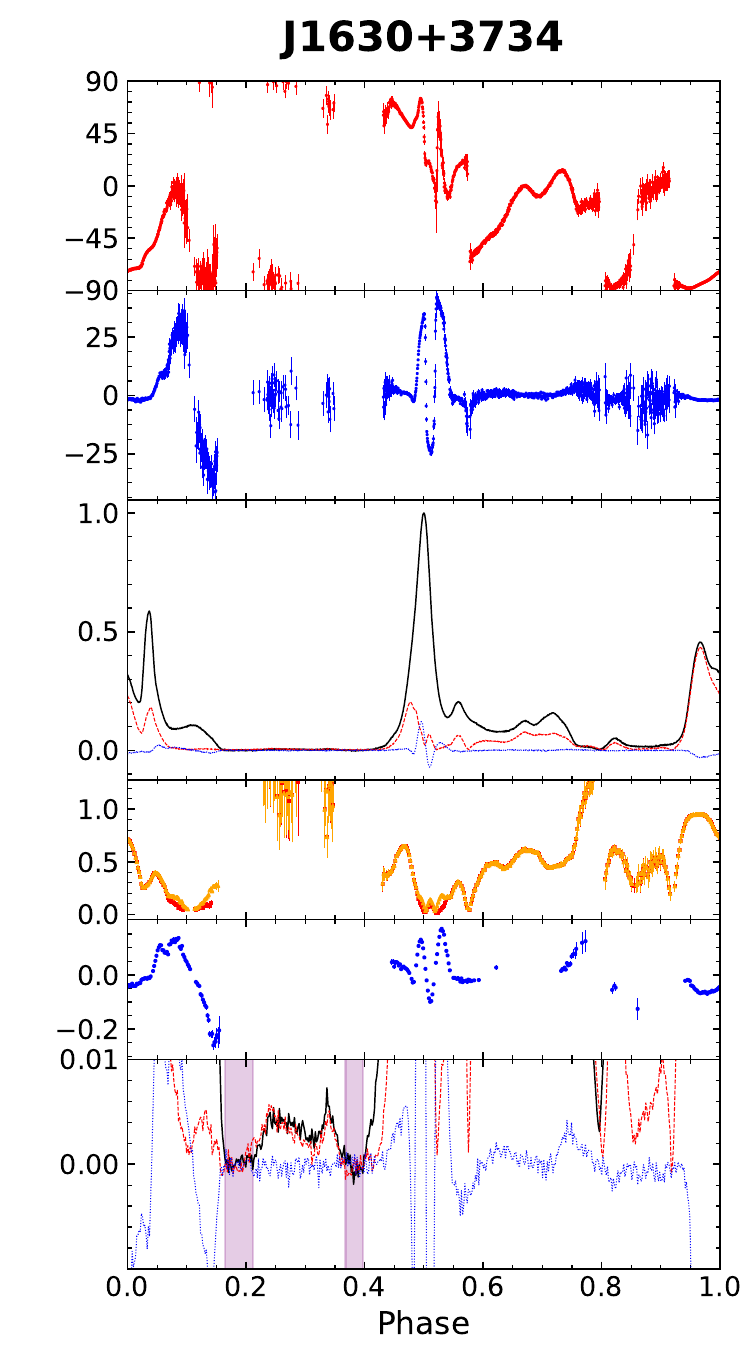}   
\end{minipage}
\begin{minipage}[t]{0.66\columnwidth}
    \centering
    \includegraphics[width=1\columnwidth]{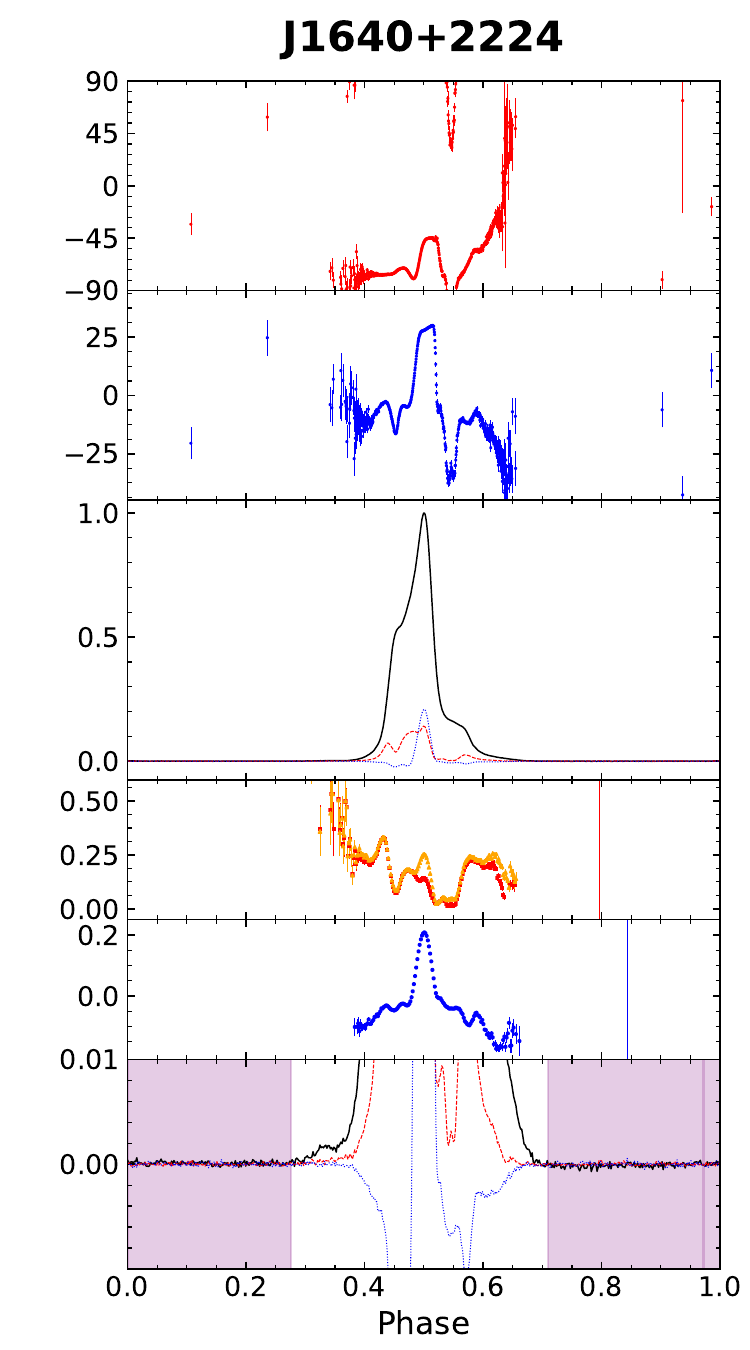}   
\end{minipage}\\
\hspace{1mm}
\begin{minipage}[t]{0.66\columnwidth}
    \centering
    \includegraphics[width=1\columnwidth]{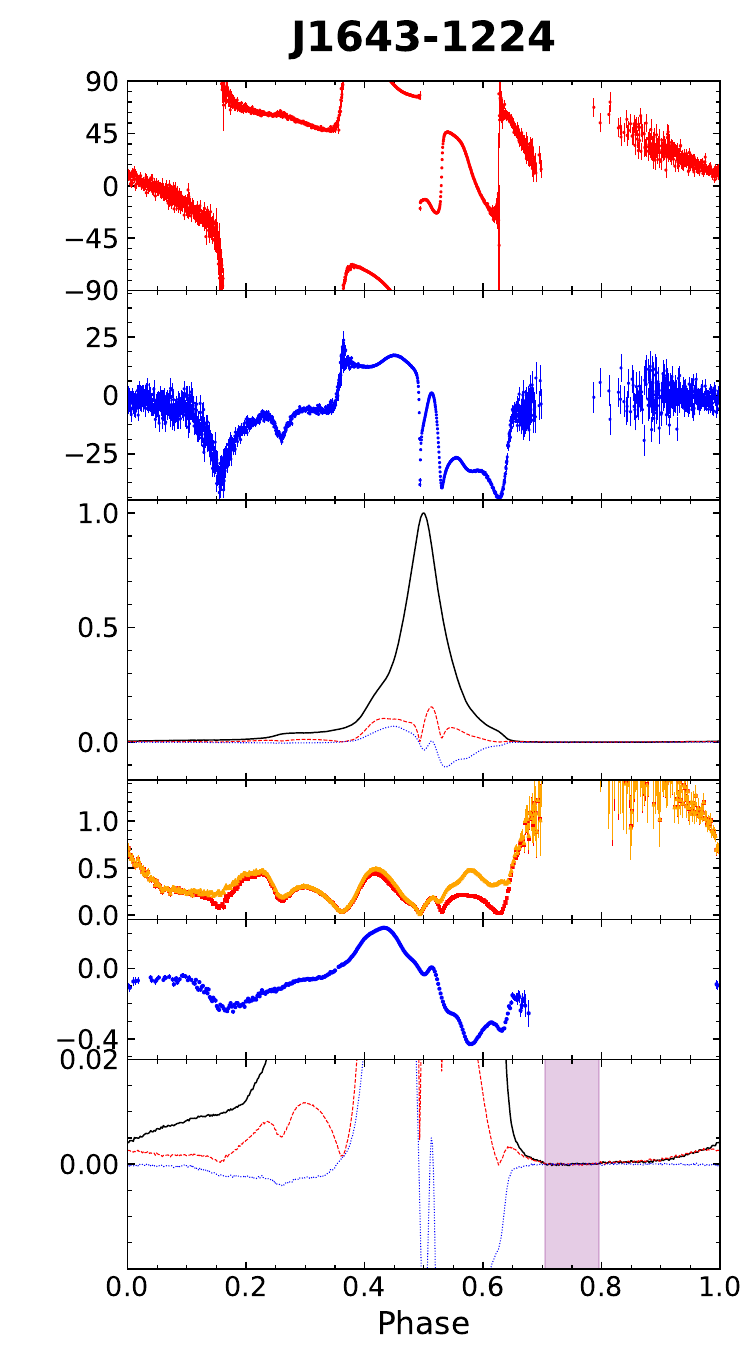}   
\end{minipage}
\begin{minipage}[t]{0.66\columnwidth}
    \centering
    \includegraphics[width=1\columnwidth]{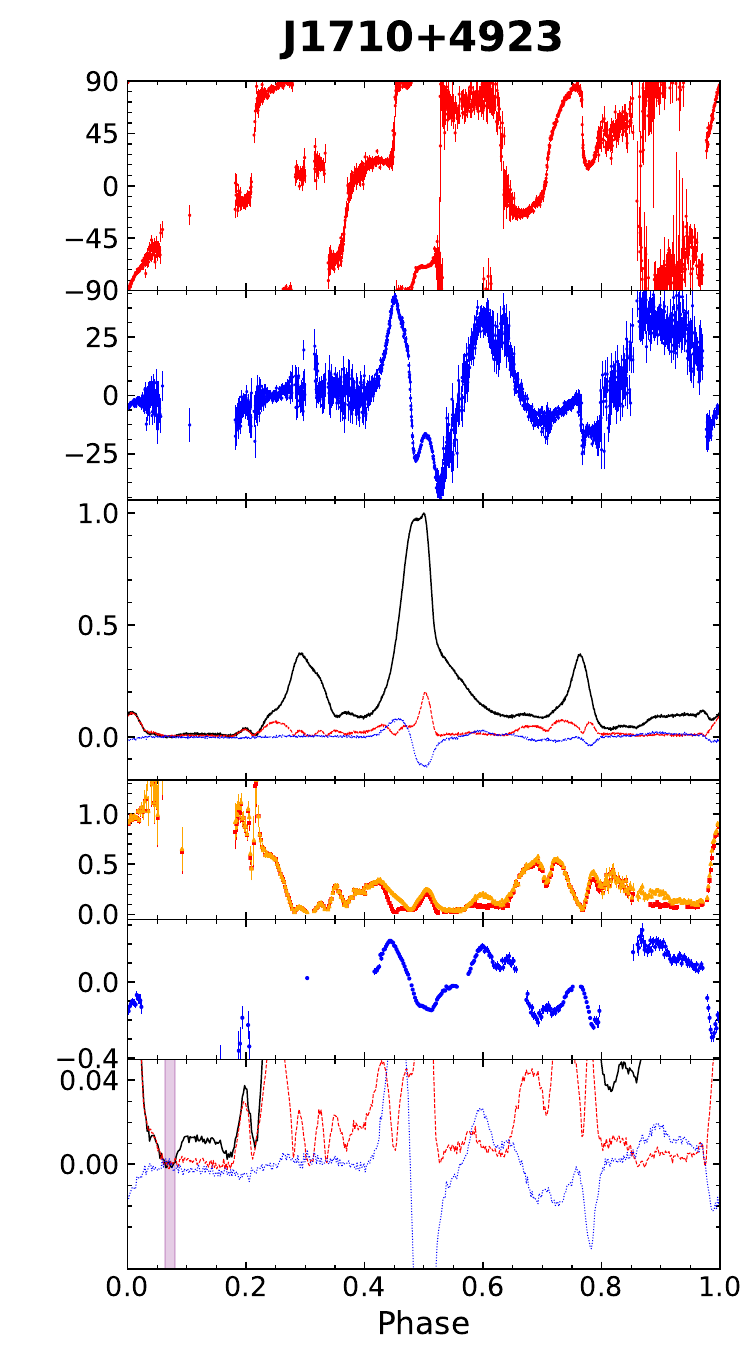}   
\end{minipage}
\begin{minipage}[t]{0.66\columnwidth}
    \centering
    \includegraphics[width=1\columnwidth]{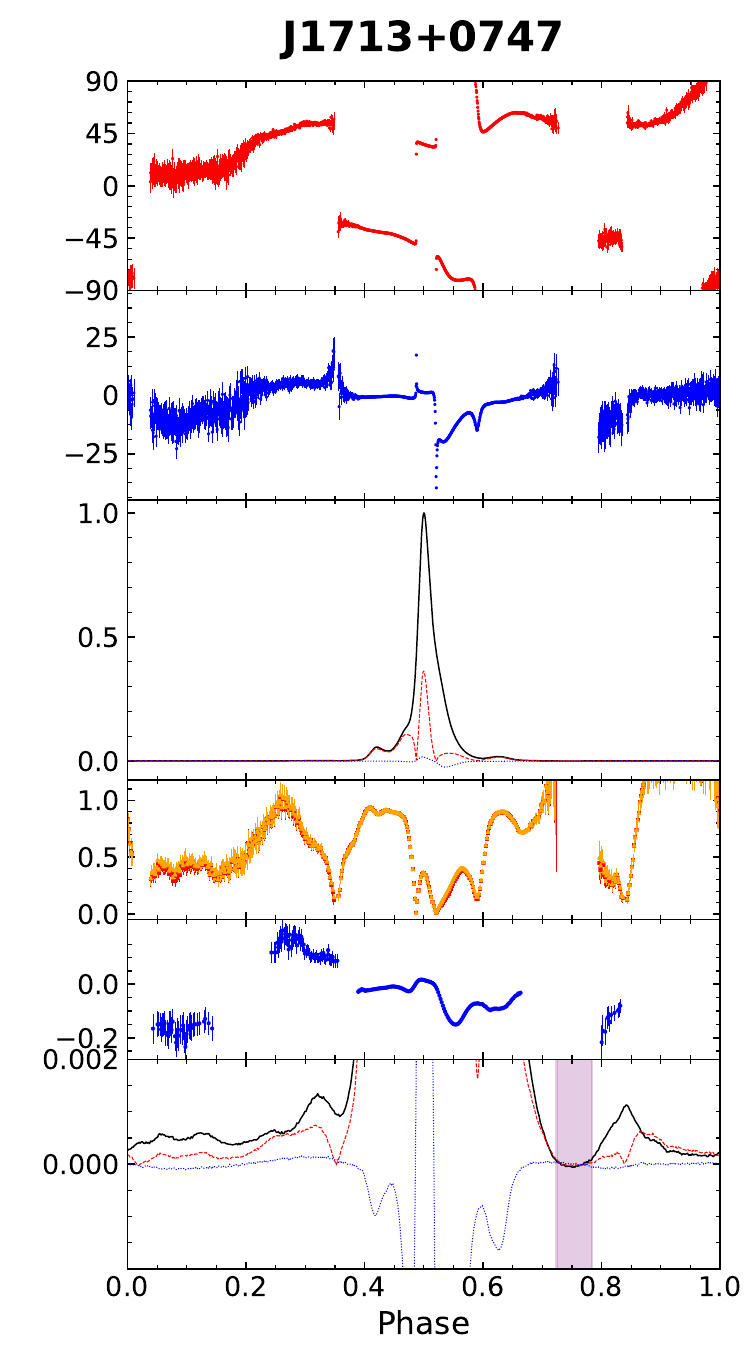}   
\end{minipage}
}
\caption{Continued.}
\end{figure*}

\begin{figure*}
\ContinuedFloat
\centering
\begin{minipage}[t]{0.66\columnwidth}
    \centering
    \includegraphics[width=1\columnwidth]{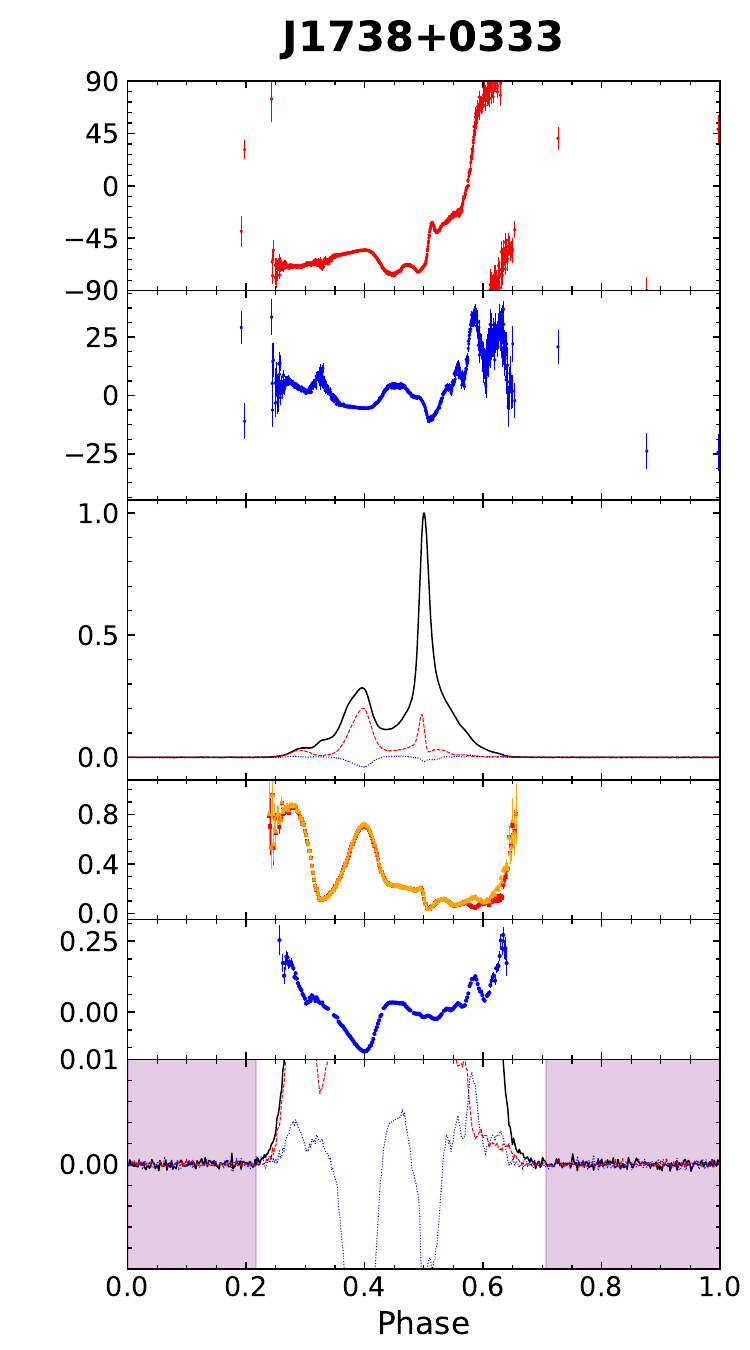}   
\end{minipage}
\begin{minipage}[t]{0.66\columnwidth}
\centering
\includegraphics[width=1\columnwidth]{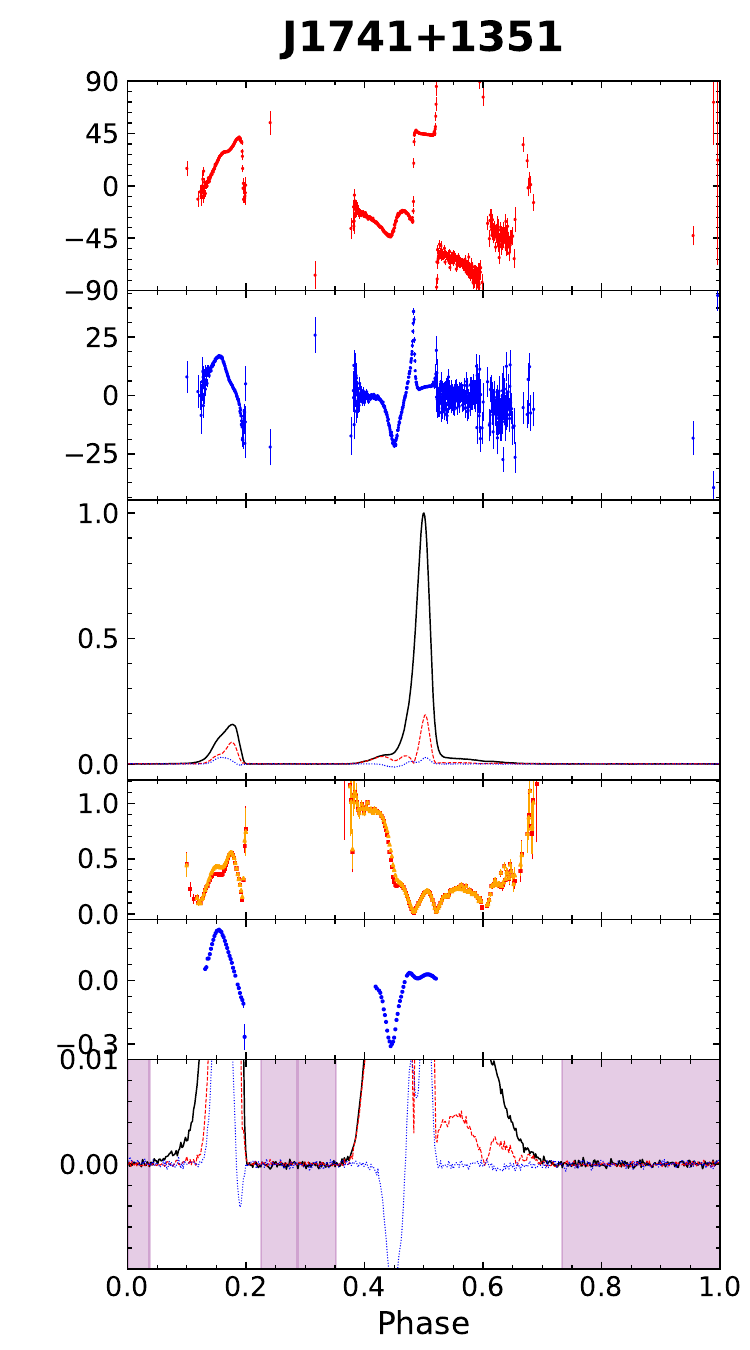}   
\end{minipage}
\begin{minipage}[t]{0.66\columnwidth}
    \centering
    \includegraphics[width=1\columnwidth]{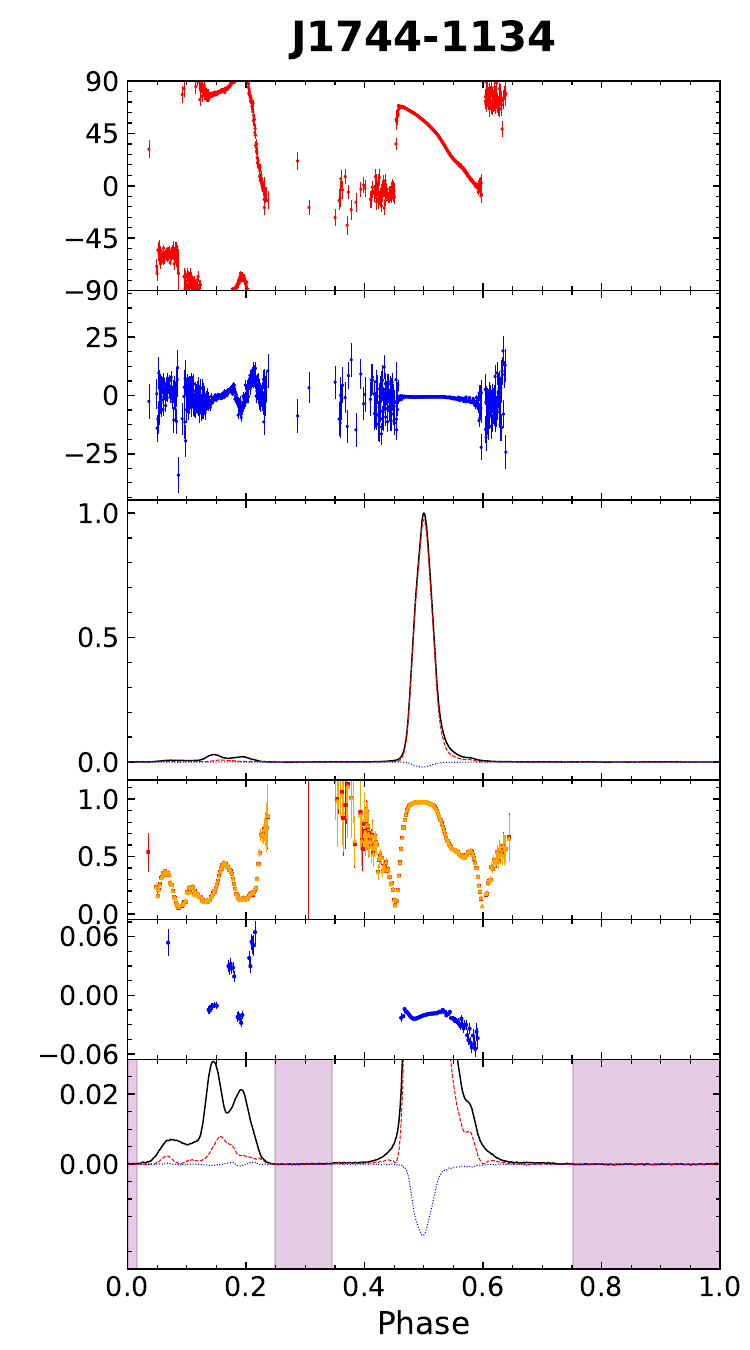}   
\end{minipage}\\
\begin{minipage}[t]{0.66\columnwidth}
\centering
\includegraphics[width=1\columnwidth]{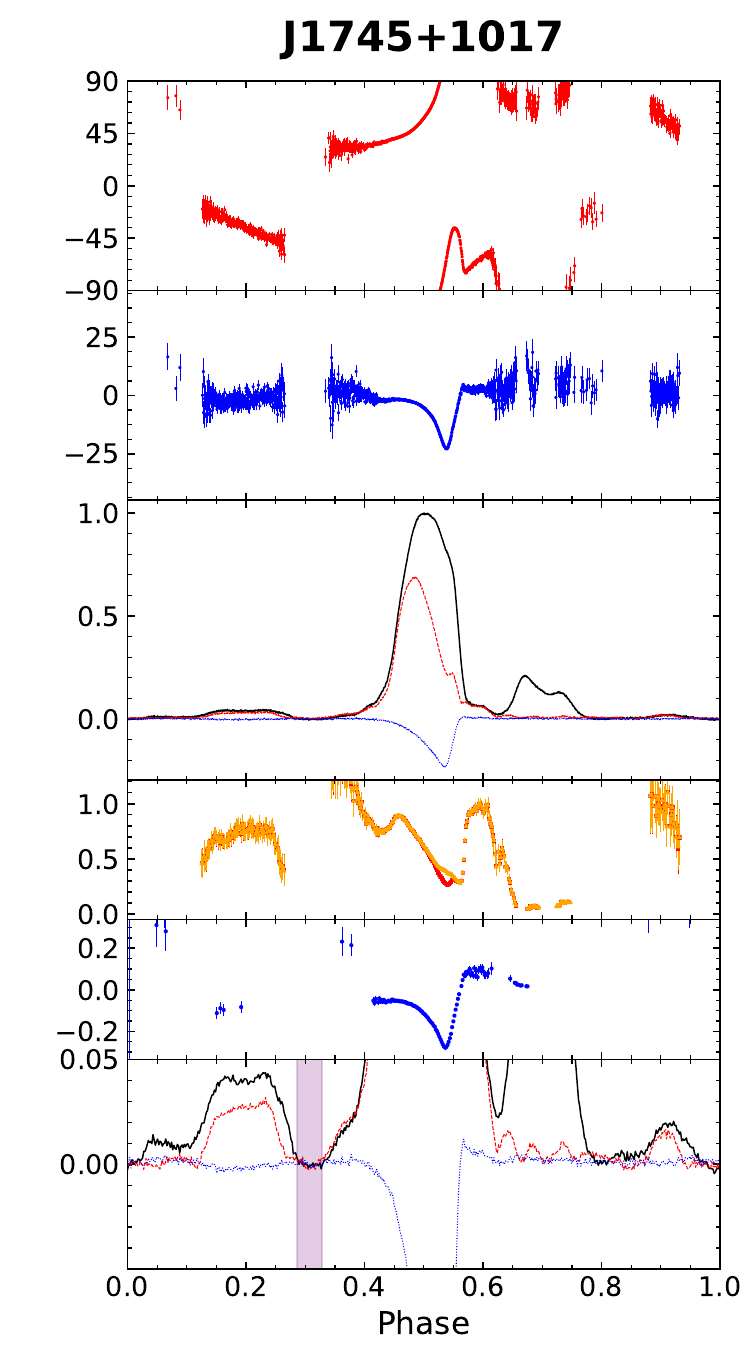}   
\end{minipage}
\begin{minipage}[t]{0.66\columnwidth}
\centering
\includegraphics[width=1\columnwidth]{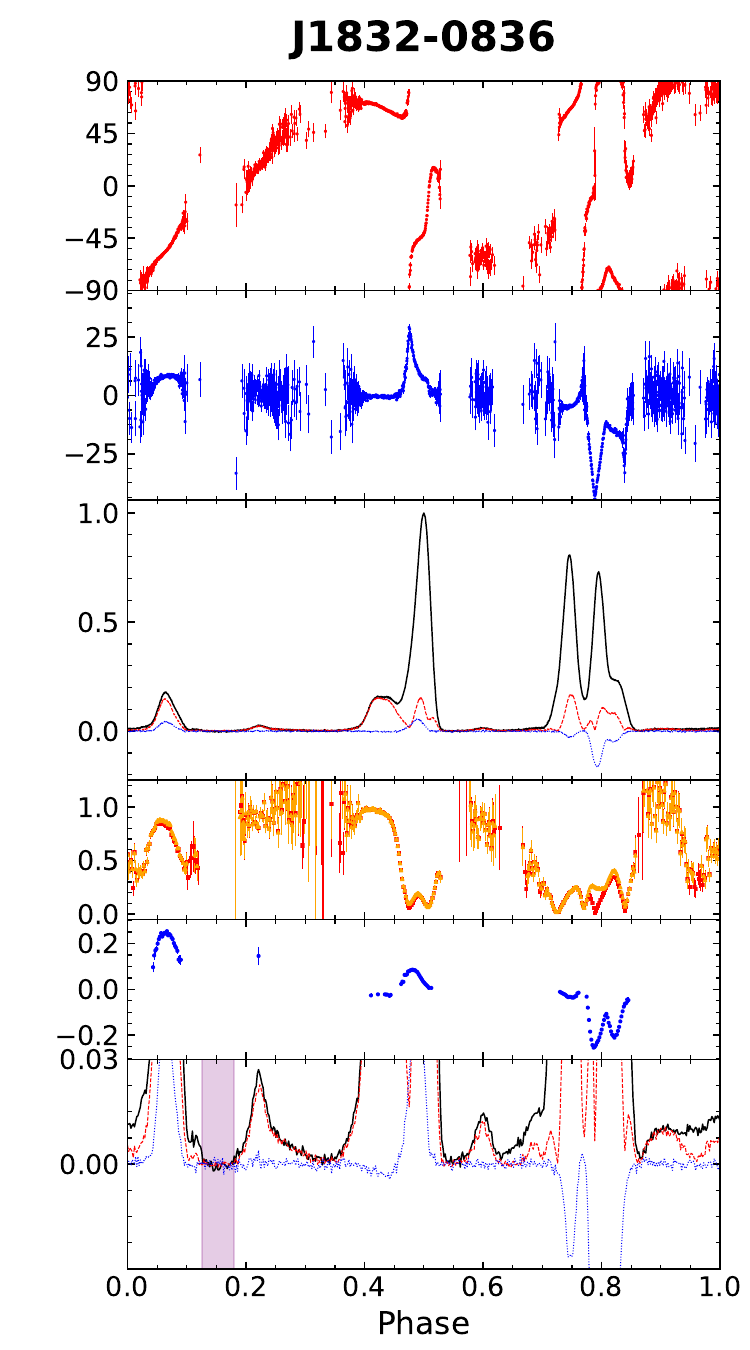}   
\end{minipage}
\begin{minipage}[t]{0.66\columnwidth}
\centering
\includegraphics[width=1\columnwidth]{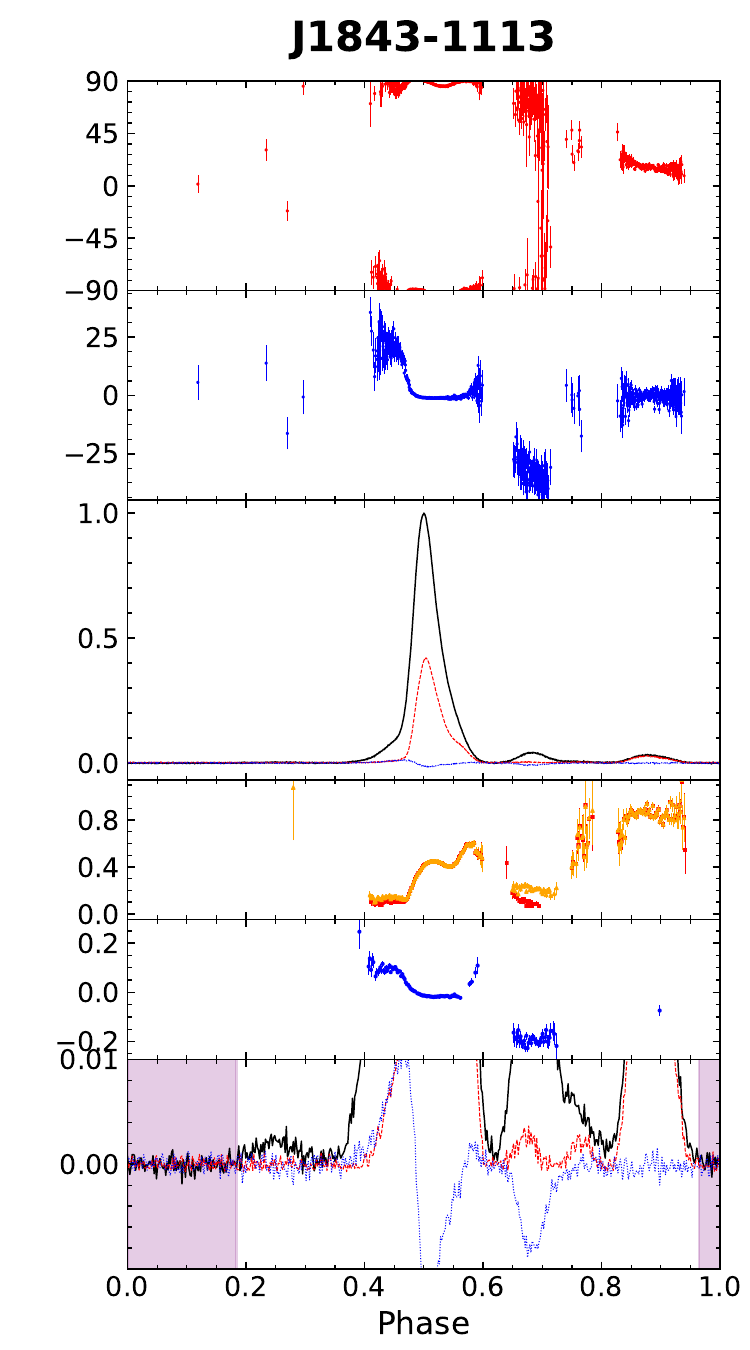}   
\end{minipage}
\caption{Continued.}
\end{figure*}

\begin{figure*}
\ContinuedFloat
\centering
\FIGSWITCH{
\begin{minipage}[t]{0.66\columnwidth}
    \centering
    \includegraphics[width=1\columnwidth]{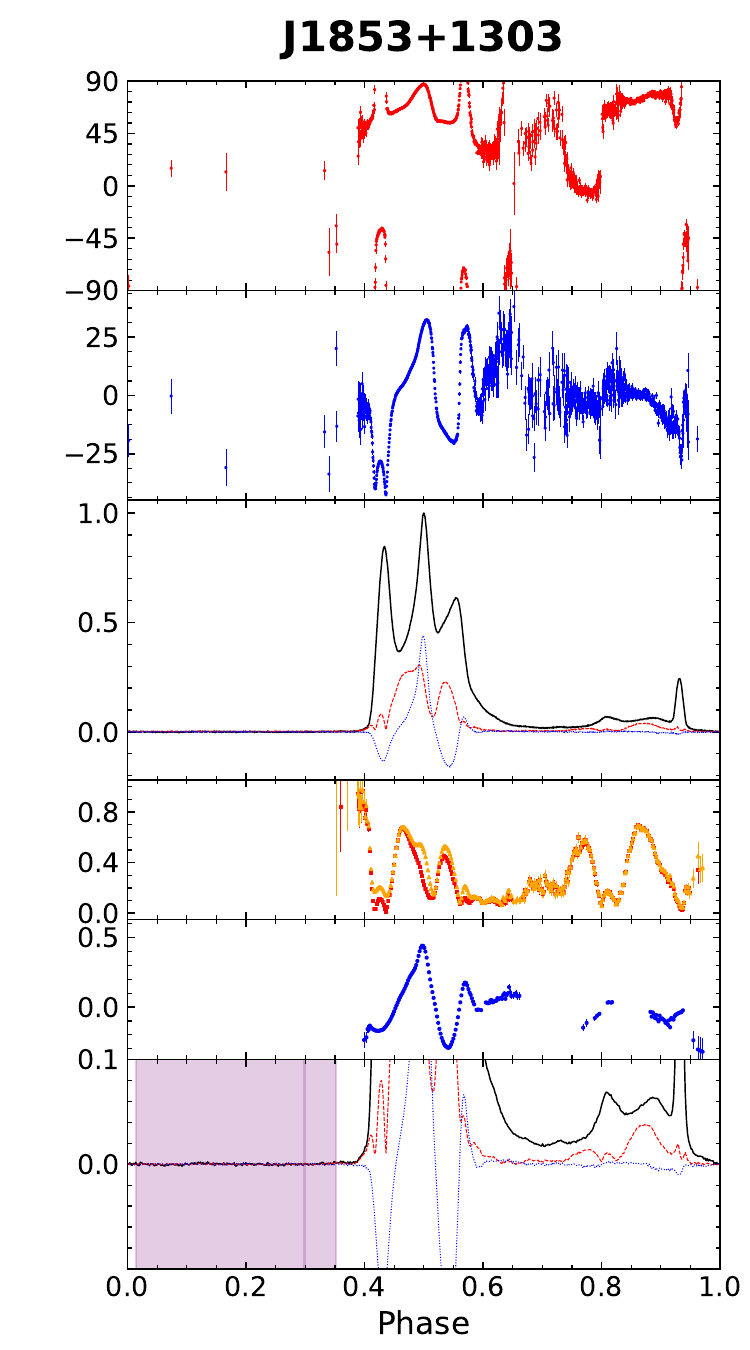}   
\end{minipage}
\begin{minipage}[t]{0.66\columnwidth}
    \centering
    \includegraphics[width=1\columnwidth]{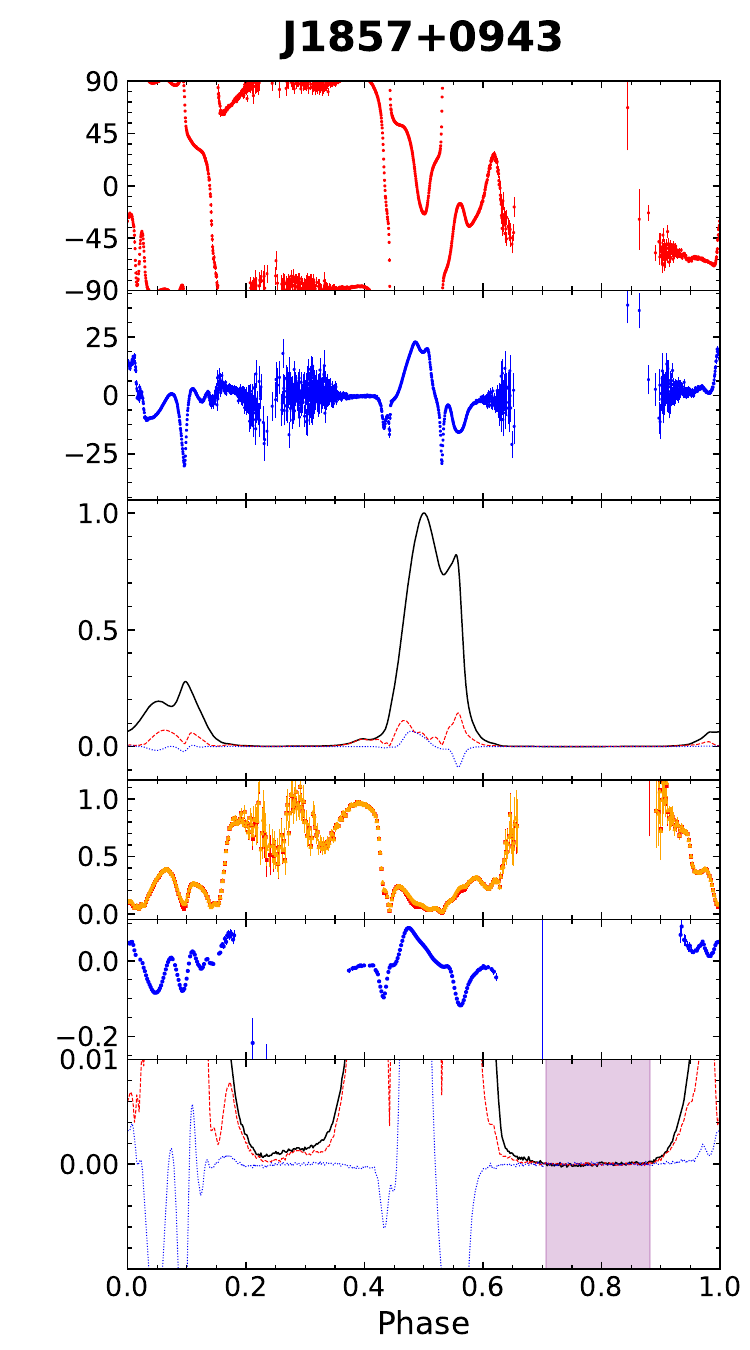}   
\end{minipage}
\begin{minipage}[t]{0.66\columnwidth}
    \centering
    \includegraphics[width=1\columnwidth]{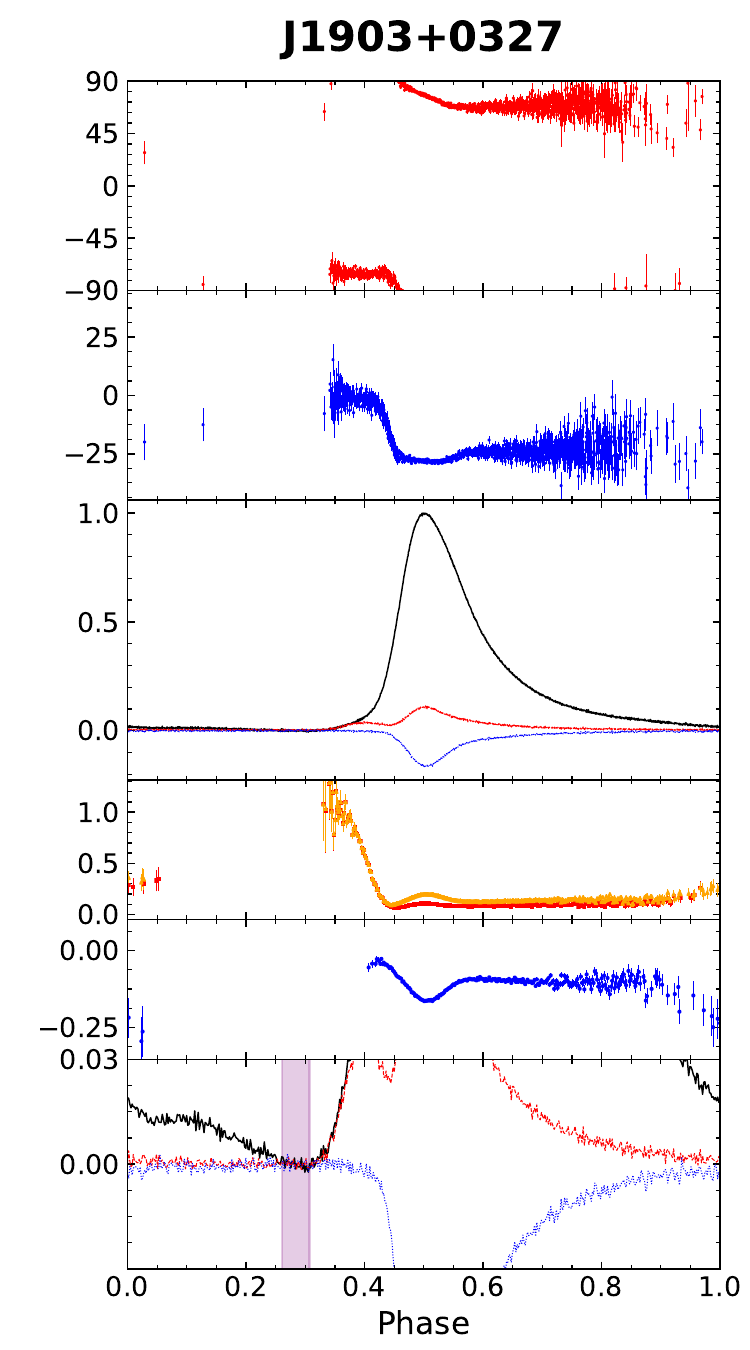}   
\end{minipage}\\
\hspace{1mm}
\begin{minipage}[t]{0.66\columnwidth}
    \centering
    \includegraphics[width=1\columnwidth]{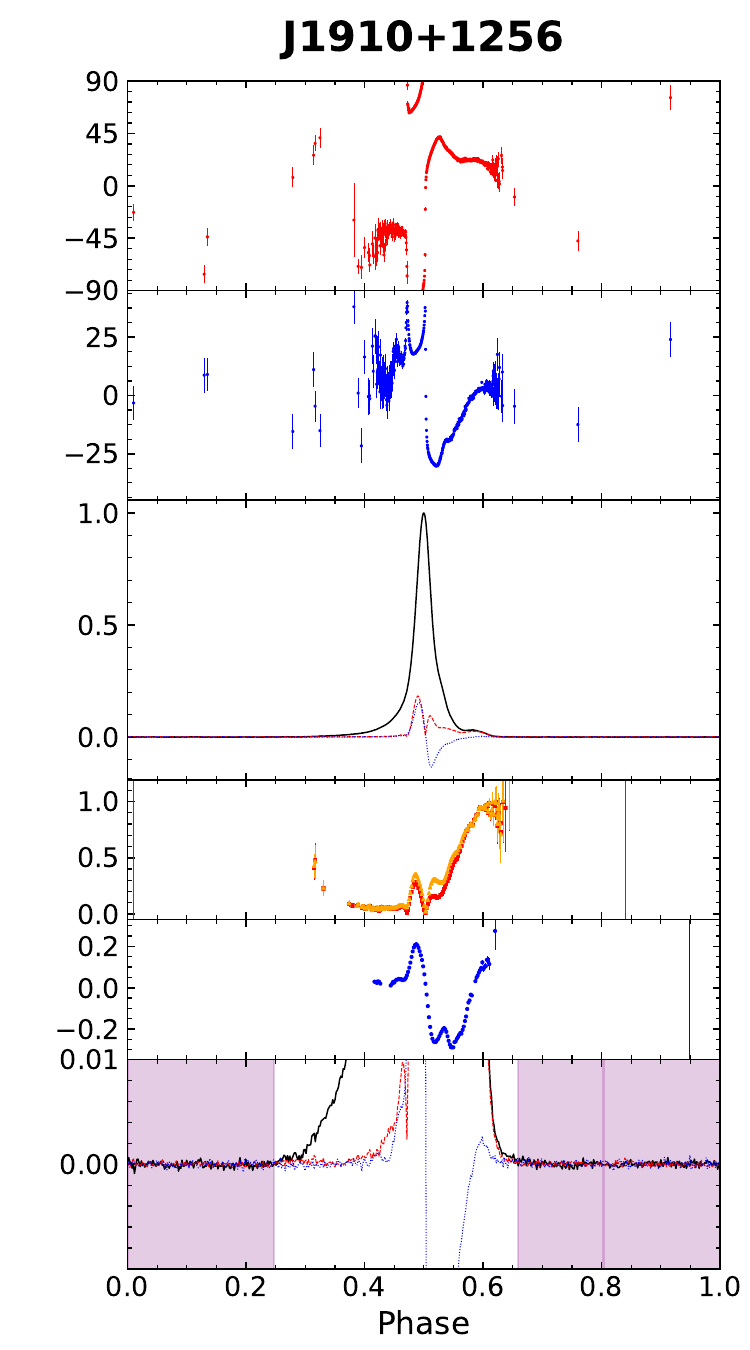}   
\end{minipage}
\begin{minipage}[t]{0.66\columnwidth}
    \centering
    \includegraphics[width=1\columnwidth]{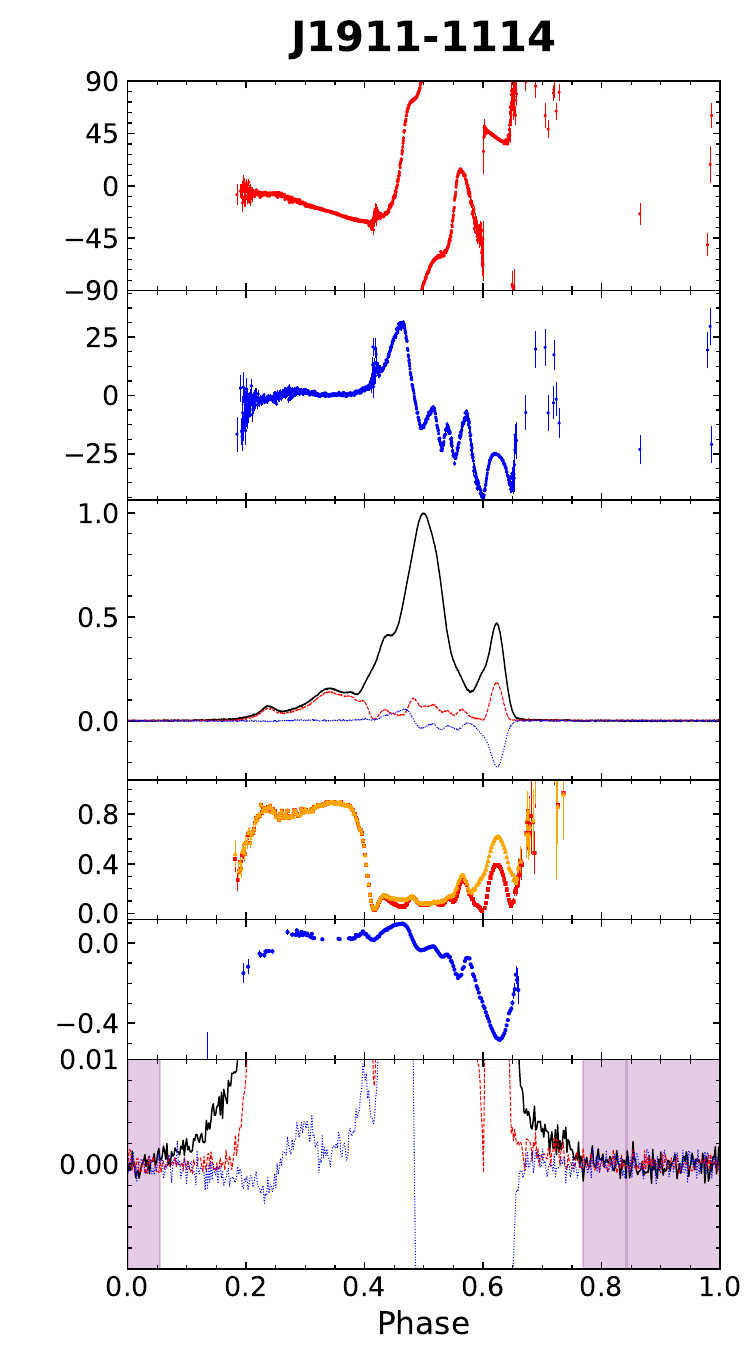}   
\end{minipage}
\begin{minipage}[t]{0.66\columnwidth}
    \centering
    \includegraphics[width=1\columnwidth]{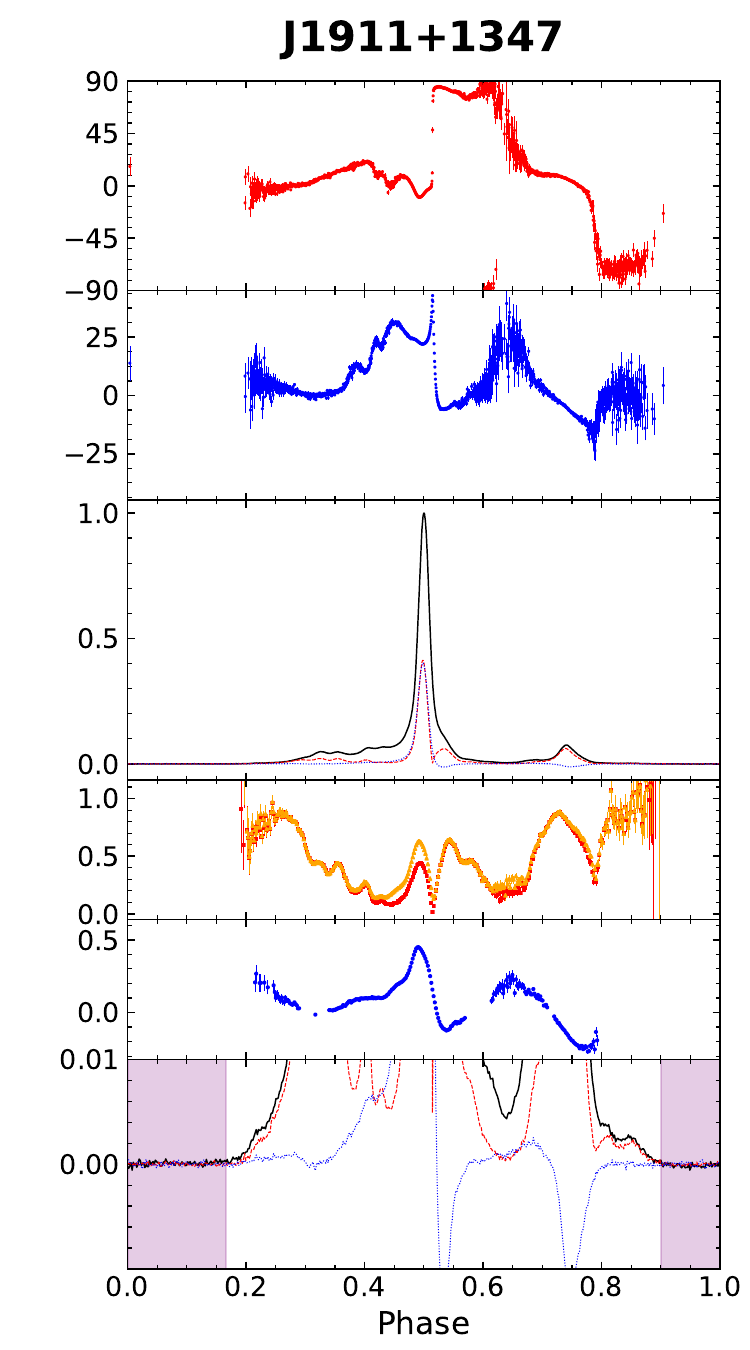}   
\end{minipage}
}
\caption{Continued.}
\end{figure*}

\begin{figure*}
\ContinuedFloat
\centering
\begin{minipage}[t]{0.66\columnwidth}
    \centering
    \includegraphics[width=1\columnwidth]{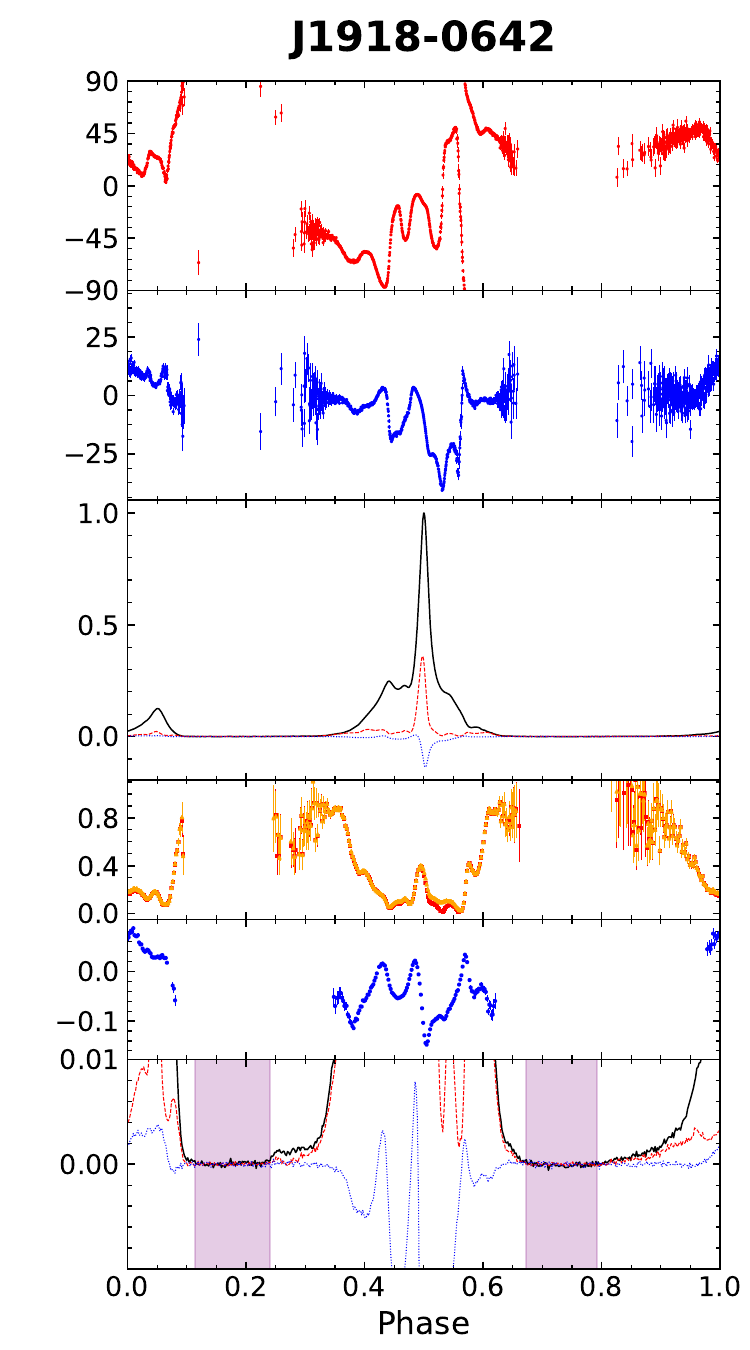}   
\end{minipage}
\begin{minipage}[t]{0.66\columnwidth}
    \centering
    \includegraphics[width=1\columnwidth]{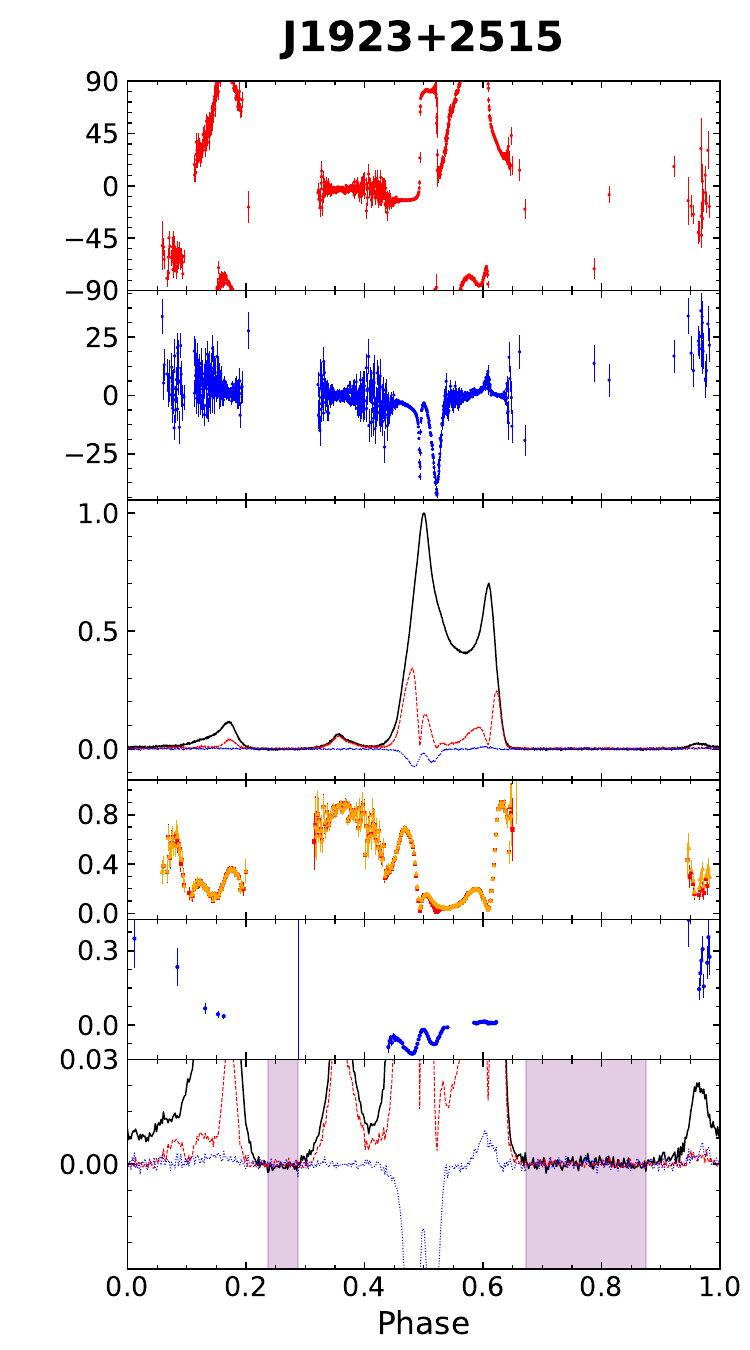}   
\end{minipage}
\begin{minipage}[t]{0.66\columnwidth}
    \centering
    \includegraphics[width=1\columnwidth]{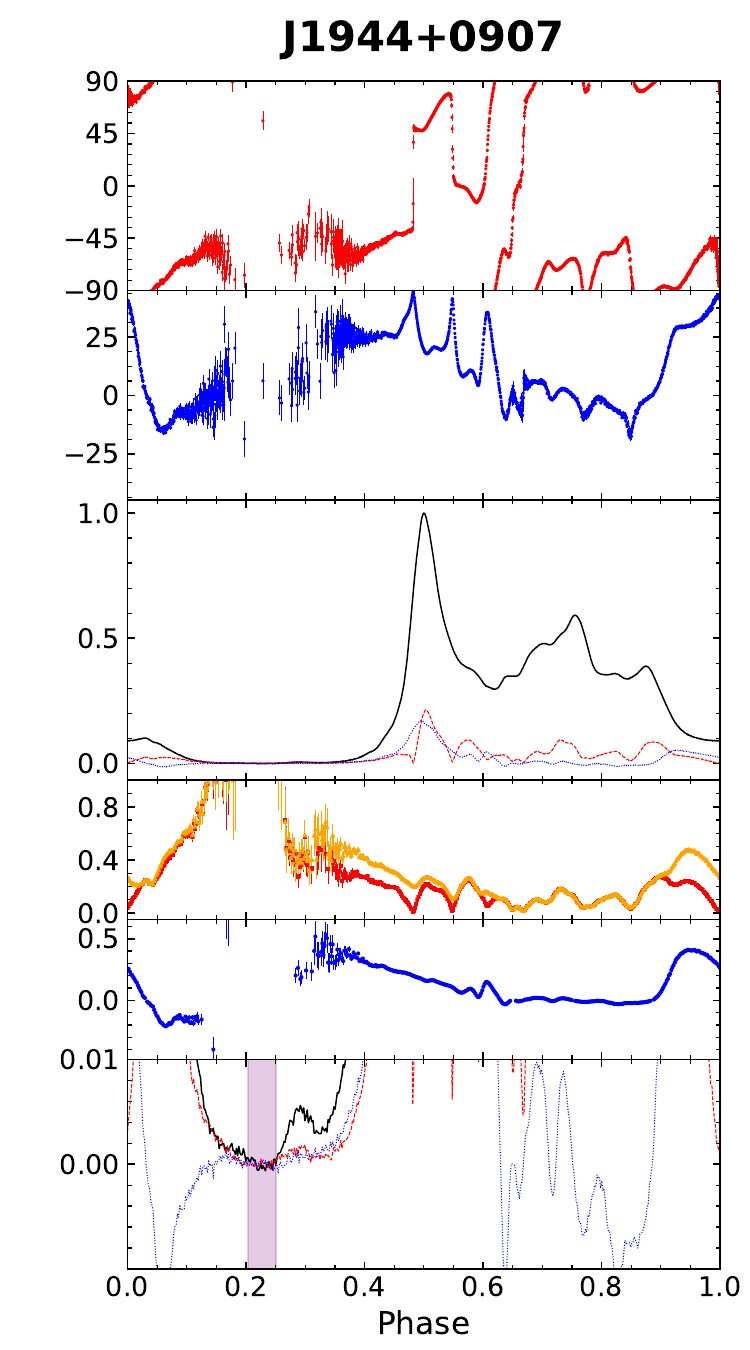}   
\end{minipage}\\
\begin{minipage}[t]{0.66\columnwidth}
    \centering
    \includegraphics[width=1\columnwidth]{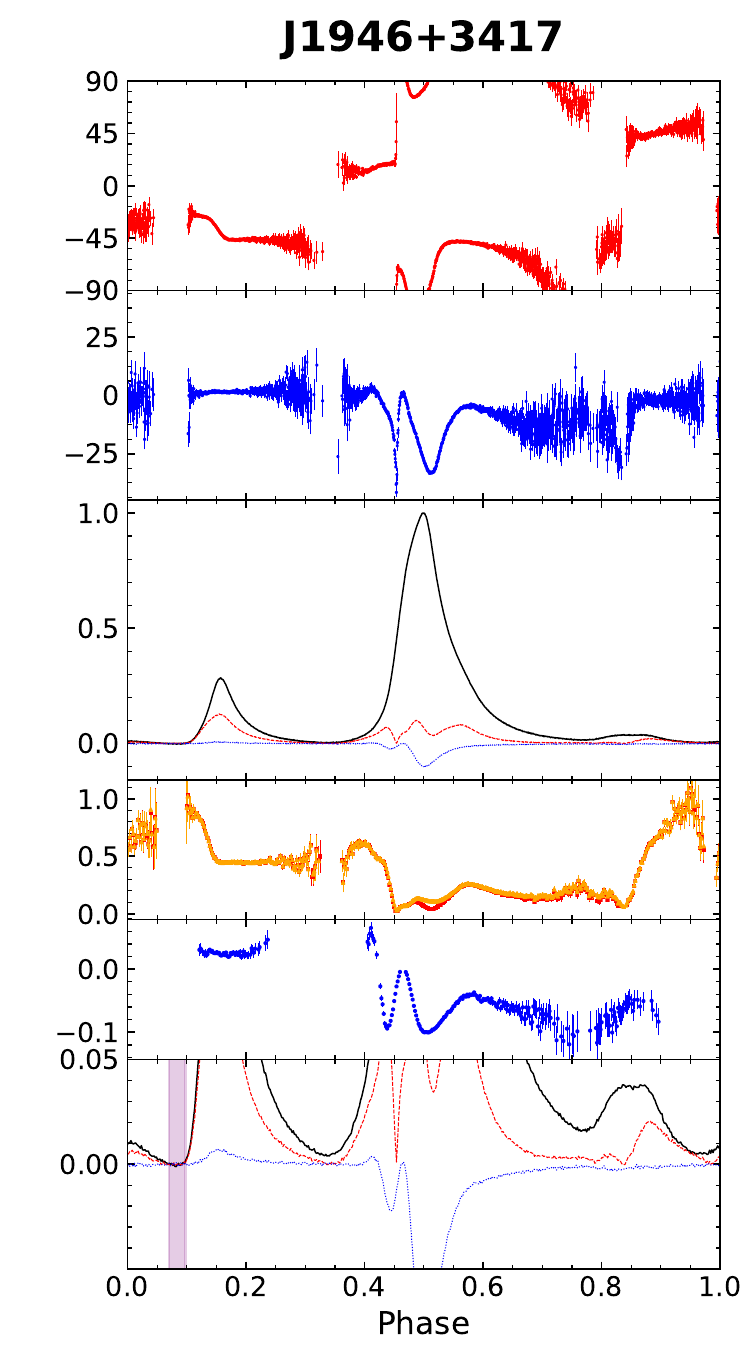}   
\end{minipage}
\begin{minipage}[t]{0.66\columnwidth}
    \centering
    \includegraphics[width=1\columnwidth]{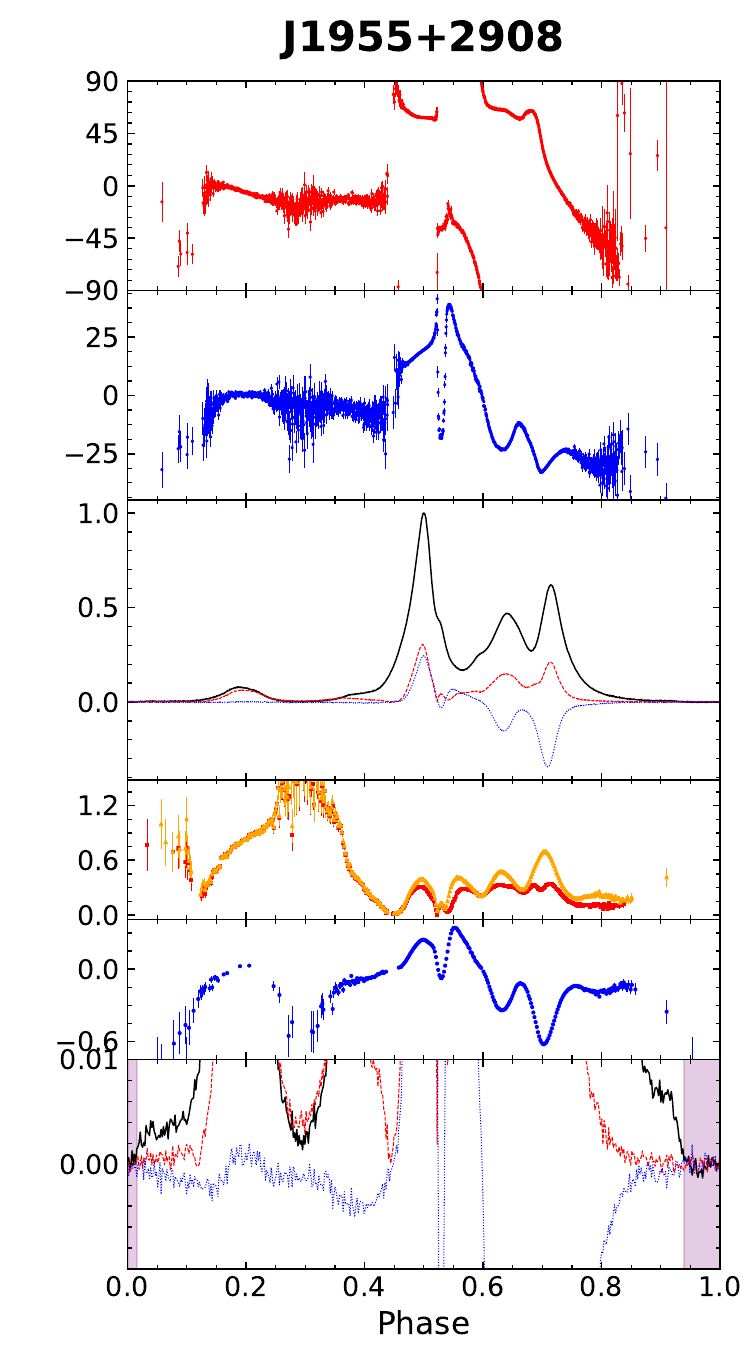}   
\end{minipage}
\begin{minipage}[t]{0.66\columnwidth}
    \centering
    \includegraphics[width=1\columnwidth]{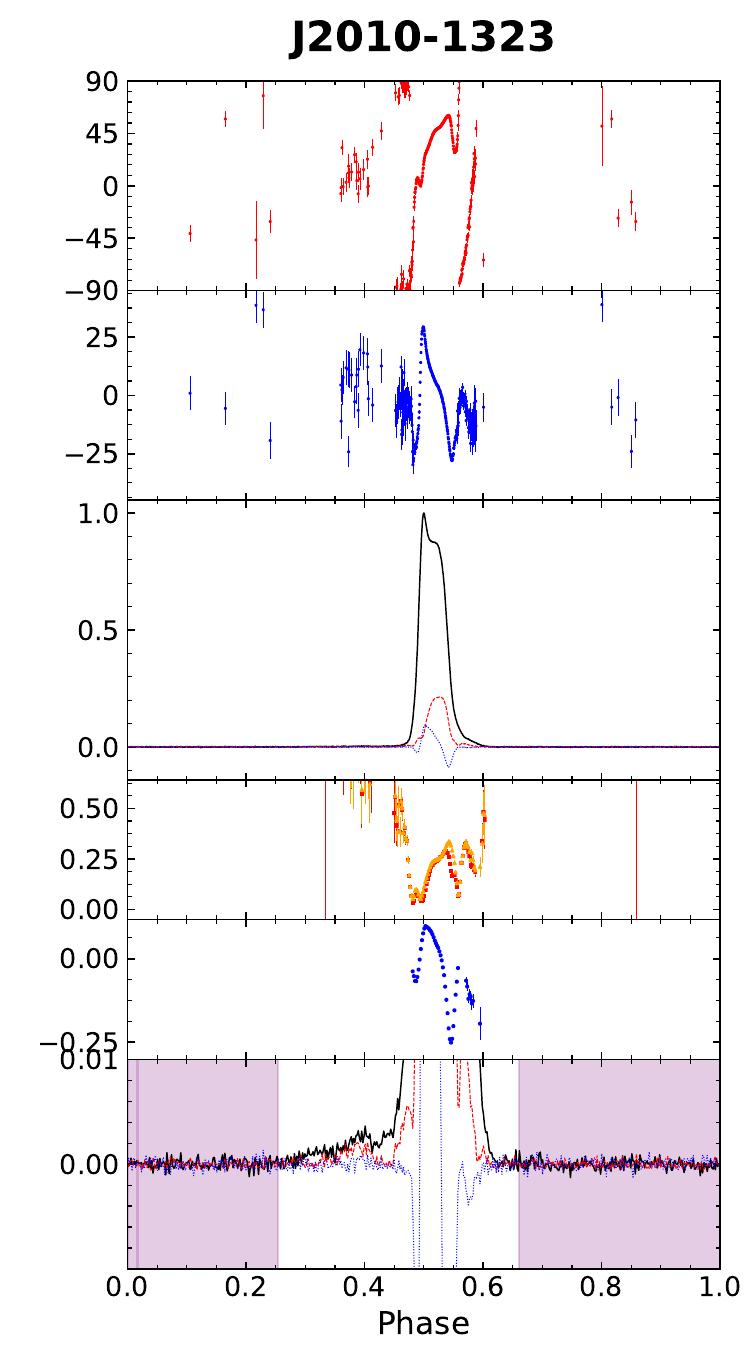}   
\end{minipage}
\caption{Continued.}
\end{figure*}

\begin{figure*}
\ContinuedFloat
\centering
\FIGSWITCH{
\begin{minipage}[t]{0.66\columnwidth}
    \centering
    \includegraphics[width=1\columnwidth]{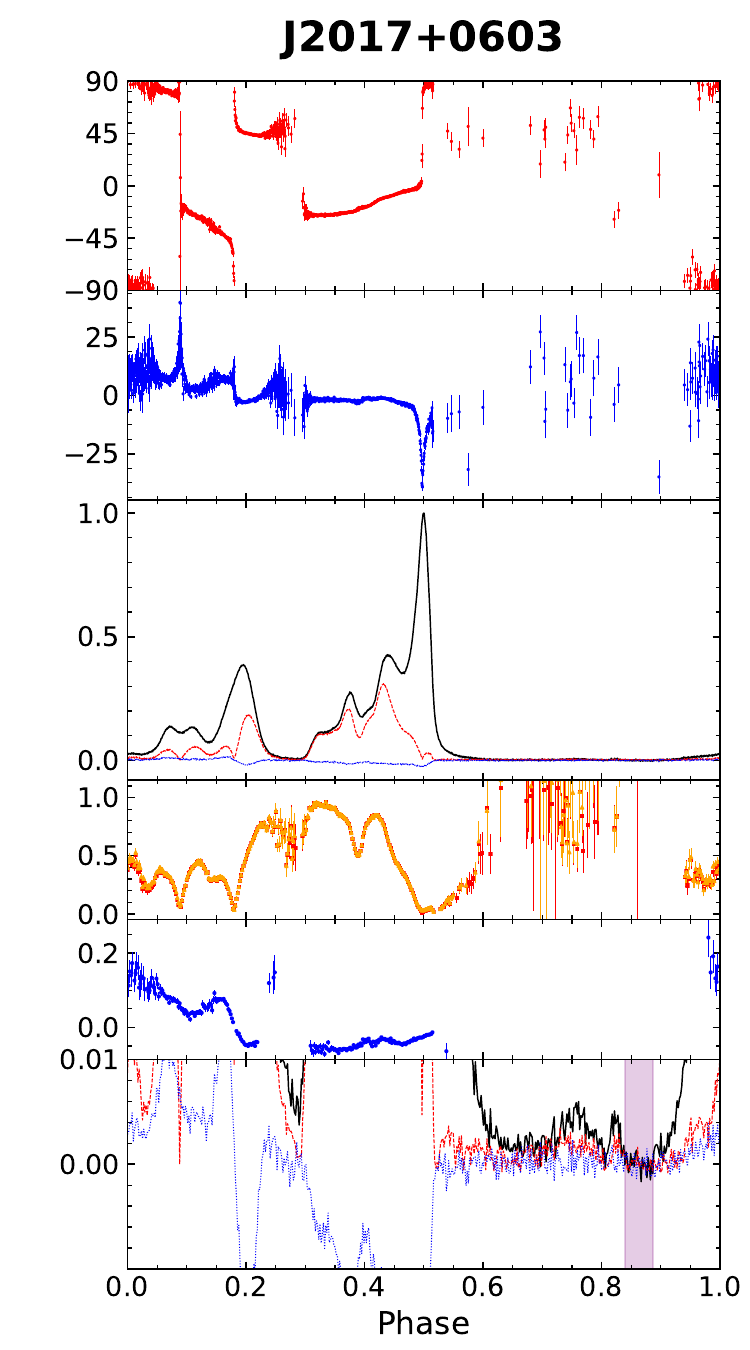}   
\end{minipage}
\begin{minipage}[t]{0.66\columnwidth}
    \centering
    \includegraphics[width=1\columnwidth]{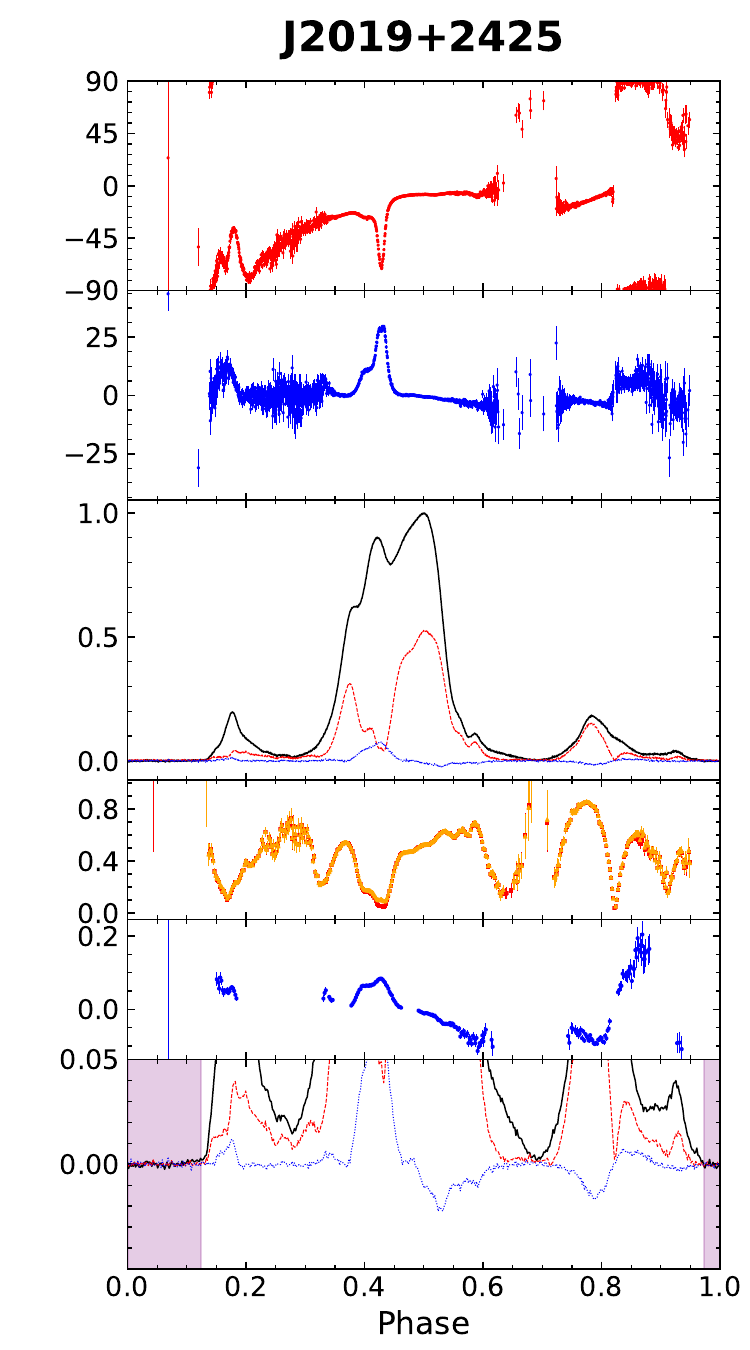}   
\end{minipage}
\begin{minipage}[t]{0.66\columnwidth}
    \centering
    \includegraphics[width=1\columnwidth]{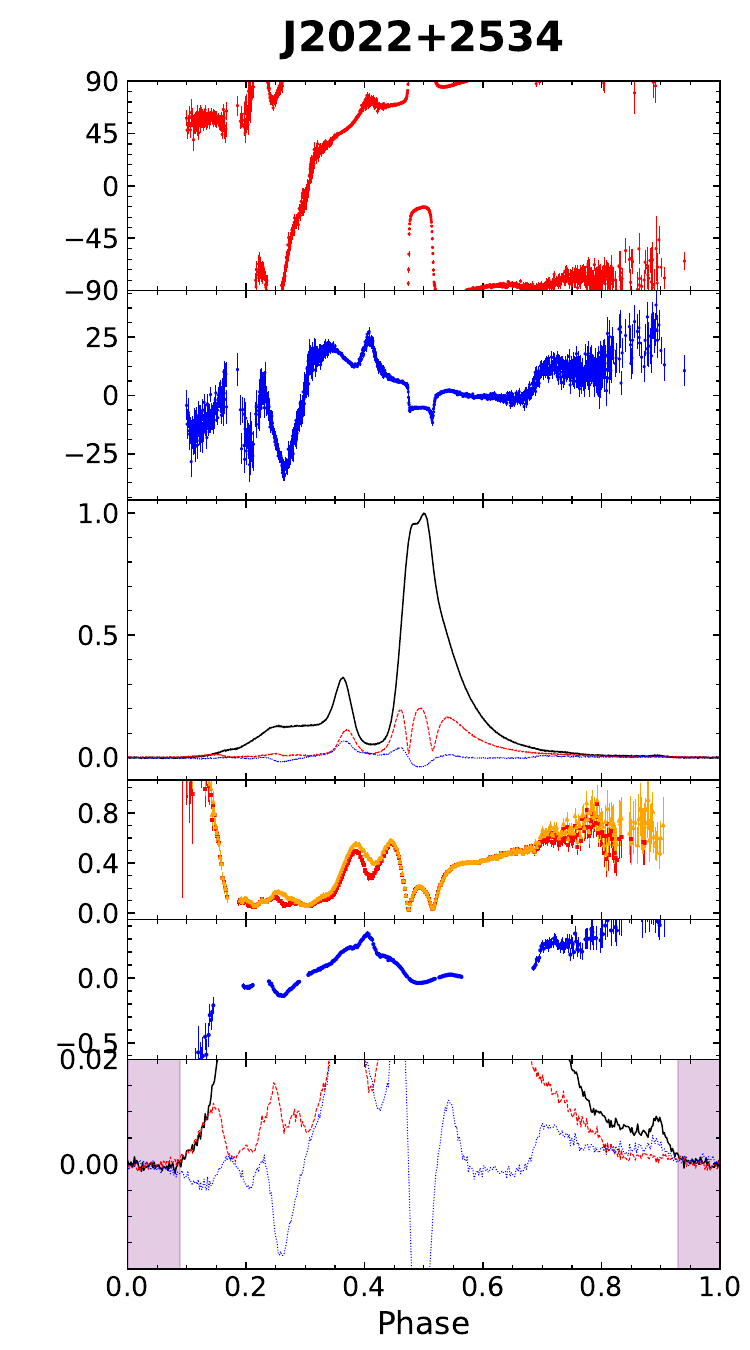}   
\end{minipage}\\
\hspace{1mm}
\begin{minipage}[t]{0.66\columnwidth}
    \centering
    \includegraphics[width=1\columnwidth]{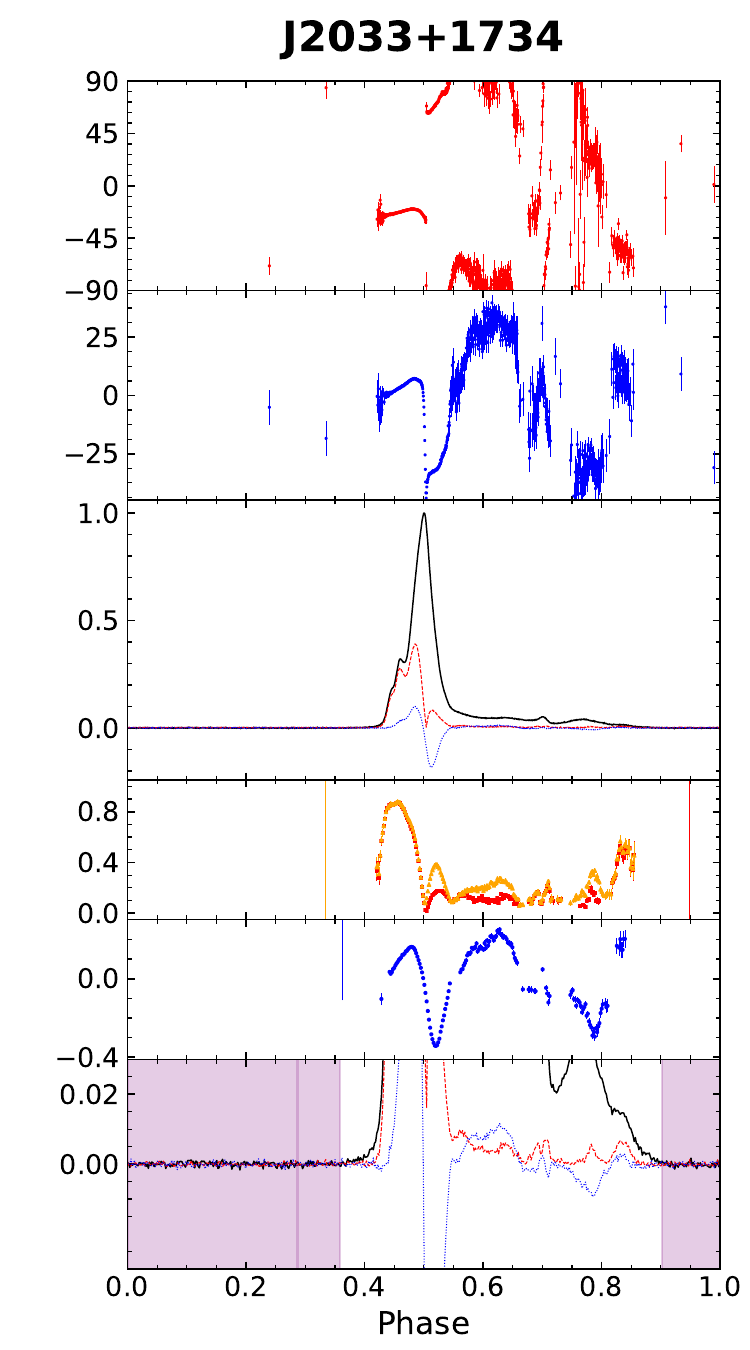}   
\end{minipage}
\begin{minipage}[t]{0.66\columnwidth}
    \centering
    \includegraphics[width=1\columnwidth]{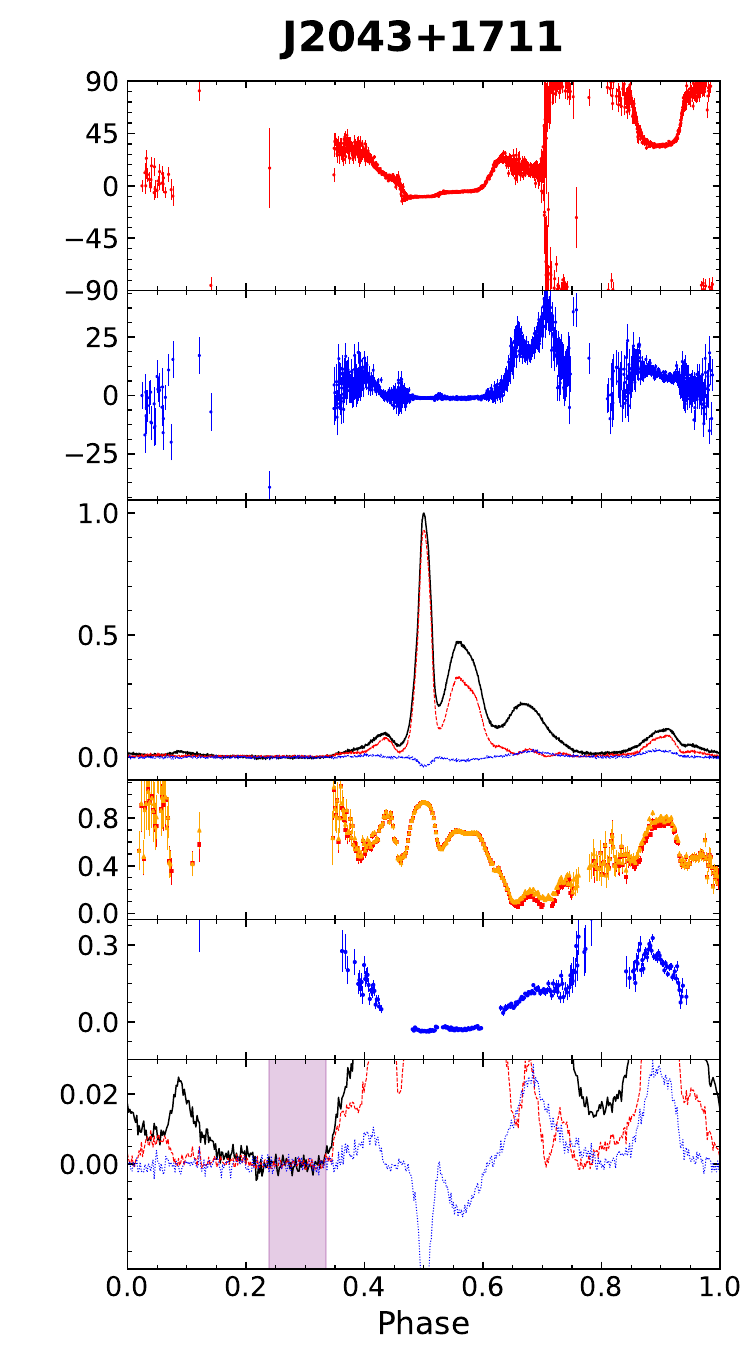}   
\end{minipage}
\begin{minipage}[t]{0.66\columnwidth}
    \centering
    \includegraphics[width=1\columnwidth]{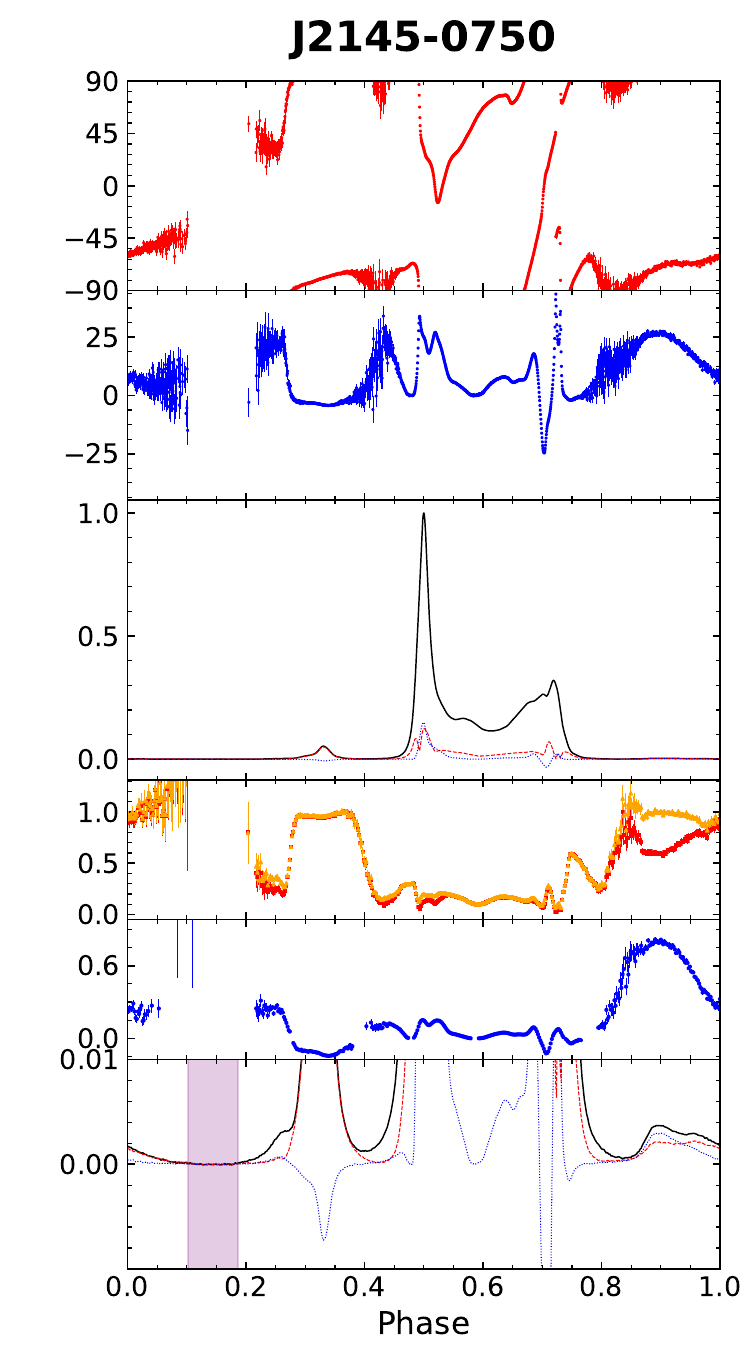}   
\end{minipage}
}
\caption{Continued.}
\end{figure*}

\begin{figure*}
\ContinuedFloat
\centering
\begin{minipage}[t]{0.66\columnwidth}
    \centering
    \includegraphics[width=1\columnwidth]{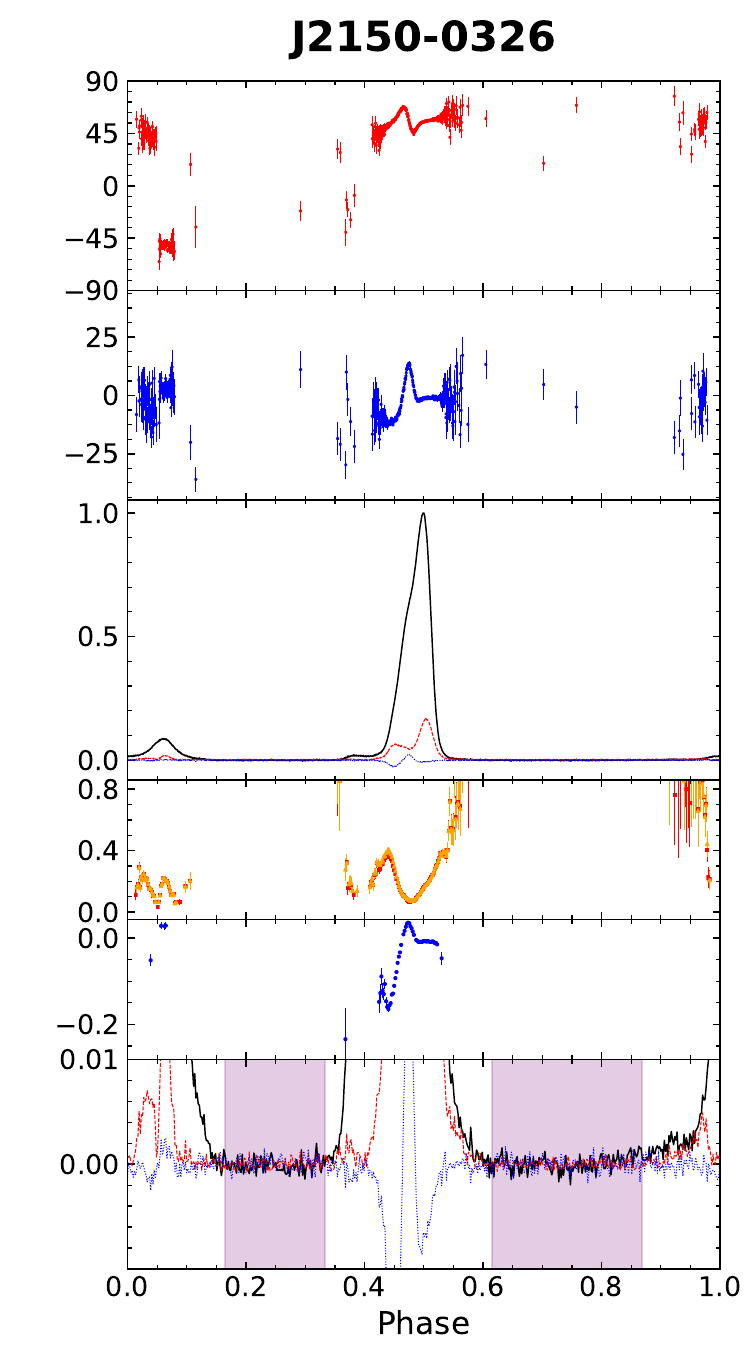}   
\end{minipage}
\begin{minipage}[t]{0.66\columnwidth}
    \centering
    \includegraphics[width=1\columnwidth]{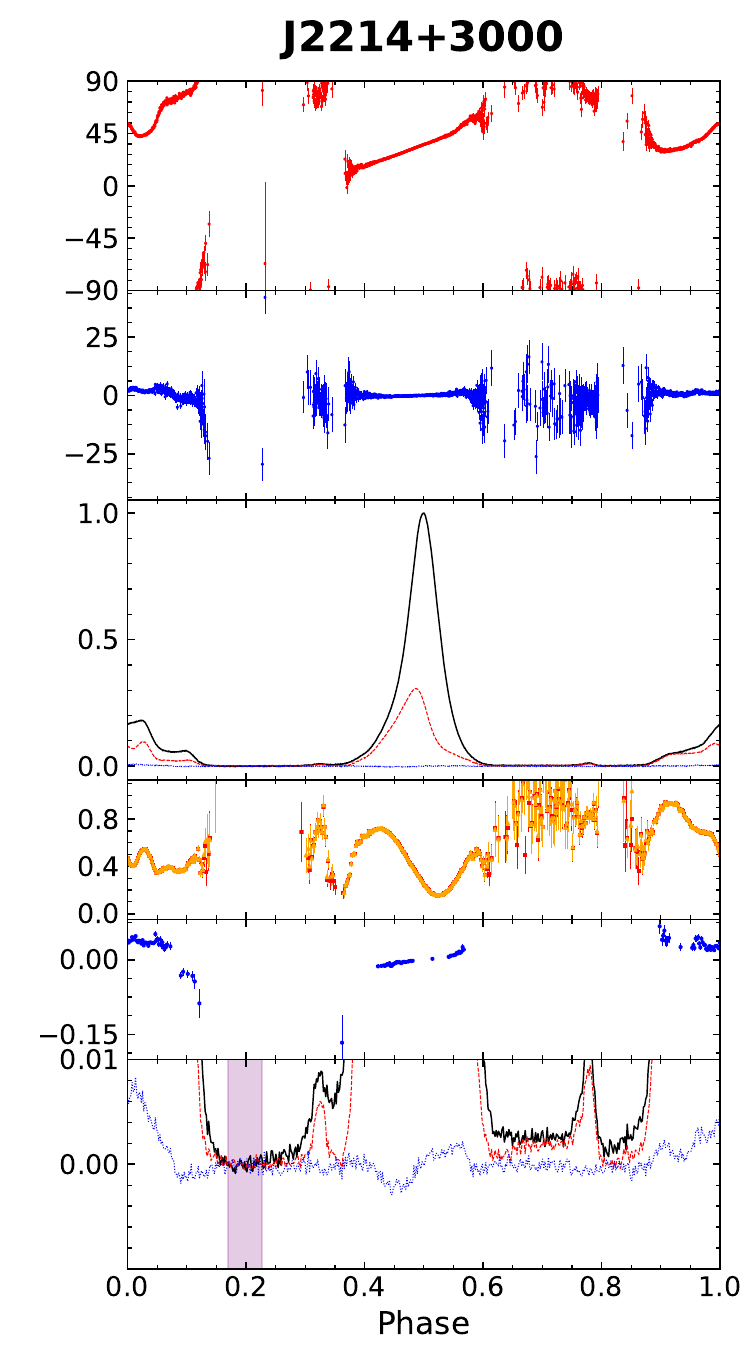}   
\end{minipage}
\begin{minipage}[t]{0.66\columnwidth}
    \centering
    \includegraphics[width=1\columnwidth]{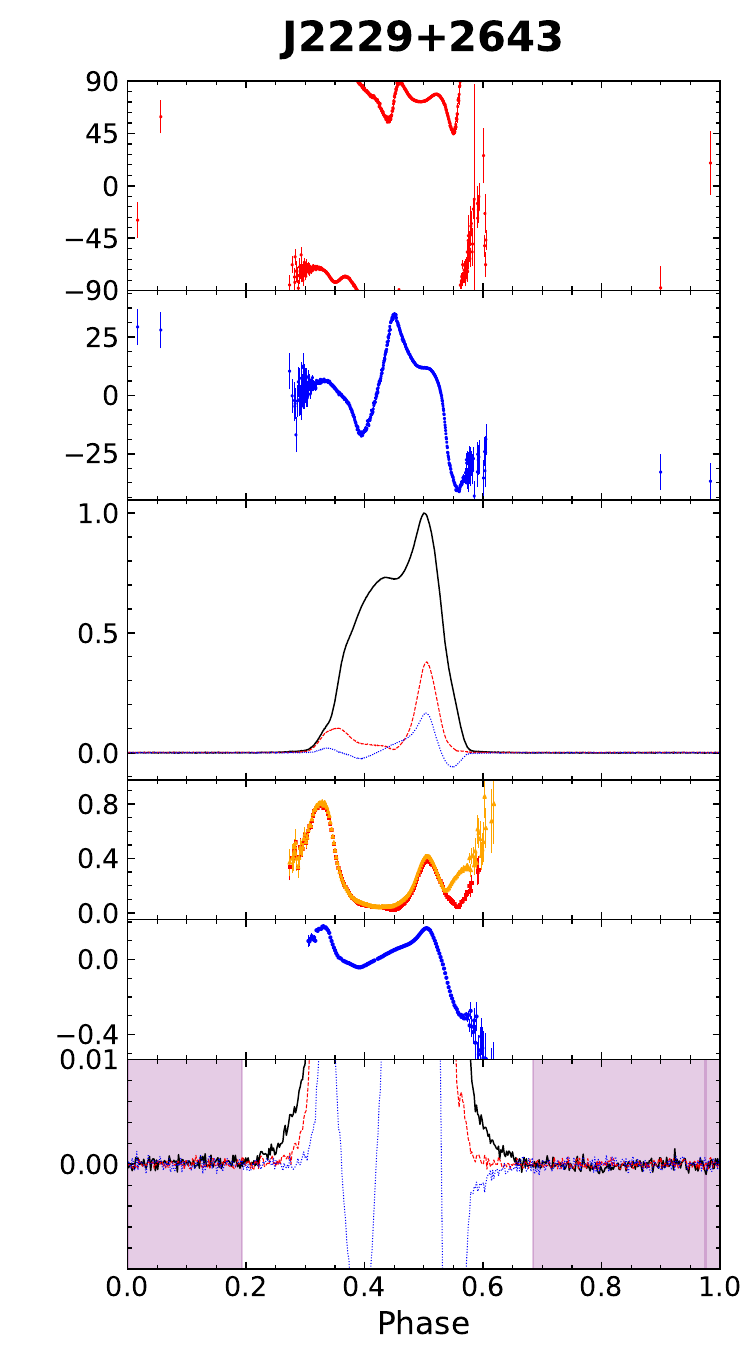}   
\end{minipage}\\
\begin{minipage}[t]{0.66\columnwidth}
    \centering
    \includegraphics[width=1\columnwidth]{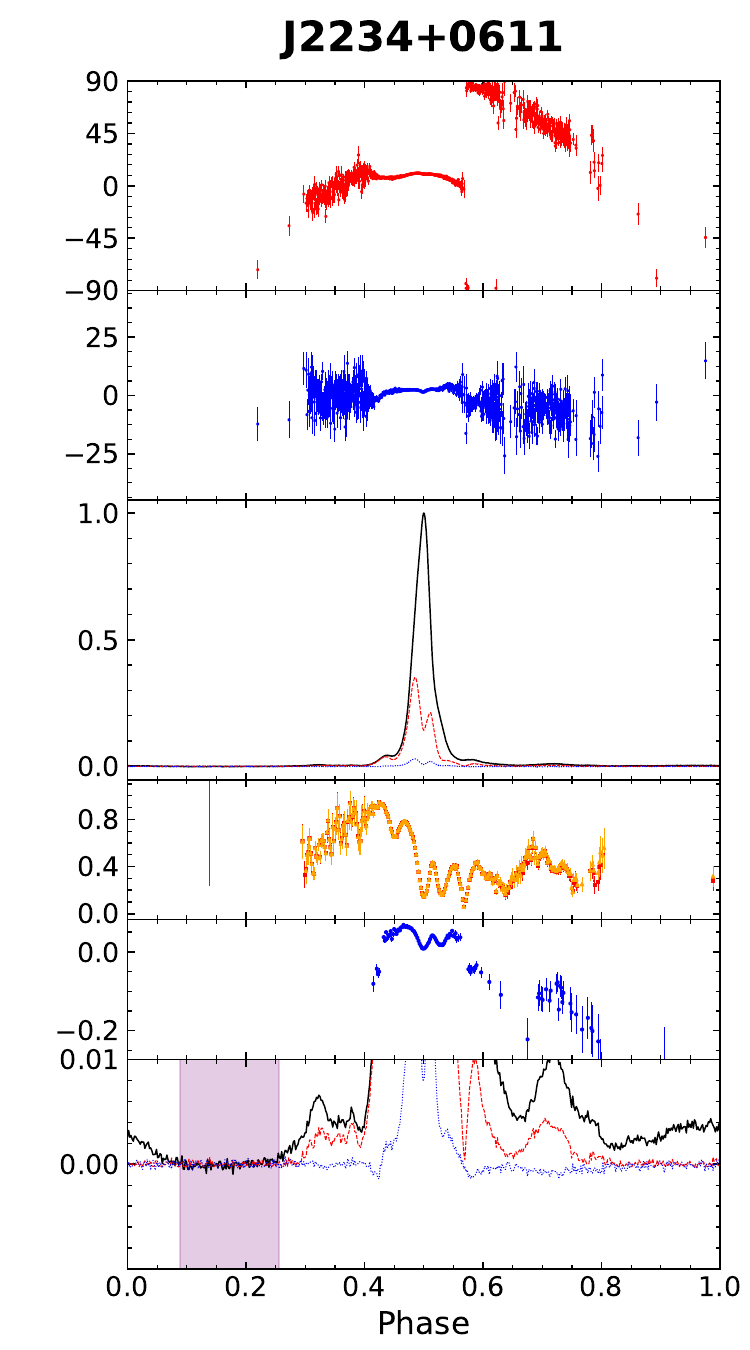}   
\end{minipage}
\begin{minipage}[t]{0.66\columnwidth}
    \centering
    \includegraphics[width=1\columnwidth]{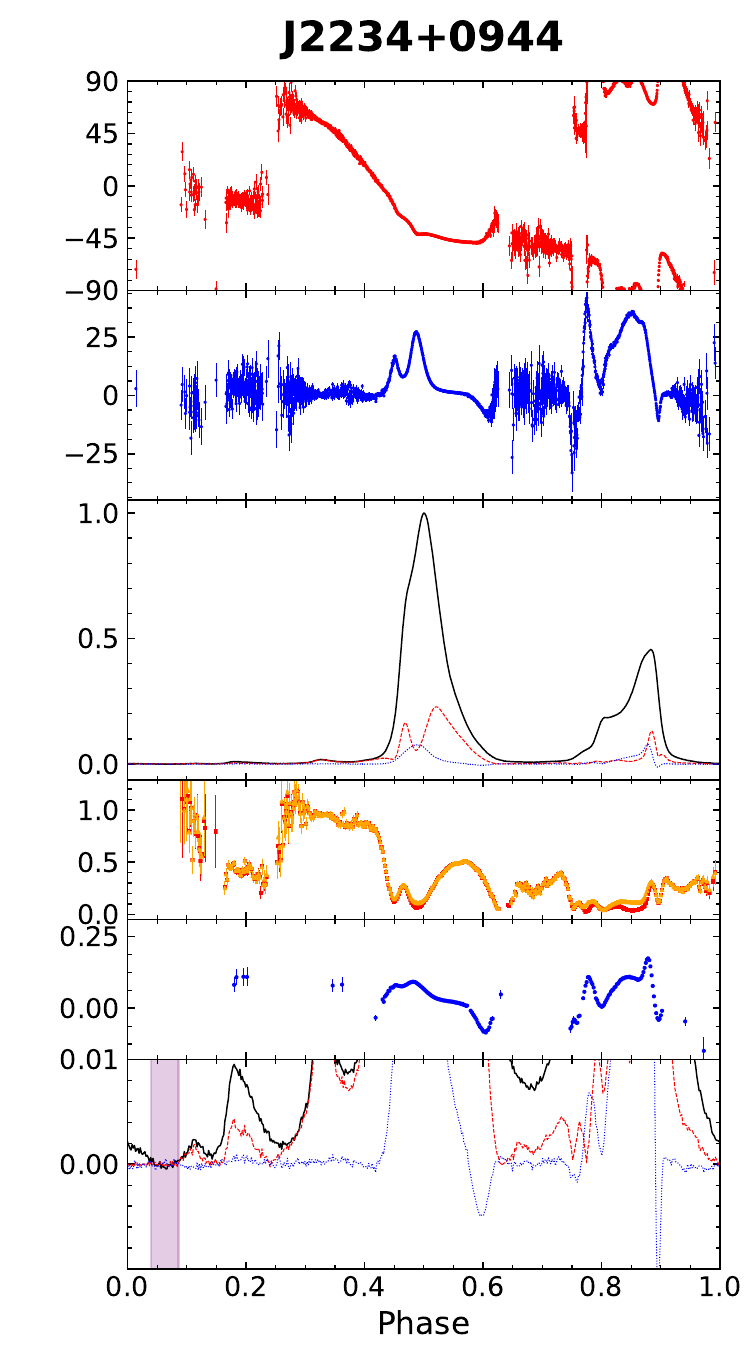}   
\end{minipage}
\begin{minipage}[t]{0.66\columnwidth}
    \centering
    \includegraphics[width=1\columnwidth]{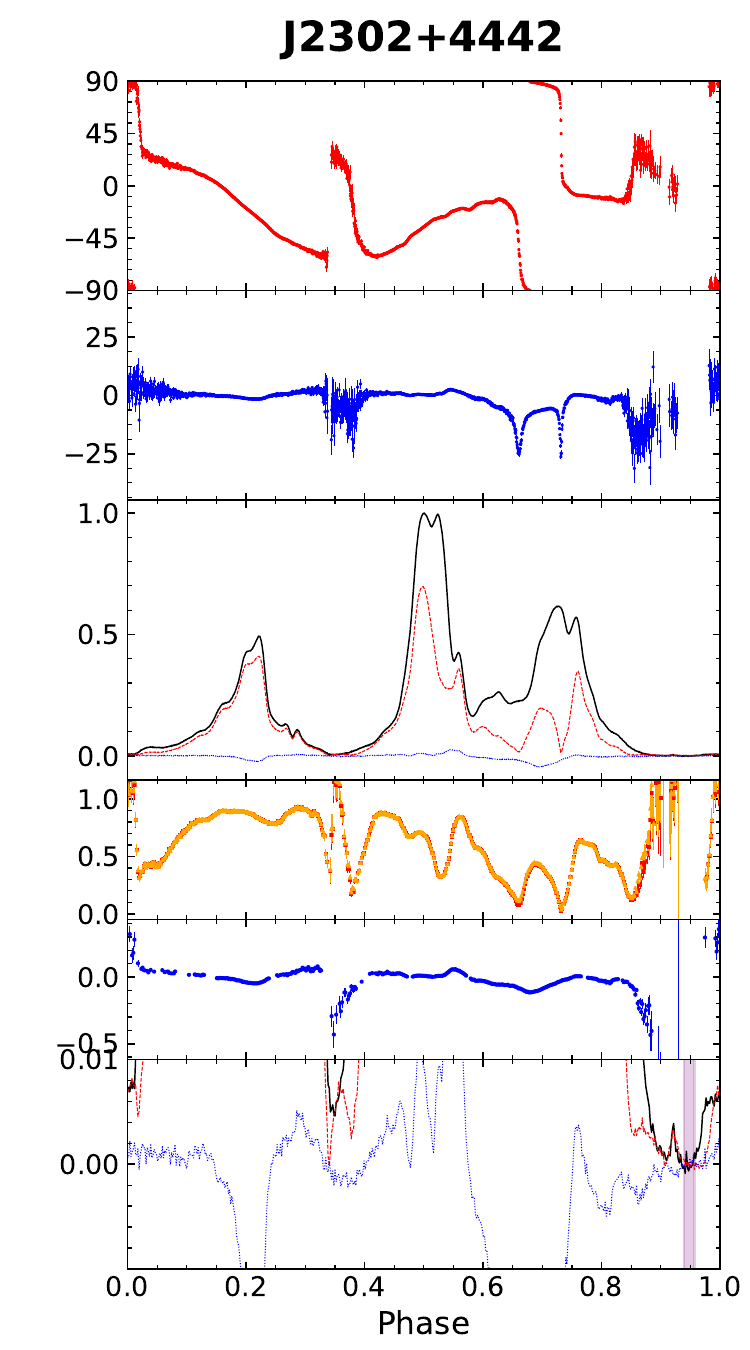}   
\end{minipage}
\caption{Continued.}
\end{figure*}

\begin{figure*}
\ContinuedFloat
\centering
\begin{minipage}[t]{0.66\columnwidth}
    \centering
    \includegraphics[width=1\columnwidth]{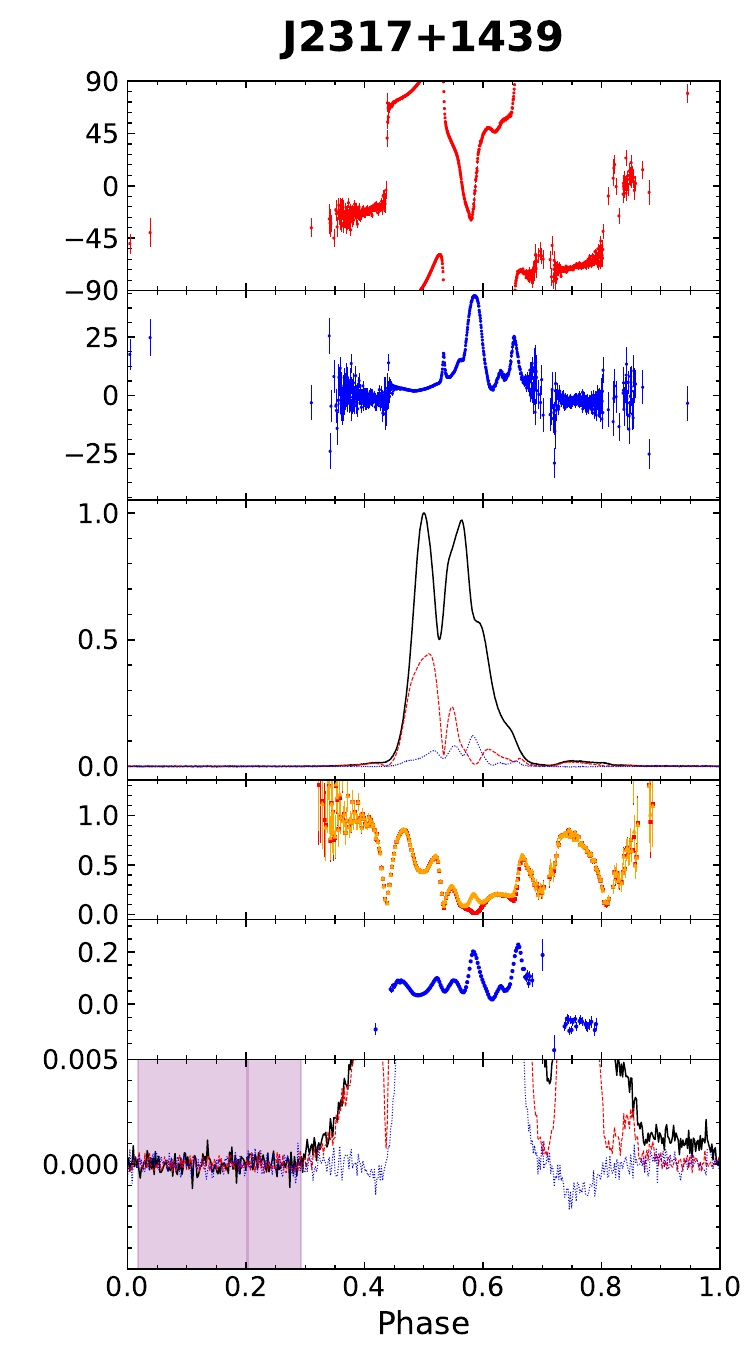}   
\end{minipage}
\begin{minipage}[t]{0.66\columnwidth}
    \centering
    \includegraphics[width=1\columnwidth]{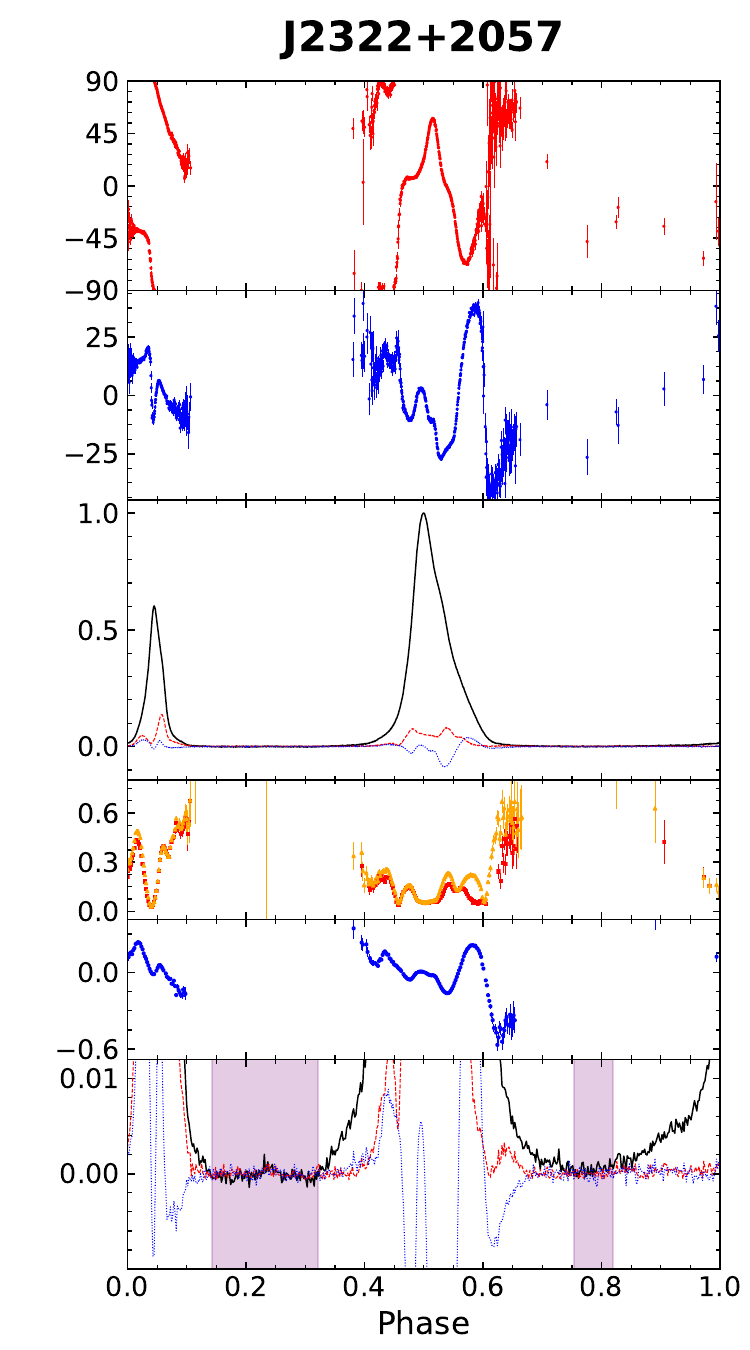}   
\end{minipage}
\caption{Continued.}
\end{figure*}

\end{appendix}

\label{lastpage}
\end{document}